\documentclass{aa}

\usepackage{txfonts}
\usepackage{graphicx}
\usepackage{amsmath}
\usepackage{pifont}
\usepackage{makecell}
\usepackage{subcaption}
\usepackage{comment}
\usepackage{lscape}
\usepackage{placeins}
\usepackage{xcolor} 

\usepackage{tabularx}
\usepackage[colorlinks=true,
            linkcolor=blue,
            citecolor=blue,
            urlcolor=blue,
            filecolor=blue]{hyperref}

\usepackage[utf8]{inputenc}
\usepackage{newunicodechar}
\newunicodechar{⁡}{}                  

\newcommand{\um}{$\mu$m}

\newcommand{\ha}{H$\alpha$}
\newcommand{\hb}{H$\beta$}

\newcommand{\paa}{Pa$\alpha$}
\newcommand{\pab}{Pa$\beta$}
\newcommand{\pag}{Pa$\gamma$}

\newcommand{\brg}{Br$\gamma$}
\newcommand{\hei}{He\,{\sc i}}

\newcommand{\oiii}{[O\,{\sc iii}]}

\newcommand{\namemiric}{{\it Pelias}} 
\newcommand{\namejades}{{\it Neleus}} 
\newcommand{\zspecmiric}{0.7050}
\newcommand{\zspecjades}{0.7474}
\newcommand{\hst}{{\it HST}}
\newcommand{\jwst}{{\it JWST}}

\begin{document}


\title{JWST Reveals Two Overmassive Black Hole Candidates in Dwarf Galaxies at $z\approx0.7$}
   \subtitle{Pushing Black Hole Searches into the Dwarf-Galaxy Regime}


     \author{E. Iani\inst{\ref{inst:1}}\fnmsep\thanks{email: edoardo.iani@ist.ac.at}
    \and P. Rinaldi\inst{\ref{inst:2}}
    \and A. Torralba\inst{\ref{inst:1}}
    \and J. Lyu \inst{\ref{inst:6}}
    \and R. Navarro-Carrera\inst{\ref{inst:5}}
    \and G. H. Rieke\inst{\ref{inst:6}}
    \and F. Sun\inst{\ref{inst:8}}
    \and C. Willott\inst{\ref{inst:7}}
    \and Y. Zhu\inst{\ref{inst:2}}
    \and A. Alonso-Herrero\inst{\ref{inst:13}}
    \and M. Annunziatella\inst{\ref{inst:14}}
    \and P. Bergamini\inst{\ref{inst:3}}
    \and K. Caputi\inst{\ref{inst:5}}
    \and M. Catone\inst{\ref{inst:15}, \ref{inst:22}}
    \and L. Colina\inst{\ref{inst:16}}
    \and R. Cooper\inst{\ref{inst:5}}
    \and L. Costantin\inst{\ref{inst:16}}
    \and A. Crespo G\'omez\inst{\ref{inst:2}}
    \and G. Desprez\inst{\ref{inst:5}}
    \and C. Di Cesare\inst{\ref{inst:1}}
    \and M. J. Hayes\inst{\ref{inst:4}}
    \and I. Jermann\inst{\ref{inst:17}, \ref{inst:18}}
    \and G. Kotiwale\inst{\ref{inst:1}}
    \and I. Kramarenko\inst{\ref{inst:1}}
    \and D. Langeroodi\inst{\ref{inst:19}}
    \and S. Mascia\inst{\ref{inst:1}}
    \and J. Matthee\inst{\ref{inst:1}}
    \and J. Melinder\inst{\ref{inst:4}}
    \and A. Muzzin\inst{\ref{inst:9}}
    \and B. Navarrete\inst{\ref{inst:1}}
    \and G. Noirot\inst{\ref{inst:2}}
    \and G. \"Ostlin\inst{\ref{inst:4}}
    \and F. Pacucci\inst{\ref{inst:8}, \ref{inst:21}}
    \and G. Rodighiero\inst{\ref{inst:15},\ref{inst:22}}
    \and M. Sawicki\inst{\ref{inst:11}}
    \and Y. Sun\inst{\ref{inst:6}}
    \and Z. Wu\inst{\ref{inst:8}}
    \and G. Yang\inst{\ref{inst:24}}
    }

\institute{
   Institute of Science and Technology Austria (ISTA), Am Campus 1, 3400 Klosterneuburg, Austria\label{inst:1}
   \and Space Telescope Science Institute, 3700 San Martin Drive, Baltimore, MD 21218, USA\label{inst:2}
   \and Steward Observatory, University of Arizona, 933 North Cherry Avenue, Tucson, AZ 85721, USA\label{inst:6}
   \and Kapteyn Astronomical Institute, University of Groningen, P.O. Box 800, 9700AV Groningen, The Netherlands\label{inst:5}
   \and Center for Astrophysics $\vert$ Harvard \& Smithsonian, 60 Garden St, Cambridge, MA 02138, USA\label{inst:8}
   \and NRC Herzberg, 5071 West Saanich Rd, Victoria, BC V9E 2E7, Canada \label{inst:7}
   \and Centro de Astrobiolog\'{\i}a (CAB), CSIC-INTA, Camino Bajo del Castillo s/n, E-28692 Villanueva de la Ca\~nada, Madrid, Spain\label{inst:13} 
   \and INAF – IASF Milano, Via A. Corti 12, I-20133 Milano, Italy\label{inst:14}
   \and INAF — OAS, Astrophysics and Space Science Observatory Bologna, Via Gobetti 93/3, 40129, Bologna, Italy\label{inst:3}
   \and Dipartimento di Fisica e Astronomia “G. Galilei”, Università di Padova, Vicolo dell’Osservatorio 3, 35122 Padova, Italy\label{inst:15}
   \and INAF, Osservatorio Astronomico di Padova, Vicolo dell’Osservatorio 5, 35122 Padova, Italy\label{inst:22}
   \and Centro de Astrobiología (CAB), CSIC-INTA, Ctra. de Ajalvir km 4, Torrejón de Ardoz 28850, Madrid, Spain\label{inst:16}
   \and Department of Astronomy, Oscar Klein Centre, Stockholm University, AlbaNova, 10691 Stockholm, Sweden\label{inst:4}
   \and Cosmic Dawn Center (DAWN), Denmark\label{inst:17}
   \and DTU-Space, Technical University of Denmark, Elektrovej 327, 2800 Kgs. Lyngby, Denmark\label{inst:18}
   \and DARK, Niels Bohr Institute, University of Copenhagen, Jagtvej 155A, 2200 Copenhagen, Denmark\label{inst:19}
   \and Department of Physics and Astronomy, York University, 4700 Keele St., Toronto, Ontario, M3J 1P3, Canada\label{inst:9} 
   \and Black Hole Initiative, Harvard University, 20 Garden St, Cambridge, MA 02138, USA\label{inst:21}
   \and Department of Astronomy \& Physics and Institute for Computational Astrophysics, Saint Mary's University, 923 Robie Street, Halifax, NS B3H 3C3, Canada\label{inst:11}
   \and Nanjing Institute of Astronomical Optics \& Technology, Chinese Academy of Sciences, Nanjing 210042, China\label{inst:24}}

   \date{Received XXX, 20XX}

    \titlerunning{Gemini cosmici distantes}
    \authorrunning{E.Iani, et al.}

   \abstract
    {We report the discovery and characterization of two compact galaxies, \namemiric\ and \namejades, at $z\approx0.71$ and $z\approx0.75$, identified in MACS J0416.1-2403 and GOODS-North. 
    Both exhibit unusual Spectral Energy Distributions (SEDs) with very blue rest-frame UV–optical emission and a steep rise toward near- and mid-infrared wavelengths. 
    JWST/NIRISS and JWST/NIRSpec spectroscopy shows strong rest-frame optical lines ([O\,{\sc iii}]$\lambda\lambda$4959,5007, H$\alpha$) with extreme equivalent widths ($\gtrsim 1000$~\AA), indicating young burst-dominated populations with low metallicities ($Z\approx0.1 - 0.4~ Z_\odot$), low dust attenuation ($A_V \approx 0.2$~mag) and stellar masses of $M_\star\approx10^7~\rm M_\odot$. 
    Nonetheless, JWST/MIRI photometry reveals a strong mid-infrared excess that cannot be explained by stellar populations or star-formation–heated dust alone, requiring a hot-dust component most naturally associated with a deeply embedded Active Galactic Nucleus (AGN). 
    SED modelling yields $L_{\rm bol}\approx10^{43.7 - 44.0}~\rm{erg~s^{-1}}$, implying $M_\bullet\approx10^{5.7 - 6.7}\rm~M_\odot$ (assuming Eddington-limited accretion $\lambda_{\rm Edd} \leq 1$). 
    Given the very low $M_{\star}$ of the hosts, this corresponds to black-hole–to–stellar mass ratios of $\approx6 - 60\%$, well above the extrapolation of local scaling relations. 
    The lack of X-ray detections suggests the accretion may be either heavily obscured or intrinsically X-ray weak. 
    Moreover, the overall SED of \namemiric\ and \namejades\ shows similarities to those of Blue Excess Hot Dust Obscured Galaxies (DOGs), suggesting that short-lived phases of buried, X-ray–weak AGN activity, where rapid black-hole growth dominates the MIR emission, can occur even in much lower-mass systems.
    Finally, their SEDs also show the characteristic “V-shaped” continuum seen in Little Red Dots (LRDs), though with the inflection occurring at significantly redder wavelengths compared to classic LRDs, consistent with recently proposed scenarios in which the turnover shifts to longer wavelengths for systems with bluer optical slopes.}

   \keywords{}

   \maketitle

\section{Introduction}
It is now widely recognized that galaxies host supermassive black holes (SMBHs) at their centres, whose growth is closely linked to that of their host galaxies. 
In the local Universe, this connection is reflected in well-established empirical relations between black-hole mass ($M_{\bullet}$) and host-galaxy properties, most notably the stellar bulge mass and stellar velocity dispersion, pointing to a long-term co-evolution between SMBHs and galaxies (e.g., \citealt{Kormendy_coevolution_2013}).

The existence of SMBHs at early cosmic times has been known for over two decades through the discovery of quasars in the distant Universe (e.g., \citealt{Fan_survey_2003}). 
Observations have revealed luminous quasars already at $z \gtrsim 6$, hosting black holes with masses exceeding $10^{8 - 9}~\rm M_{\odot}$ (e.g., ULAS1120+0641 at $z=7.085$; \citealt{Mortlock_luminous_2011}). 
Such extreme masses imply that highly efficient black-hole growth must have taken place within the first $\approx$1 Gyr after the Big Bang (e.g., \citealt{Volonteri_rapid_2005, Pacucci_growth_2015}). 
The presence of these massive objects at such early epochs poses significant challenges to models of black-hole seeding and early accretion, raising fundamental questions about how these systems formed and grew so rapidly (e.g., \citealt{Volonteri_formation_2012, Inayoshi_assembly_2020}).

Despite decades of observational and theoretical progress (see \citealt{Alexander_what_2025} for a recent review), SMBH–galaxy scaling relations remain constrained almost exclusively to the massive-galaxy regime ($M_{\star}>10^{9.5 - 10}~\rm M_{\odot}$), making the incidence of such black holes in dwarf galaxies (DGs, $M_{\star}\lesssim 10^{9.5}~\rm M_{\odot}$) almost unknown. 
Extending these studies to low-mass galaxies is essential for a comprehensive understanding of black-hole growth, as dwarf systems are expected to preserve the clearest imprints of black-hole seeding physics (e.g., \citealt{Volonteri_formation_2012, Greene_low-mass_2012, Reines_observational_2016, Reines_hunting_2022, Scoggins+25}). 
In massive galaxies, black holes have typically undergone substantial growth via repeated mergers and prolonged accretion (e.g., \citealt{DiMatteo_energy_2005, Springel_simulations_2005}). 
In contrast, DGs typically assemble most of their mass through {\it in situ} star formation and are subject to strong stellar feedback (e.g., \citealt{Behroozi_UNIVERSEMACHINE_2019}), conditions that limit subsequent black hole growth and may preserve information about the initial seed population (e.g., 
\citealt{Volonteri_formation_2010, Greene_low-mass_2012, Reines_observational_2016, Habouzit_blossoms_2017}).

Identifying accreting black holes in DGs has nevertheless proven exceptionally difficult in the past. 
As emphasized by \citet{Reines_observational_2016}, traditional optical spectroscopic searches are strongly biased against low-mass, highly obscured, and actively star-forming systems, while X-ray and radio searches suffer from severe contamination by stellar-mass X-ray binaries, supernova remnants, and compact star-forming regions (e.g., \citealt{Reines_dwarf_2013, Trump_biases_2015, Mezcua+18, Cann_limitations_2019}). 
As a consequence, the known population of accreting black holes in DGs is heavily incomplete and dominated by nearby, relatively unobscured systems, with only a handful of confirmed detections beyond the local Universe.

The infrared provides a powerful and complementary avenue for identifying accreting black holes, as dust heated by the central engine efficiently reprocesses the primary radiation and re-emits it as a characteristically red, power-law continuum (e.g., \citealt{Barvainis_hot_1987, Lacy_obscured_2004, Stern_mid-infrared_2005, Alonso-Herrero_infared_2006, Caputi_generalized_2013}). 
Infrared emission is comparatively insensitive to line-of-sight obscuration, enabling the detection of heavily dust-enshrouded active galactic nuclei (AGNs) that are difficult to identify at shorter wavelengths. 
In moderate- to high-mass galaxies, mid-IR color selection has proven highly effective (e.g., \citealt{Donley_spitzer_2008, Assef_low-resolution_2010, Hainline_spectroscopic_2014, Lyu+24}). 
However, careful multi-wavelength studies have demonstrated that mid-infrared AGN selection in DGs at the resolution of pre-\jwst\ facilities can be uncertain.
Compact, low-metallicity starburst galaxies can heat dust to temperatures higher than normal star-forming galaxies, yielding red IR colors that can overlap with classical AGN selection criteria, while the coarse spatial resolution of surveys (e.g., \emph{WISE}; \citealt{Wright_wise_2010}) led to frequent source confusion and mis-association (e.g., \citealt{Izotov_star-forming_2011, Izotov_dust_2014, Remy-Ruyer_linking_2015, Hainline+16, Reines_observational_2016, Lupi_difficulties_2020, Sturm_star-forming_2025}).

Nonetheless, in the first years of \jwst\ operations (\citealt{Gardner_jwst_2023}), the combination of Near Infrared Camera \citep[NIRCam;][]{Rieke+23} and Near Infrared Spectrograph’s Micro-shutter Array (NIRSpec/MSA; \citealt{Jakobsen_NIRSpec_2022}) has enabled a qualitatively new view of black-hole growth. 
By extending sensitivity to faint systems, \jwst\ has revealed accreting black holes in regimes of luminosity, host mass, and obscuration that were largely inaccessible to previous facilities, especially at early cosmic epochs (e.g., \citealt{Maiolino_jades_2024}). 
Within this newly revealed population of faint, high-redshift sources, a particularly distinctive class has emerged: the so-called \textit{Little Red Dots} (LRDs; \citealt{Matthee+24}).
These sources are compact systems characterized by steeply rising optical continua and blue rest-frame UV colors (\citealt{Labbe+23, Matthee+24, Barro+24, Furtak+23, Greene+24, Kokorev+24, Kocevski+25}) that defy straightforward classification within standard AGN or galaxy populations, and in several cases show broad permitted emission lines (e.g., \citealt{Hviding_RUBIES_2025}) consistent with accretion onto massive black holes, opening a new window on black hole growth in a regime that was previously unexplored. 
The puzzling nature of LRDs, and the physical origin of their unusual “V”-shaped (\citealt{Setton_little_2025}) spectral energy distributions (SEDs), has become a central topic of debate, not only in the high-$z$ community but also in light of the recent discovery of low-redshift and even local analogues (e.g., \citealt{Juodzbalis_JADES_2024, Chen+25, Ji_lord_2025, Ma_counting_2025, Lin_discovery_2025, Loiacono_big_2025, Rinaldi+25b}).

Complementing these discoveries, the Mid-Infrared Instrument \citep[MIRI;][]{Rieke+15, Wright+23}, and in particular its imager \citep[MIRIM;][]{Bouchet+15, Dicken+24}, extends \jwst’s\ spectral coverage into the 5.6–25.5~\um\ mid-infrared regime, providing access to obscured nuclear emission that is {\it invisible} at shorter wavelengths. 
Crucially, \jwst’s\ angular resolution at these wavelengths enables the separation of compact nuclear components from host-galaxy star formation, overcoming the blending and confusion that limited previous infrared surveys, especially in low-mass systems. 
In this broader effort to uncover previously hidden phases of black-hole growth, MIRI has already demonstrated its transformative role by revealing sources that would otherwise remain entirely missed and/or misclassified.  
A striking example is \textit{Virgil}, a low-mass Lyman-$\alpha$ emitter at $z \approx 6.6$, whose extremely red rest-frame optical–near-infrared color, ${\rm (F444W - F1500W)} \approx 3$~mag, emerges only when MIRI is considered and points to the presence of a heavily dust-obscured (type~II) AGN \citep{Iani+25, Rinaldi+25}.

In this work, we exploit the unique capability of \jwst\ to report the discovery of two heavily dust-obscured accreting black holes hosted by DGs at $z \approx 0.7$. 
These galaxies have extremely low stellar masses, placing them among the least massive systems known to host actively growing black holes. 
They therefore probe a region of parameter space--low stellar mass, significant dust obscuration, and low-to-intermediate redshift--that has remained poorly explored to date.  
We demonstrate that, once mid-infrared constraints from MIRI are included, the observed red infrared continua cannot be reproduced by star formation alone. Instead, the data require the presence of a deeply embedded AGN, providing direct evidence that supermassive black-hole growth is already underway in very low-mass galaxies. These systems thus offer a rare glimpse into black hole growth in the dwarf-galaxy regime.

Throughout this paper, we assume a flat $\Lambda$-CDM cosmology with $\Omega_{\rm M} = 0.3$, $\Omega_\Lambda = 0.7$, and a Hubble constant $H_0 = 70\rm~ km~ s^{-1}~ Mpc^{-1}$. 
We adopt the AB magnitudes \citep{Oke+83}. 
All stellar mass and SFR estimations assume a universal Chabrier initial mass function \cite[IMF;][]{Chabrier+03}.

\section{Targets properties}
\label{sec:data}
In the following sections, we describe our targets, briefly outline the dataset used in this work, and present the extraction of the multi-wavelength photometry of \namemiric\ and \namejades, together with their morphological analysis. For a more detailed discussion of the dataset collection for both objects and the estimate of their spectroscopic redshift, we refer the reader to \S~\ref{app:dataset}. 

Due to their remarkable similarity but great distance in the sky, we refer to them as \namemiric\ and \namejades\footnote{According to Greek Mythology, Pelias and Neleus were twin sons of Tyro and Poseidon who were abandoned at birth but survived to adulthood. After avenging their mother by killing her cruel stepmother, Sidero, the brothers fought over the throne of Iolcus. Pelias won, banishing Neleus to Messenia.}.

\subsection{\namemiric\ and \namejades}
The first object we discovered, \namemiric, was found looking for potential LRD-like objects within the deep MIRIM 7.7~$\mu$m and 10~$\mu$m observations carried out by the \jwst\ Cycle 3 program \textit{MIRI in Clusters} \cite[MIRIC, JWST PID 5578; Iani et al. \textit{in prep.},][]{Iani+24jwst.prop.5578} that covered the central 5.9 arcmin$^2$ region of MACS J0416.1-2403 \cite[hereafter M0416; e.g.][]{Ebeling+01}.
\namemiric\ was identified based on its very red F277W - F1000W color ($\approx 4.8$~mag), blue F090W - F277W color ($\approx -1.1$~mag) and compact morphology. 

\namemiric\ is located at $z=$~\zspecmiric\ (see \S~\ref{app-sec:miric_redshift}) behind the lensing cluster M0416 at $\rm RA = 04^h:16^m:13.^s5711$, $\rm DEC = -24^\circ:05':01.''124$. 
Due to its peripheral position on the sky-plane with respect to the centre of the lensing cluster, \namemiric\ is minimally magnified by M0416 having a magnification factor $\mu = 1.18 \pm 0.01$ (inferred adopting the M0416 lensing model by \citealt{Bergamini+23}). 
We note that this object was already reported in the \textit{CAnadian NIRISS Unbiased Cluster Survey} \cite[CANUCS;][]{Willott+22}. 

To properly characterise \namemiric, we collected a panchromatic dataset spanning from the X-rays to the radio domain.
In particular, we gathered X-ray data from the {\it Chandra} observatory \citep{Ogrean+15}, \hst\ ACS imaging in 7 bands \citep{Postman+12,Lotz+17,Steinhardt+20}, \jwst\ NIRISS Wide Field Slitless Spectroscopy (WFSS) in 4 filters \citep{Sarrouh+25}, \jwst\ NIRCam imaging in 15 bands \citep{Windhorst+17jwst.prop.1176, Sarrouh+25, Fu+25, Iani+23jwst.prop.3538} and WFSS in 6 medium bands \citep{Sun+23jwst.prop.2883, Iani+23jwst.prop.3538}, \jwst\ MIRI imaging at 7.7~$\mu$m and 10~$\mu$m (Iani et al., {\it in prep}.), {\it Herschel} imaging at 100~$\mu$m and 160~$\mu$m \citep{Egami+10}, ALMA band 4 and band 6 observations \cite[e.g.][]{Gonzalez-Lopez+17a} and JVLA observations at 6 cm and 13 cm \cite{Heywood+21}.

The second object, \namejades, was found while searching for similar objects to \namemiric\ in the \textsc{Dawn JWST Archive} (\textsc{DJA})\footnote{\url{https://dawn-cph.github.io/dja/}}.
\namejades\ is located in the {\it Great Observatories Origin Deep Survey - North} \cite[GOODS-N,][]{Giavalisco+04}
at $\rm RA = 12^h:36^m:57.^s0409$, $\rm DEC = 62^\circ:08':16.''839$ with a reported spectroscopic redshift of $z_{\rm spec} = $~\zspecjades.
As in the case of \namemiric, \namejades\ is characterised by a strong red F277W - F1280W color ($\approx 5.1$~mag), a blue F090W - F277W color ($\approx -1.2$~mag) and appears compact.
The object was also listed in the JADES catalog \cite[ID 1036453,][]{D'Eugenio+25}. 

Also in the case of \namejades, we collected multi-wavelength dataset including {\it Chandra} observations \cite[e.g.][]{Xue+16}, \hst\ imaging in 9 filters \cite[e.g.][]{Illingworth+16, Whitaker+19}, \jwst\ NIRSpec MSA observations \citep{2017jwst.prop.1211F, Maseda+24}, \jwst\ NIRCam imaging in 9 filters \citep{Eisenstein+17jwst.prop.1181}, \jwst\ MIRI imaging at 7.7~$\mu$m and 12.8~$\mu$m \cite{Eisenstein+17jwst.prop.1181}, {\it Spitzer} imaging at 16~$\mu$m and 24~$\mu$m \cite[e.g.][]{Liu+18, Dickinson+03}, {\it Herschel} imaging at 100~$\mu$m and 160~$\mu$m \citep{Magnelli+13} and JVLA observations at 20 cm \citep{Owen+18}.

\subsection{Photometry}
\label{sec:photometry}
We extracted the photometry of our targets using \textsc{sep} \citep{Barbary+16}, a \textsc{Python} version of the \textsc{SourcExtractor} software \citep{Bertin+96}.  
We measured our targets photometry within Kron apertures \citep{Kron+80}, replicating standard \textsc{SourcExtractor} settings, i.e. $K = 2.5$ and a minimum radius $r = 1.75$~pixels below which a standard circular aperture of $r=0.''1$ is adopted instead. 

To estimate the errors, we drew random apertures (with the same Kron parameters of the apertures used to extract the flux) in a $5''\times5''$ sky region surrounding our targets after masking all close-by sources.
We adopted as error the standard deviation of all the fluxes estimated within the random sky apertures.
To properly account for systemic uncertainties in the absolute flux calibration of the \hst~and \jwst~data, we imposed a minimum error of 5\% on the photometry in all bands. 

Then, we applied aperture correction to the extracted fluxes (and corresponding errors). 
To do so, we computed the fraction of the missing
light outside the Kron apertures \cite[e.g.,][]{Whitaker+11, Weaver+23, Kokorev+24} in comparison to the curve of growth of the different PSFs.
For the \jwst~ filters, we employed the PSF models by \textsc{STPSF} \citep{Perrin+14}, while for \hst~ we resorted to the curve of encircled energy reported by \textsc{STScI}\footnote{\url{https://www.stsci.edu/hst/instrumentation/}}.

Finally, we corrected the derived photometry for Galactic extinction adopting an $R_V = 3.1$ and assuming the color excess $E(B-V)$ at the targets coordinates  \citep{Schlafly+11} as reported by the IRSA\footnote{\url{https://irsa.ipac.caltech.edu/applications/DUST/}} dust maps.
For \namemiric\ we found an average $E(B-V)= 0.0356 \pm 0.0004$, while for \namejades\ we inferred $E(B-V) = 0.0097 \pm 0.0004$.   
Knowing $E(B-V)$, we derived the correction per passband considering the extinction curve by \cite{Gordon+23} by means of the \textsc{dust\_extinction}\footnote{\url{https://dust-extinction.readthedocs.io/en/latest/}} Python library.
We report the corrections applied $f_{\rm ext}$ in Table~\ref{app-tab:photometry_combined} along with the observed galaxy photometry for both sources. 
We note that for \namemiric, the reported photometry is not corrected for the lensing magnification factor $\mu$ (see \S~\ref{sec:z_mu}).

\subsection{Morphology}
\label{sec:compactness}
\begin{figure*}
    \centering
    \includegraphics[width=.9\linewidth]{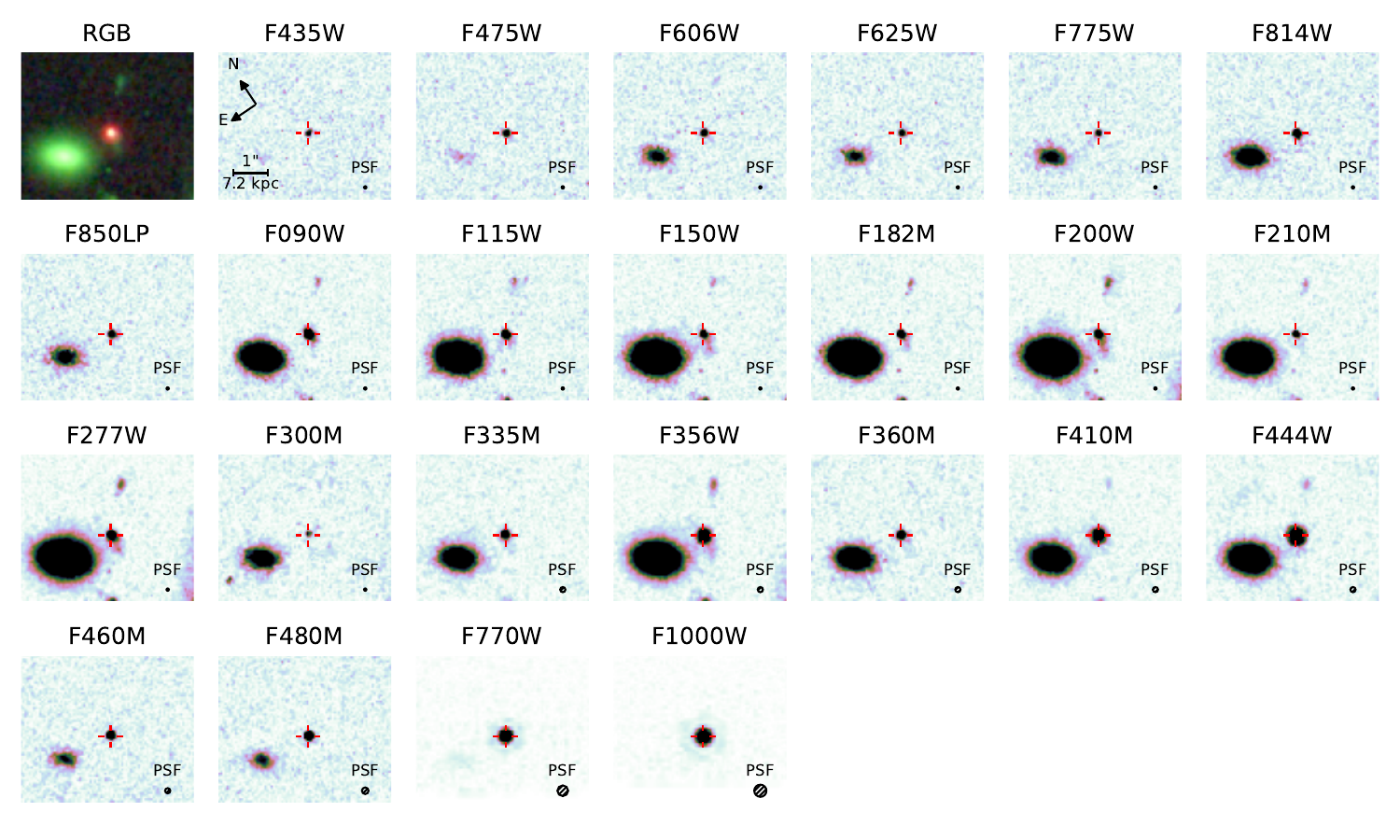}\\
    \includegraphics[width=.8\linewidth]{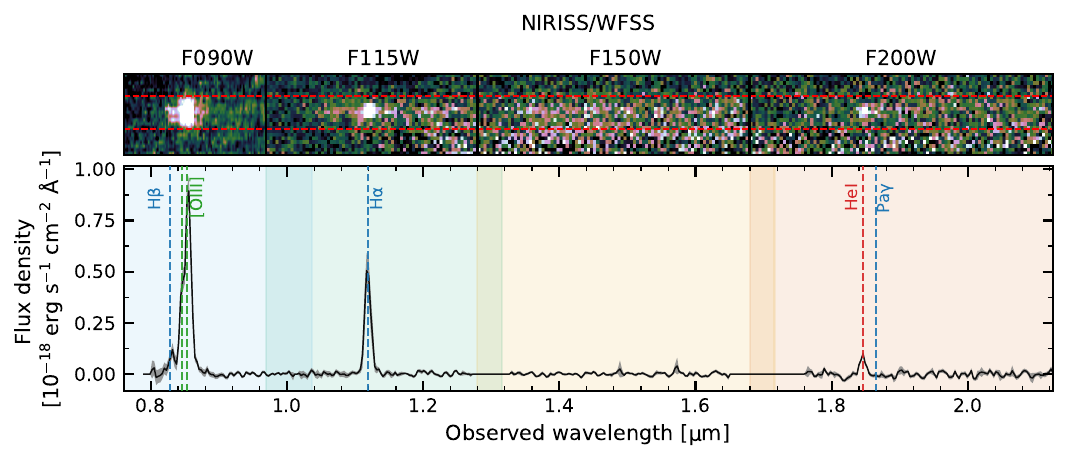}
    \caption{\textbf{Top panel}: $5'' \times 4''$ cutouts of the 24 \hst/ACS and \jwst/NIRCam and MIRI available filters for this study centred at the coordinates of \namemiric\ (red dot marker, $RA = 64.05654$~deg, $DEC = -24.08365$~deg). The RGB cutout was made combining all filters with effective wavelength below $1.2~\mu$m (observed) for the B channel, between $1.2~\mu$m and $3~\mu$m for the G channel, and $> 3~\mu$m for the R channel. In the bottom right corner of each cutout we show the size of the corresponding PSF. \textbf{Bottom panel}: 2D (top) and 1D (bottom) spectra of \namemiric\ as obtained from NIRISS WFSS observations in the F090W, F115W, F150W and F200W filters \citep{Willott+22, Sarrouh+25}. In the 1D spectrum, we highlight the wavelength range covered by the different NIRISS filters with different background colors. We also point out the position of the main rest-frame optical features found in the spectrum. }
    \label{fig:sed_cutouts_miric}
\end{figure*}

\begin{figure*}
    \centering
    \includegraphics[width=.9\linewidth]{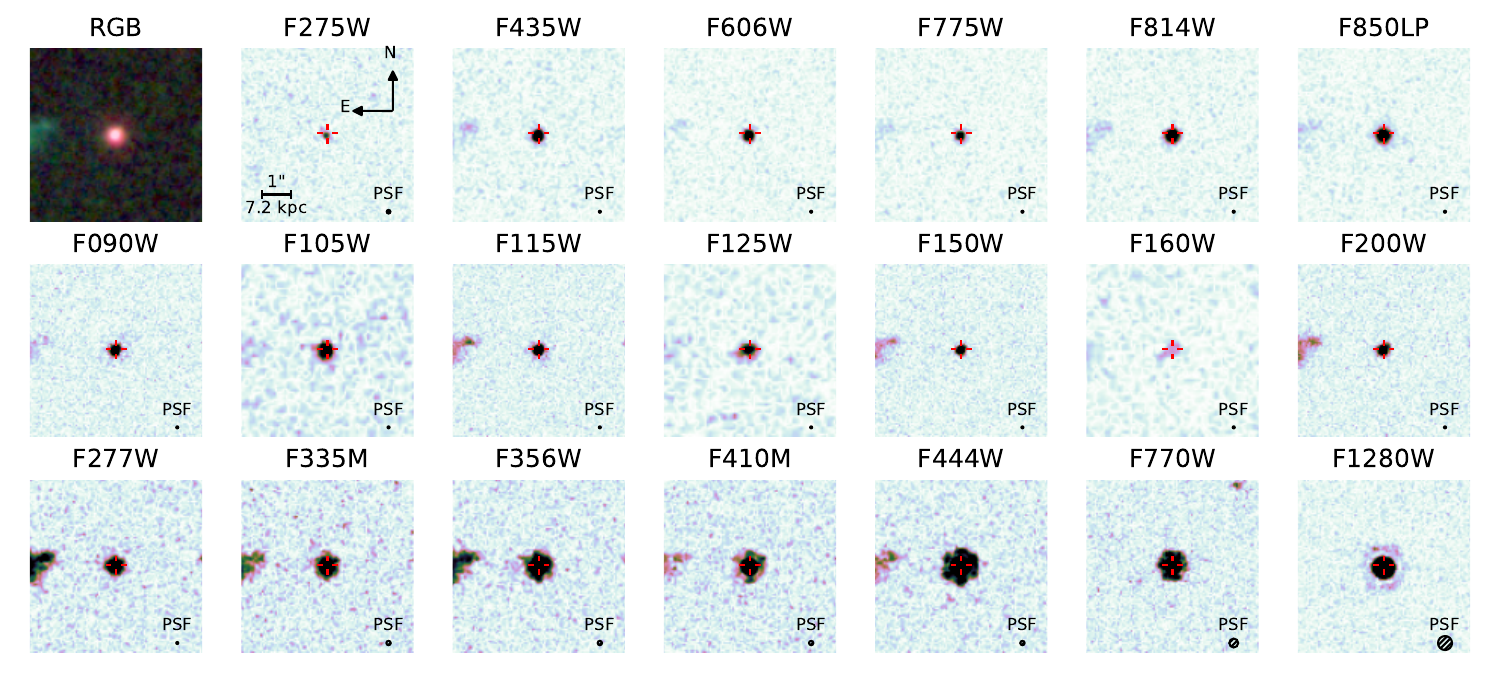}\\
    \includegraphics[width=.8\linewidth]{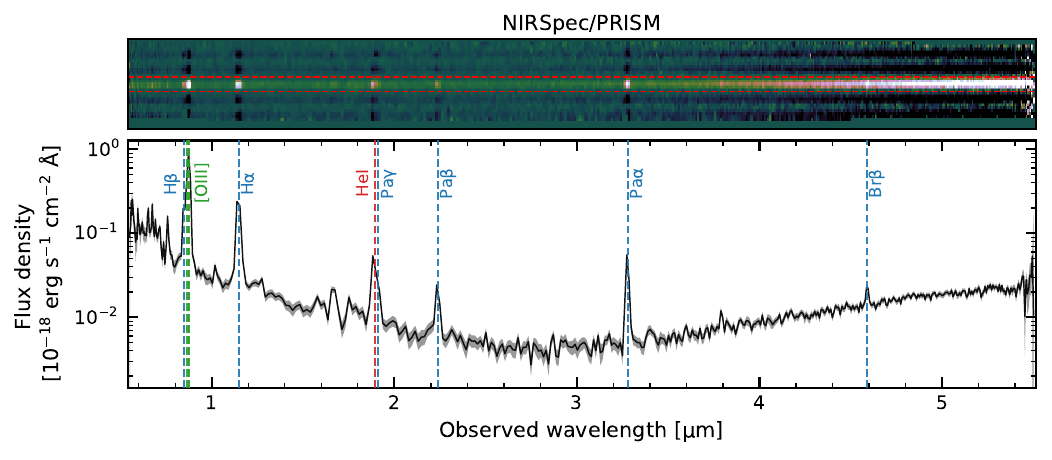}
    \caption{Similar to Figure~\ref{fig:sed_cutouts_miric} but for \namejades\ (red dot marker, $\rm RA = 189.23767$~deg, $\rm DEC = +62.13801$~deg). The cutouts size is $5'\times 5'$ while the 2D and 1D spectra shown are from the publicly available NIRSpec PRISM observations \citep{Maseda+24}.}
    \label{fig:sed_cutouts_jades}
\end{figure*}

As revealed by the cutouts of the available filters presented in Figures~\ref{fig:sed_cutouts_miric},\ref{fig:sed_cutouts_jades}, \namemiric\ and \namejades\ appear compact at most wavelengths. 

We performed 2D-modelling of the surface brightness (SB) of our targets in all available JWST bands using {\sc pysersic} \citep{Pasha+23}, testing several parametric descriptions including a point-like source, a single S\'ersic profile, a S\'ersic + point-like source and a double-S\'ersic model. 
Model selection was based comparing the $\chi^2_{\rm red}$ and the Bayesian Information Criterion \cite[BIC, see][]{Schwarz+78}, enabling a consistent comparison between models of different complexity \cite[e.g.,][]{Liddle+07}.
We report the best-fit parameters in Table~\ref{app-tab:2d_morph}.
For a selection of filters we also show their best-fit models and residuals in Figure~\ref{app-fig:2d_morph_fit}.

At short NIRCam wavelengths, where the angular resolution is highest, the surface-brightness distributions of both targets are best described by a double-S\'ersic model. 
This decomposition reveals two structural components: a compact central component with circularized effective radius $r_{e,\rm circ} \simeq 0.4 - 0.8$~pkpc, and a more extended component with characteristic sizes of $r_{e,\rm circ} \simeq 1 - 3$~pkpc. 
In both galaxies, the two S\'ersic components exhibit low S\'ersic indices ($n \sim 1$), indicating exponential light profiles rather than a classical bulge--disk configuration. 
The extended component shows a flattened morphology ($q \sim 0.6 - 0.7$), consistent with an inclined stellar disk, while the compact component appears significantly rounder ($q \gtrsim 0.8$). 
The presence of exponential profiles at both spatial scales therefore suggests disk-dominated systems hosting a centrally concentrated exponential component, rather than a prominent spheroidal bulge.

Toward longer wavelengths (observed-frame $\gtrsim 3~\mu$m), the preferred description transitions to a single S\'ersic profile with circularized effective radii of order $r_{e,\rm circ} \sim 0.5 - 0.8$~pkpc. 
At these wavelengths the emission is increasingly dominated by hot dust continuum rather than by direct stellar light (see \S~\ref{sec:sed_fitting}). 
The apparent morphological simplification therefore reflects both the progressively broader point-spread function and the fact that the dust emission traces the unresolved combination of the central and disk components. 
Indeed, when the NIRCam images are convolved to a MIRI-like resolution, the fits similarly favour a single S\'ersic model, confirming that the loss of multi-component structure is primarily driven by resolution effects.

Overall, the results indicate that both galaxies are intrinsically composed of a compact exponential component embedded within a disk-like structure.
The structural properties derived closely resemble those reported for star-forming galaxies at similar redshift presented in recent literature \cite[e.g.,][]{Calabro+17, Genin+25}.

\section{On the Nature of \namemiric\ and \namejades : insights from photometry, spectroscopy and SED-fitting}

The striking feature of our targets is their SED, showing blue colors at rest-frame short wavelengths (NUV-to-optical) and a steep red rest-frame NIR-to-MIR slopes, see Figures~\ref{fig:sed_cutouts_miric}, ~\ref{fig:sed_cutouts_jades}.
In the case of \namemiric, we measured F090W - F277W~$ \simeq -1.1$~mag while F277W - F1000W~$\simeq 4.8$~mag. 
We found similar colors also for \namejades, with F090W - F277W~$ \simeq -1.2$~mag and F277W - F1280W~$\simeq 5.1$~mag.
Such a combination of very blue optical colors and extremely red NIR–MIR emission suggests a transition between stellar-dominated and dust-dominated regimes, potentially indicating the coexistence of unobscured stellar populations and strong hot-dust emission. 

The far-UV emission (FUV; $\lesssim 2000$~\AA\ rest-frame) would provide a direct probe of the unobscured recent star formation and dust attenuation. 
However, due to their low-$z$ nature and limited short-wavelength coverage, we do not have sufficient information to constrain the FUV properties of our targets. 
In the case of \namemiric, despite that M0416 was observed with \hst/WFC3 UVIS filters (F225W, F275W, F336W, F390W), the region where our target is located is missing such coverage, thus preventing us from constraining its rest-frame FUV emission. 
For \namejades, we only have one FUV band (F275W), probing the rest-frame emission at about 1550~\AA\ and implying an absolute UV magnitude $M_{\rm UV} = -17.1\pm 0.23$ mag, but the lack of additional bands prevented us from constraining the UV continuum slope.

The available photometry allowed us to investigate their rest-frame optical ($\beta_{\rm opt}$) and NIR ($\beta_{\rm NIR}$) slopes. 
The rest-frame optical slope primarily traces the stellar continuum and the relative contribution of young versus evolved stellar populations, while the NIR slope is sensitive to old stellar populations and the emergence of hot dust emission associated with obscured star formation and/or AGN activity. 
Measuring and comparing $\beta_{\rm opt}$ and $\beta_{\rm NIR}$ therefore provides first key diagnostics of the dominant emission mechanisms shaping the observed SEDs.

\subsection{Optical and NIR slopes}
\label{sec:slopes}
We measure the optical and near-infrared spectral slopes, $\beta_{\rm opt}$ and $\beta_{\rm NIR}$, by fitting power laws of the form $f(\lambda) \propto \lambda^{\beta}$ to the photometry.
Following the operative definition commonly adopted for the UV continuum \cite[e.g.,][]{Calzetti+94, Castellano+12}, we perform a $\chi^2$ minimization using the relation
\begin{equation}
m_i = -2.5(\beta+2)\log_{10}(\lambda_i) + cost. 
\end{equation}
where $m_i$ is the observed magnitude in the filter with effective wavelength $\lambda_i$.
The fits are performed over the rest-frame wavelength ranges $0.4-1.6~\mu$m and $1.5-3~\mu$m for the optical and NIR slopes, respectively.

For \namemiric, several filters sampling the rest-frame optical emission are affected by strong emission lines (e.g., \hb+\oiii\ and \ha). 
We therefore correct the photometry by estimating the contribution of these lines from the NIRISS spectrum and subtracting the inferred flux from the broadband measurements (details are provided in \S~\ref{app:slopes}).
From the corrected photometry we obtain $\beta_{\rm opt} = -2.18\pm0.14$ and $\beta_{\rm NIR} = 2.05\pm0.29$.

For \namejades, the NIRSpec PRISM spectrum directly detects the continuum emission.
We therefore measure the slopes on the spectrum after masking emission lines, obtaining $\beta_{\rm opt} = -2.00\pm0.15$ and $\beta_{\rm NIR} = 3.12\pm0.04$.
Also for \namejades, a detailed discussion of the emission-line corrections and of the differences between spectral and photometric measurements is presented in \S~\ref{app:slopes}.

\subsection{Comparison with known galaxy SEDs}
\label{sec:known_templates}
We decided to compare the overall SED of our targets with observed empirical galaxy templates to see if already known galaxies show similar SED shapes.
To this purpose, we collected the SED of nearby/low-$z$ quiescent, star-forming, starburst and AGN-hosting galaxies by \cite{Brown+14}, \cite{Brown+19} and \cite{Polletta+07}, a selection of quasar (QSO) templates with different levels of dust extinction\footnote{The QSO models are available at \url{https://github.com/karlan/AGN_templates}.} \citep{Lyu+17, Lyu+18}, the average SED (photometric) of $\approx$ 36000 blue QSOs by \cite{Trefoloni+25}, the maximal LRD SED model by \cite{Akins+25}, the best-fit SED of {\it Virgil} from \cite{Iani+25} and three best-fit templates of Blue Hot Dust Obscured Galaxies \cite[DOGs; e.g.][]{Noboriguchi+19} by \cite{Assef+20}.
We also utilised 50 single stellar population templates of different age, metallicity and dust extinction by means of the \textsc{bpass} models \cite[v. 2.2;][]{Stanway+18}, see \S~\ref{app:empirical_templates} for more details.
After applying redshift, all models were compared to the photometry of our targets convolving their SED with the NIRCam/F277W filter and matching the inferred F277W flux to the one of our target (see Figure~\ref{app-fig:empirical_templates}).

\begin{figure}
    \centering
    \includegraphics[width=\linewidth]{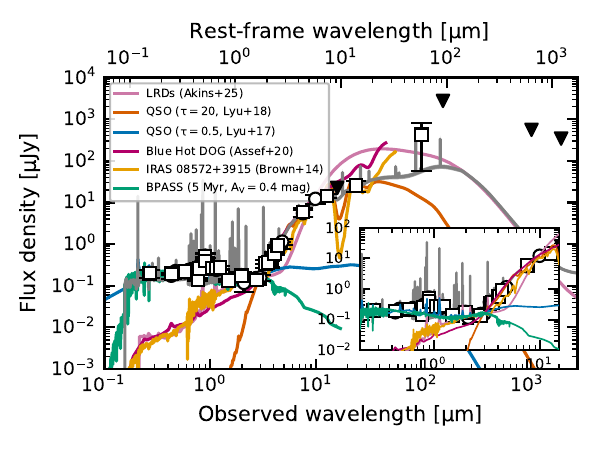}
    \caption{Comparison between our targets photometry and galaxy empirical templates: IRAS 08572+3915 (orange) from \cite{Brown+14}, a QSO with little dust attenuation ($\tau = 0.5$, light blue) by \cite{Lyu+17}, a heavily obscured QSO ($\tau = 20$, red) from \cite{Lyu+18}, the maximal SED template of LRDs/"V"-shaped objects (purple) by \cite{Akins+25}, a young SSP (5 Myr, green) from the {\sc bpass} models by \cite{Eldridge+17} and the the Blue Hot DOG W0204-0506 by \cite{Assef+20} (magenta). \namemiric\ and \namejades\ data points are reported with open circles and open squares, respectively. Upperlimits are shown with black downward triangles. In grey, we present the best-fit model obtained by {\sc cigale} (see \S~\ref{sec:sed_fitting}).}
    \label{fig:template_comparison}
\end{figure}

In general, no templates were found to fully reproduce the extreme SED of our targets since unable to properly reproduce at the same time the blue shape of the rest-frame NUV-to-optical continuum and its red steep rest-frame 2 - 3~\um\ emission, see Figure~\ref{fig:template_comparison}.

Among all models normalised to the observed F277W filter, the best agreements at short wavelengths come from a QSO template with low dust obscuration (optical depth $\tau = 0.5$) and a \textsc{bpass} model of a 5~Myr old single stellar population having a total stellar mass of $\log_{10}(M_\star~[\rm M_\odot]) = 6.7$, a solar metallicity and color excess $E(B-V) = 0.1$ (assuming the \cite{Calzetti+00} attenuation law).

Similarly, only three templates succeeded in reproducing the red MIR slope: the heavily obscured (optical depth $\tau = 20$) QSO template, the Blue Hot DOG W0204-0506 from \cite{Assef+20} and the ULIRG IRAS 08572+3915 \cite[$z \approx 0.06$, e.g. ][]{Zhang+24} by \cite{Brown+14}.
Compared to the dusty QSO template, W0204-0506 and IRAS 08572+3915 have also a blue component (even though not as extreme as in our targets). 
While the AGN presence in Blue Hot DOGs is well-known, also IRAS 08572+3915 was found to be very young starburst hosting an AGN with a fairly smooth torus viewed almost edge-on \citep{Efstathiou+14}. 

Despite the wide range of ages and extinction adopted, no \textsc{bpass} template was able to reproduce the reddest wavelengths excluding the possibility that part of the emission at such red wavelengths is due only to an old and/or strongly attenuated stellar population.

\subsection{First insights on galaxy properties from emission lines}
\label{sec:eml_properties}
The detection of emission lines in the available NIRISS and NIRCam WFSS data (\namemiric) and NIRSpec MSA (\namejades) allowed us to further investigate the physical properties of our targets among which the impact of dust on the nebular emission, their on-going star formation activity and the metallicity of their ISM.
We highlight that the main assumption in the estimates presented in this section is that the mechanism at the origin of the observed line fluxes is purely star formation, i.e. we excluded any contribution from AGN activity that, even in the case of (edge-on) AGN obscured geometry, would otherwise contribute to the observed optical/NIR line fluxes via potential emission from an extended narrow-line region (NLR). 

We report the main results derived under these assumptions in Table~\ref{tab:properties_lines}.

\begin{table}
\centering
\caption{Physical properties of \namemiric\ and \namejades\ derived from emission-line diagnostics (see \S~\ref{sec:eml_properties}).}
\begin{tabular}{lcc}
\hline
\hline
Property & \namemiric & \namejades \\
\hline
EW$_0$(\hb) [\AA] & $108 \pm 46$ & $367 \pm 76$ \\
EW$_0([\ion{O}{iii}]\lambda5007$) [\AA] & $990 \pm 121$ & $2035\pm148$ \\
EW$_0$(\ha) [\AA] & $1015 \pm 129$ & $1791 \pm 173$ \\
$E(B-V)_{\rm neb}$ (Calzetti) & $\simeq 0.06$ & $\simeq 0.05$ \\
$E(B-V)_{\rm neb}$ (SMC Bar) & $\simeq 0.08$ & $\simeq 0.07$ \\
$\log_{10}(L_{\rm H\alpha,int})$ [erg s$^{-1}$]  & $41.20 \pm 0.09^\dagger$ & $41.34 \pm 0.05$ \\
$\log_{10}({\rm SFR(H\alpha)})$ [$M_\odot$ yr$^{-1}$] & $-0.09 \pm 0.09^\dagger$ & $0.03 \pm 0.05$ \\ 
$\log_{10}({\rm SFR(NUV)})$ [$M_\odot$ yr$^{-1}$] & $-0.53 \pm 0.15^\dagger$ & $-0.49 \pm 0.07$ \\
$\log_{10}({\rm SFR(FUV)})$ [$M_\odot$ yr$^{-1}$] & - & $-0.46 \pm 0.13$ \\
$\log_{10}({\rm SFR(H\alpha)/SFR(UV)})$ & $0.44 \pm 0.17$ & $0.49 \pm 0.20$ \\
R3 $=\log_{10}([$\ion{O}{iii}]$\lambda5007$/\hb) & $0.98 \pm 0.18$ & $0.71 \pm 0.09$ \\
O32 $=\log([$\ion{O}{iii}]/[\ion{O}{ii}]) & - & $1.45 \pm 0.15$ \\
Ne3O2 $=\log([$\ion{Ne}{iii}]/[\ion{O}{ii}]) & - & $0.22 \pm 0.19$ \\
$12+\log(\mathrm{O/H})$  & $\sim 7.8{-}8.2$ & $7.66 \pm 0.12$ \\
$Z~[Z_\odot]$ & $\sim 0.1- 0.4$ & $\sim 0.1$ \\
\hline
\hline
\end{tabular}
\label{tab:properties_lines}
\tablefoot{For \namemiric, quantities highlighted with $^\dagger$ are not corrected for lensing magnification $\mu = 1.18$.Values assume that nebular emission is powered purely by star formation.}
\end{table}

\subsubsection{Lines Equivalent Widths}
\label{sec:ew0}
We estimated the rest-frame equivalent widths (EW$_0$) of the \hb, [\ion{O}{iii}]$\lambda5007$, and \ha\ emission lines following the definition:
\begin{equation}
    {\rm EW}_0 = \frac{f_{\rm line}}{(1+z) \cdot f_{\lambda, \rm cont}}
\end{equation}
For \namejades, the continuum level at the observed wavelength of each line was obtained directly from the analysis of the NIRSpec PRISM spectrum while, for \namemiric, because of the non detection of the continuum, we scaled the estimated F150W flux (corrected for line contamination) assuming the optical slope $\beta_{\rm opt} = -2.18 \pm 0.14$, see \S~\ref{sec:slopes}. 
This approach accounted for the wavelength dependence of the continuum between the F150W band ($\lambda_{\rm piv} \approx 1.5~\mu$m) and the observed wavelengths of the optical emission lines. 
For \namemiric, we derived rest-frame equivalent widths EW$_0({\rm H}\beta) \approx 108 \pm 46$~\AA, EW$_0([$\ion{O}{iii}$]\lambda5007) \approx 990 \pm 121$~\AA, and EW$_0({\rm H}\alpha) \approx 1015 \pm 129$~\AA. 
For \namejades, we inferred EW$_0({\rm H}\beta) \approx 367 \pm 76$~\AA, EW$_0([$\ion{O}{iii}$]\lambda5007) \approx 2035 \pm 148$~\AA, and EW$_0({\rm H}\alpha) \approx 1791 \pm 173$~\AA.\footnote{We note that the reported values assume no differential dust attenuation between the nebular emission and the stellar continuum, which would otherwise lead to even larger intrinsic equivalent widths.}
These very large equivalent widths placed both objects firmly in the regime of Extreme Emission-Line Galaxies \cite[EELGs; e.g.][]{Amorin+14}, indicating a young ($\lesssim 20$~Myr), burst-dominated stellar population undergoing an episode of star formation with high sSFR \cite[][]{Leitherer+99, Eldridge+17}.

\subsubsection{Dust reddening}
\label{sec:dust}
Given the detection of multiple hydrogen transitions (\hb, \ha, \paa, \brg), we estimated the impact of dust on the nebular component by comparing the observed relative strength of some transitions (\ha/\hb, \paa/\ha, \brg/\ha) to the theoretical expectations from quantum physics.
In particular, we used as reference the theoretical line ratios expected in standard Case~B recombination conditions \citep[e.g.][]{Osterbrock+06}, i.e. assuming an ISM electron temperature $T_e = 10^4~{\rm K}$ and density $n_e = 100~{\rm cm^{-3}}$ and that all ionising photons are reprocessed by the gas (``on the spot'' approximation).\footnote{We note that variations in both $T_e$ and $n_e$ affect the expected line ratios.}
According to Case B, the intrinsic ratios are (\ha/\hb)$_0 = 2.863$ (the so-called {\it Balmer Decrement}), (\paa/\ha)$_0 = 0.118$ and  (\brg/\ha)$_0 = 0.010$\footnote{In the case of \paa/\ha\ and \brg/\ha, we derived the Case B values via {\sc pyneb} \citep{Luridiana+15}.}. 

For \namemiric\ the observed ratios are (\ha/\hb)~$ =4.890 \pm 2.082$, (\paa/\ha)~$ = 0.136 \pm 0.016$ and (\brg/\ha)~$ = 0.014\pm0.004$. 
For \namejades, (\ha/\hb)~$=2.880 \pm 0.588$, (\paa/\ha)~$ = 0.133 \pm 0.008$ and (\brg/\ha)~$ = 0.008 \pm 0.003$.

To quantify the implied nebular reddening, we derived $E(B-V)_{\rm neb}$ from each ratio using \cite[e.g.][]{Momcheva+13}:
\begin{equation}
\label{eq:ebv_solution}
E(B-V)_{\rm neb} =
\frac{2.5}{k(\lambda_2)-k(\lambda_1)}
\log_{10}\left[\frac{(f(\lambda_1)/f(\lambda_2))_{\rm obs}}{(f(\lambda_1)/f(\lambda_2))_0}\right].
\end{equation}

We considered two prescriptions for the extinction/attenuation law $k(\lambda)$: (i) the starburst attenuation law of \citet{Calzetti+00}, and (ii) an SMC Bar extinction curve (with $R_V\simeq 2.74$) from \citet{Gordon+03}\footnote{According to the attenuation curve by \citet{Calzetti+00}: $k({\rm H\beta}) = 4.598$, $k({\rm H\alpha}) = 3.326$, $k({\rm Pa\alpha}) = 0.587$, $k({\rm Br\gamma}) = 0.389$. Assuming the extinction law by \cite{Gordon+03}: $k({\rm H\beta}) = 3.273$, $k({\rm H\alpha}) = 2.166$, $k({\rm Pa\alpha}) = 0.658$, $k({\rm Br\gamma}) = 0.385$.}. 
Since our comparison with the {\sc bpass} models suggested a possible dwarf galaxy nature for our targets (see \S~\ref{sec:known_templates}), we tested an SMC Bar extinction curve to account for the possibility of steeper dust attenuation in low-metallicity systems, as expected for DGs, and to verify that the inferred attenuation remains low independently of the adopted dust prescription.

Using the \citet{Calzetti+00} law, the \paa/\ha\ ratios imply low reddening, $E(B-V)_{\rm neb}\simeq 0.06$ (i.e. $A_V\simeq 0.23$~mag) for \namemiric\ and $E(B-V)_{\rm neb}\simeq 0.05$ (i.e. $A_V\simeq 0.19$~mag) for \namejades; the $Br\gamma/H\alpha$ ratios yielded values consistent with similarly small attenuation (and with zero within the uncertainties for \namejades). 
The Balmer Decrement yielded $A_V\simeq 1.9$~mag for \namemiric, but with very large uncertainty due to the noisy ${\rm H\beta}$ measurement, while \namejades\ was consistent with the intrinsic Case~B ratio.
Adopting an SMC Bar curve, we obtained consistent conclusions: \paa/\ha\ implies $E(B-V)_{\rm neb}\simeq 0.08$ ($A_V\simeq 0.22$~mag) for \namemiric\ and $E(B-V)_{\rm neb}\simeq 0.07$ ($A_V\simeq 0.19$~mag) for \namejades, while \brg/\ha\ remained consistent with small attenuation and with zero for \namejades. 
We report all the different nebular reddening estimates in Table~\ref{tab-app:Av_estimates}.

Overall, independently of the adopted attenuation prescription, the optical–to–NIR hydrogen line ratios remained close to their Case B expectations, indicating that dust plays only a minor role in shaping the observed recombination-line fluxes of our targets.

\subsubsection{Star formation rates and burstiness}
\label{sec:sfr_burstiness}
Having derived the dust attenuation, we estimated the intrinsic \ha\ luminosities of our targets.
Assuming the reddening derived using the \cite{Calzetti+00} law\footnote{The adoption of the SMC Bar extinction curve would have produced negligible differences between the intrinsic \ha\ luminosities, on the order of 0.01 dex. In the case of the intrinsic UV luminosities, the SMC Bar extinction curve would have increased the NUV SFR estimates by $\approx 0.2$ dex while $\approx 0.3$ dex for the FUV.}, we obtained $\log_{10}(\mu\cdot L({\rm H\alpha})_{\rm int}~[{\rm erg~s^{-1}}]) = 41.20 \pm 0.09$ for \namemiric\ and $\log_{10}(L({\rm H\alpha})_{\rm int}~[{\rm erg~s^{-1}}]) = 41.34 \pm 0.05$ for \namejades.
Using the $L({\rm H\alpha})_{\rm int}$–SFR conversion from \cite{Kennicutt+12}, we derived $\log_{10}(\mu\cdot{\rm SFR(H\alpha)}~[{\rm M_\odot~yr^{-1}}]) = -0.09 \pm 0.09$ and $\log_{10}({\rm SFR(H\alpha)}~[{\rm M_\odot~yr^{-1}}]) = 0.03 \pm 0.05$ for \namemiric\ and \namejades, respectively.
Adopting instead the calibration from \cite{Theios+19} for low-metallicity galaxies ($Z=0.2~Z_\odot$, IMF cut-off at $100~M_\odot$ including binaries) would have lowered these estimates by $\approx 0.3$ dex.
These values trace the instantaneous star formation activity on timescales of $\sim 10$ Myr.

We also estimated the SFR from the rest-frame UV stellar continuum, which probes star formation over longer ($\gtrsim 100$ Myr) timescales assuming a continuous SFH.
For \namemiric, in the absence of FUV coverage, we derived the SFR from the NUV luminosity at $\approx 2500$\,\AA\ (HST/ACS F435W) using the \cite{Kennicutt+12} conversion, obtaining $\log_{10}(\mu\cdot\nu L_{\rm NUV}~[{\rm erg~s^{-1}}]) = 42.64 \pm 0.15$ and $\log_{10}(\mu\cdot{\rm SFR(NUV)}~[{\rm M_\odot~yr^{-1}}]) = -0.53 \pm 0.15$.
For \namejades\ we obtained $\log_{10}(\nu L(\rm NUV)~[{\rm erg~s^{-1}}]) = 42.68 \pm 0.07$ and $\log_{10}({\rm SFR(NUV)}~[{\rm M_\odot~yr^{-1}}]) = -0.49 \pm 0.07$.

For \namejades, we additionally estimated the FUV SFR from the conversion of the HST/WFC3 UVIS F275W flux, deriving $\log_{10}(\mu\cdot\nu L(\rm FUV)~[{\rm erg~s^{-1}}]) = 42.89 \pm 0.13$ and $\log_{10}(\mu\cdot{\rm SFR(FUV)}~[{\rm M_\odot~yr^{-1}}]) = -0.46 \pm 0.13$.
Assuming the \cite{Theios+19} calibration would have reduced the estimate by $\approx 0.1$ dex.
We note that in correcting the UV continuum, we assumed comparable stellar and nebular colour excess given the low attenuation; adopting a differential reddening relation \cite[e.g., $E(B-V)_\star = 0.58\cdot E(B-V)_{\rm neb}$;][]{Steidel+14, Rodriguez-Munoz+22} would have decreased the UV SFR by $\sim 0.1$ dex.

The ratio between the \ha- and UV-based SFR estimates allowed us to probe the burstiness of the galaxies \cite[e.g.][]{Atek+22, Navarro-Carrera+24}, with systems typically classified as bursty when $\log_{10}({\rm SFR(H\alpha)/SFR(UV)}) \gtrsim 0.3$.
For \namemiric\ and \namejades\, we obtained ratios of $0.44 \pm 0.17$ and $0.49 \pm 0.20$, respectively, suggesting that both systems are consistent with being in a starburst phase at the time of observation.
However, we highlight that uncertainties in dust attenuation and metallicity-dependent SFR calibrations may reduce the significance of this result.

\subsubsection{ISM metallicity}
\label{sec:metallicity}
Thanks to the detection of several metal transitions in the NIRISS WFSS (\namemiric) and NIRSpec MSA (\namejades) spectra, we constrained the ISM metallicity of our targets using strong-line diagnostics.

Since both sources display the [\ion{O}{iii}] optical doublet, we first used the R3 indicator ($\equiv \log_{10}([\ion{O}{iii}]\lambda5007/{\rm H}\beta)$).
For \namemiric\ we measured [\ion{O}{iii}]$\lambda5007/{\rm H}\beta = 8.6 \pm 3.6$ (R3 $= 0.98 \pm 0.18$), while for \namejades\ we obtained [\ion{O}{iii}]$\lambda5007/{\rm H}\beta = 5.1 \pm 1.1$ (R3 $= 0.71 \pm 0.09$).
Using the calibration of \cite{Curti+20}, the latter corresponds to two metallicity solutions, $12+\log_{10}({\rm O/H}) = 7.75 \pm 0.20$ and $8.20 \pm 0.15$, i.e. $Z \approx 0.1$–$0.4\,Z_\odot$ \cite[assuming $12+\log_{10}({\rm O/H})_\odot = 8.69 \pm 0.05$;][]{Allende-Prieto+01}.
For \namemiric, the measured ratio lies significantly above the relation reported by \cite{Curti+20} and is only marginally consistent within the uncertainties ($12+\log_{10}({\rm O/H}) = 7.8$–$8.2$), possibly indicating the presence of a different mechanism powering the line emission, e.g. an AGN.

For \namejades, additional constraints were obtained from the [\ion{O}{ii}]$\lambda\lambda3727,3729$ doublet ([\ion{O}{ii}] hereafter).
We measured [\ion{O}{iii}]$\lambda\lambda4959,5007/[\ion{O}{ii}] = 28.5 \pm 9.6$ (uncorrected for dust), corresponding to O32 $= 1.45 \pm 0.15$, indicative of a highly ionized medium.
Using the calibration of \cite{Bian+18}, this ratio implies $12+\log_{10}({\rm O/H}) \approx 7.7$, favouring the low-metallicity branch of the R3 solutions.
This estimate was further corroborated by the ratio [\ion{Ne}{iii}]$\lambda3870/[\ion{O}{ii}] = 1.67 \pm 0.75$ (Ne3O2 $= 0.22 \pm 0.19$), which corresponds to $12+\log_{10}({\rm O/H}) = 7.66 \pm 0.12$ ($\approx 0.1\,Z_\odot$) according to \cite{Bian+18}.

\begin{figure}
    \centering
    \includegraphics[width=\linewidth]{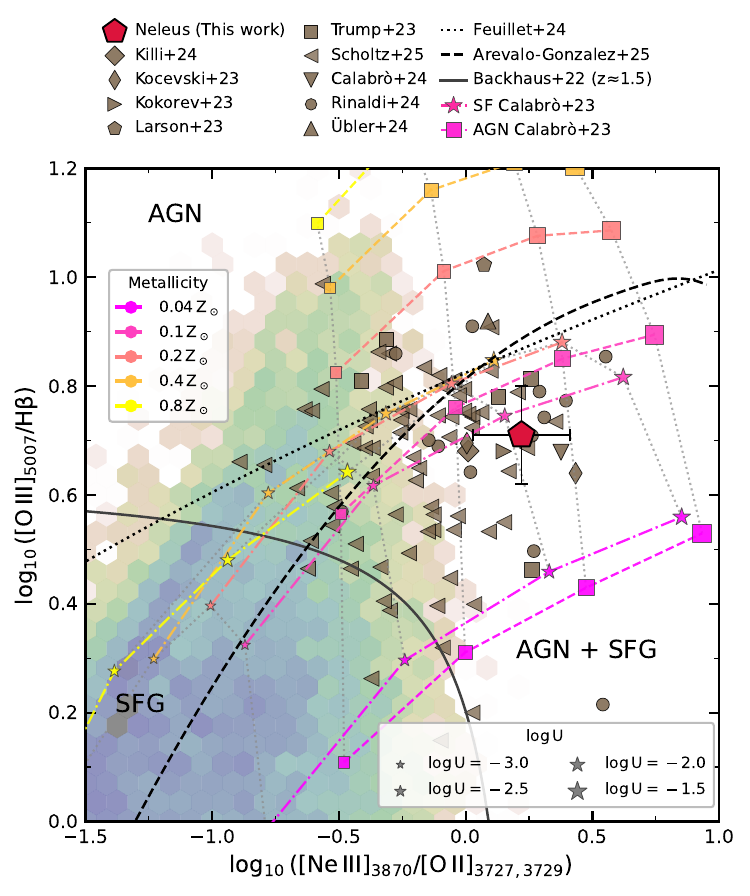}
    \caption{The OHNO diagram. The solid, dashed and dotted black demarcation lines are based on the work of \cite{Backhaus+22}, \cite{Feuillet+24} and \cite{Arevalo-Gonzalez+25}, respectively. 
    In the background, the 2D density distribution is representative of galaxies from the Sloan Digital Sky Survey \cite[SDSS;][]{York+00, Kauffmann+03}. The brown data-points show the location of recent literature at $z \gtrsim 4$ from \cite{Killi+24, Kocevski+23, Kokorev+23, Larson+23, Trump+23, Ubler+24b, Scholtz+25, Calabro+24a, Rinaldi+24} while the red pentagon highlight the position of our target \namejades. We also present a grid of SFG and AGN models by \cite{Calabro+23}, showing with different colors and sizes the representative points of models of different metallicity ($Z = 0.04, 0.1, 0.2, 0.4, 0.8~Z_\odot$) and ionisation parameter ($\log_{10}(U) = -3.0, -2.5, -2.0, -1.5$).}
    \label{fig:OHNO}
\end{figure}

Finally, the detection of [\ion{Ne}{iii}]$\lambda3870$, [\ion{O}{ii}], [\ion{O}{iii}]$\lambda5007$, and H$\beta$ allowed us to place \namejades\ on the ``OHNO'' diagnostic diagram \citep{Zeimann+15, Backhaus+22, Backhaus+23, Trump+23, Feuillet+24}, see Figure~\ref{fig:OHNO}.
The measured O32 and Ne3O2 ratios place the source within the AGN locus according to the redshift-dependent demarcation of \citet{Backhaus+22}.
However, compact, low-mass, metal-poor starbursts with very high ionization parameters or density-bounded \ion{H}{ii} regions can reach similarly extreme excitation ratios and populate the AGN region \citep[e.g.][]{Larson+23, Tripodi+24, Scholtz+25}.
In fact, as shown in Figure~\ref{fig:OHNO}, \namejades\ is located in a region populated by both AGN and SF models \citep{Calabro+23} also according to the demarcation lines by \cite{Feuillet+24} and \cite{Arevalo-Gonzalez+25}.
Interestingly, this region of the OHNO diagram is significantly populated by LRDs (with some of them showing clear broad permitted lines; \citealt{Rinaldi+25c}) and by sources recently claimed to host AGNs in the intermediate- and high-redshift literature (\citealt{Scholtz+25}).

\subsection{Do \namemiric\ and \namejades\ host an obscured AGN?}
\label{sec:agn_or_dust}
Our comparison between the SEDs of our targets and empirical galaxy templates (see Sect.~\ref{sec:known_templates}) indicated that their overall spectral shape could be consistent with a galaxy hosting an obscured AGN. 
At the same time, {\sc bpass} models suggested that our targets could be DGs, with stellar masses potentially of order of $M_\star \simeq 10^7~M_\odot$.
Based on our estimated SFR and gas-phase metallicities, similar mass values would also be inferred when extrapolating the Fundamental Metallicity Relation \cite[FMR,][]{Mannucci+10} as formulated by \cite{Curti+20} towards the low stellar mass regime.
However, in low-metallicity systems dust grains can be heated to temperatures significantly higher than the canonical $20$–$30$~K typical of solar metallicity environments \cite[e.g.][]{Draine+07}.
In this case, a steep rise at rest-frame MIR wavelengths may arise from very hot dust without requiring an AGN contribution \cite[e.g.][]{Hainline+16, Sturm+25, Rieke+25}.
We therefore tested whether the steep MIR rise observed in our photometry could be explained purely by dust emission or whether an AGN component was required.

\begin{figure}
    \centering
    \includegraphics[width=\linewidth]{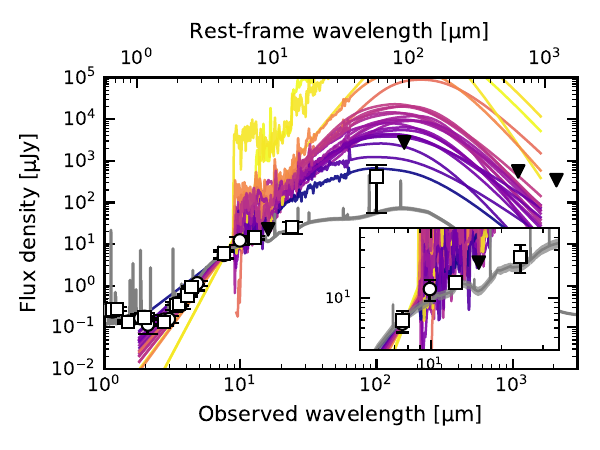}
    \includegraphics[width=\linewidth]{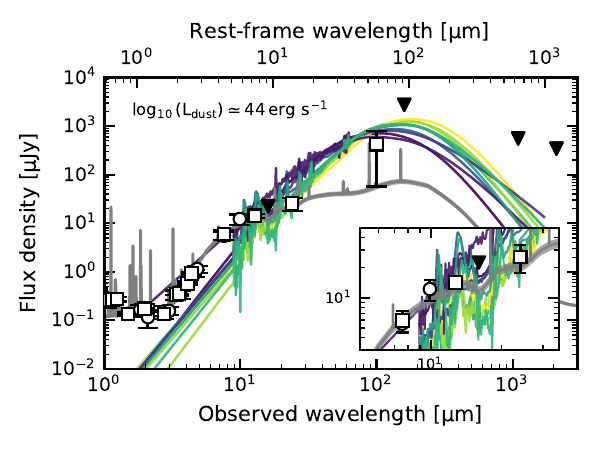}
    \caption{{\bf Top panel:} DGs templates from \cite{Lyu+24} normalised to the MIRI F770W photometry. \namemiric\ and \namejades\ photometry is reported with open circles and open squares, respectively. Upperlimits are reported with black downward triangles. In grey, we show the best-fit model obtained by {\sc cigale} (see \S~\ref{sec:sed_fitting}). The inset displays a zoom-in of the 6 - 30~$\mu$m spectral region (observed frame). {\bf Bottom panel:} as above, DGs templates (without smoothing) fixing the total dust luminosity to the one predicted by {\sc cigale} based on the available rest-frame UV-to-optical photometry. }
    \label{fig:dg_templates}
\end{figure}

To do so, we compared the reddest-wavelength emission of our SEDs with templates of DGs \citep{Lyu+24}. 
The DG templates were redshifted to the match the $z_{\rm spec}$ of our targets and normalised to reproduce their observed flux density in the F770W filter (see top panel Figure~\ref{fig:dg_templates}). 
By construction, this procedure assumes that the entire observed $\sim7~\mu$m emission originates from dust re-radiation.

While several DG templates successfully reproduced the steep SED slope down to $\simeq1.8~\mu$m (rest-frame), they systematically over-predicted the emission at longer MIR and FIR wavelengths. 
In particular, the normalised templates implied detectable fluxes at \textit{Herschel} wavelengths that were inconsistent with the available ancillary data. 
For \namejades, the discrepancy was even stronger, as the templates failed to reproduce the measured \textit{Spitzer}/MIPS 24~$\mu$m and \textit{Herschel}/PACS 100~$\mu$m fluxes reported by \citet{Liu+18}.

The inconsistency became more evident when converting the total infrared luminosities, $L({\rm TIR})$, of the normalised DG templates into star-formation rates using the calibration of \citet{Kennicutt+12}. 
Depending on the adopted template, we obtained $\log_{10}({\rm SFR(TIR)}~[M_\odot~{\rm yr}^{-1}]) \simeq 1.5-4$, exceeding the SFRs inferred from both the \ha\ emission and the rest-frame UV continuum (see \S~\ref{sec:eml_properties}) by $\approx2-5$~dex. 
Such large discrepancies would violate energy balance, as the infrared luminosities would require far more absorbed stellar radiation than available from the observed UV-to-optical emission. 
Indeed, enforcing energy balance by equating $L({\rm TIR})$ to the dust-absorbed luminosity inferred from the rest-frame NUV-to-optical SED results in predicted dust emission that falls short of the observed MIR fluxes by approximately one order of magnitude.

If significant dust-obscured star formation were responsible for the MIR emission, a substantial fraction of the nebular radiation would arise from regions opaque at optical wavelengths but still observable in the NIR. 
In this scenario, NIR recombination lines such as \paa\ and \brg\ would trace star formation missed by H$\alpha$, leading to observed NIR-to-optical line ratios exceeding their intrinsic Case~B values.

However, as discussed in \S~\ref{sec:eml_properties}, the measured \paa/\ha\ and \brg/\ha\ ratios are consistent with Case~B expectations and imply only modest extinction ($A_V \approx 0.2$~mag). 
Therefore, the NIR recombination lines do not reveal a significant additional component of ongoing star formation hidden at optical wavelengths, unless it is so deeply embedded that even Paschen and Brackett emission is fully suppressed.

As an additional test, we fixed the total dust luminosity of the DG templates to the luminosity absorbed in the rest-frame UV–optical under the energy balance assumption.
To estimate this quantity, we fitted the photometry at rest-frame wavelengths $\lesssim 1.8\,\mu$m with the SED fitting code \textsc{cigale} v.2025.1 \citep{Burgarella+05, Noll+09, Boquien+19, Yang+20, Yang+22}, fixing the redshift to the spectroscopic value ($z_{\rm spec}$).
For this run we included only the stellar population, dust attenuation, and nebular emission modules, adopting the same configuration described in \S~\ref{sec:sed_fitting}.
From the \textsc{cigale} fits we derived $\log_{10}(L_{\rm dust}~[{\rm erg~s^{-1}}]) \approx 44$ for both targets.
However, when the DG templates were normalised to this $L_{\rm dust}$, they complied with the long-wavelength constraints but systematically under-predicted the observed flux at rest-frame $ 2- 8\mu$m by $\sim1$ dex (see Fig.~\ref{fig:dg_templates}, bottom panel).
Adding the dust emission module to our {\sc cigale} run, we reached acceptable best-fits ($\chi^2_{red} = 1.8 - 4.6$) only when assuming extreme parameters, i.e. extremely large radiation field intensities ($U_{\rm min⁡}=50$, $<U> = 10^4$) and an almost complete dominance of photodissociation region (PDR) -like dust ($\gamma \approx 1$).
Such values are far above those typically observed in star-forming galaxies and even in extreme starbursts \cite[e.g.,][]{Draine+07, Magdis+12}, implying that nearly all the dust would be exposed to very intense radiation fields. 
Even in such extreme dust conditions, however, the resulting best-fit models would systematically under-predict the observed fluxes at $\gtrsim 2~\mu$m (rest-frame), while simultaneously requiring excessive dust attenuation in the rest-frame NUV-to-optical regime. 

For \namejades, the availability of MIR-to-FIR constraints provides strong evidence against a purely star-forming origin of the infrared emission. In particular, the observed SED shows a turnover at $\sim10~\mu$m (rest-frame), deviating from the behaviour of extreme dusty star-forming galaxies such as Haro~11 and similar systems \citep[e.g.][]{Rieke+25}: such a turnover is difficult to reproduce with star-formation-powered dust emission alone.
Furthermore, DG templates simultaneously fail to reproduce the observed MIR fluxes and the FIR constraints, either over-predicting long-wavelength emission or violating energy balance. Together, these results strongly favour the presence of an additional heating source, consistent with AGN activity.

For \namemiric, the lack of detections beyond $\sim10~\mu$m prevents a similar assessment of the FIR behaviour, and therefore the SED alone does not provide equally strong constraints on the dust-only scenario. 
However, this source exhibits an exceptionally high [\ion{O}{iii}]/\hb\ ratio ($\gtrsim 10$), which is rarely observed in star-forming galaxies, both locally and at high redshift \citep[e.g.][]{Amorin+14, Tang+25}. 
Even in extreme systems such as Haro~11, significantly lower values are typically found \citep{Cairos+15}.
Such extreme excitation conditions are more naturally explained by the presence of an AGN, particularly when considered in combination with the steep MIR SED and the difficulties in reproducing it with purely stellar-powered dust emission.

Taken together, the evidence points toward AGN activity in both sources, albeit supported by different diagnostics. 
For \namejades, the infrared SED shape and FIR constraints strongly disfavour a purely star-forming origin of the dust emission. 
For \namemiric, the case relies primarily on its extreme nebular excitation, as traced by the very high [O III]/H$\beta$ ratio, combined with the inability of dust-only models to reproduce the observed MIR emission.

\subsection{Panchromatic SED fitting}
\label{sec:sed_fitting}
Having found no galaxy template to describe at the same time both the rest-frame blue NUV-to-optical and red NIR-to-MIR colors of \namemiric\ and \namejades\ (see \S~\ref{sec:known_templates}), we ran \textsc{cigale} fixing the redshifts to the spectroscopic estimates and considering the available multi-wavelength information from the X-rays-to-radio wavelengths.
We also complement the panchromatic constraints on the SED with the fluxes of the main emission lines detected in the spectra of our targets.  

For the run, we assume the recently implemented stochastic star formation history (SFH) module based on power-spectrum densities presented by \cite{Carvajal+25}. 
We decided to adopt a stochastic SFH in light of our direct estimates of the burstiness parameter and since our comparison with the {\sc bpass} models (see \S~\ref{sec:known_templates}) highlighted the possible low-mass nature of our targets. 
In fact, DGs are known to be characterised by bursty SFHs \cite[e.g.,][]{Weisz+12, Amorin+14}. 
We adopt the {\sc bpass} stellar population models with sub-solar metallicities ($Z = 0.1, 0.2Z_\odot$, in light of what found in \S~\ref{sec:metallicity}), and the Chabrier IMF with a cut-off mass of 300~$M_\odot$. 
We include nebular continuum and emission lines using sub-solar metallicities and allowing the electron density $n_e$ and ionisation parameter $U$ to assume values of $n_e = 100, 1000\rm~ cm^{-3}$ and $\log_{10}(U) = -3, -2, -1$, respectively. 
For the dust attenuation, we adopt the modified version of the \cite{Calzetti+00} attenuation curve by \cite{Leitherer+02} for the continuum emission while we allow for the SMC law to attenuate the emission lines setting $E(B-V)_{\rm neb} = 0, 0.05, 0.1, 0.3, 0.5$ and allowing $E(B-V)_\star / E(B-V)_{\rm neb} = 0.3, 0.4, 0.7, 1$. 
For the dust IR emission, we resort to the models by \cite{Draine+14}. 
We add the AGN emission using the \textsc{skirtor} models \citep{Stalevski+12, Stalevski+16}, allowing for the presence of both Type I (unobscured) and II (obscured) AGNs.
Finally, we implement the radio module. 
Despite having upper-limits in the X-ray domain thanks to {\it Chandra} observations, we do not implement {\sc cigale} X-ray module \citep{Yang+20, Lopez+24} since it is not currently available in combination with the BPASS models. 
Based on our input file configurations, we tested over 2 billion models.
We report {\sc cigale}'s Bayesian best-fit estimates of our targets physical properties in Table~\ref{tab:cigale_bestfit}.

For \namemiric, {\sc cigale} returns a low-mass system with total stellar mass (not corrected for lensing magnification) $\log_{10}(\mu\cdot M_\star~[{\rm M_\odot}]) = 7.07 \pm 0.09$ and a recent star-formation rate on a 10~Myr timescale of $\log_{10}(\mu\cdot{\rm SFR}_{10{\rm Myr}}~[{\rm M_\odot~yr^{-1}}]) = -0.52 \pm 0.06$.
The corresponding specific SFR is $\log_{10}({\rm sSFR}~[{\rm yr^{-1}}]) \simeq -7.59 \pm 0.11$, at the limit of the starburst regime \cite[see, e.g.,][]{Caputi+17, Rinaldi+22} and implying that, at the current SFR, the stellar mass would double in $\approx 30 - 40$ Myr, indicating rapid ongoing assembly.
In addition, {\sc cigale} independently supports a strongly time-variable SFH, yielding $\log_{10}({\rm SFR}_{10{\rm Myr}}/{\rm SFR}_{100{\rm Myr}}) = 0.75 \pm 0.14$, in qualitative agreement with our line-based burstiness indicator (see \S~\ref{sec:eml_properties}).
The attenuation affecting emission lines is low, with $E(B-V)_{\rm neb}\simeq 0.10$, consistent with the modest nebular extinction inferred from recombination-line ratios (see \S~\ref{sec:dust}).
Finally, the MIR emission is best reproduced with a dominant AGN contribution, with an AGN fractional contribution $f_{\rm AGN}\simeq 0.70$ (between rest-frame 1.5 - 7 $\mu$m) and a nearly edge-on viewing angle ($i\simeq 90^\circ$).
The inferred AGN accretion power (not corrected for lensing magnification) is $\log_{10}(\mu\cdot L_{\rm bol}~[{\rm erg~s^{-1}}]) = 43.77 \pm 0.04$.

Similarly, \namejades\ is also found to be a low-mass galaxy with $\log_{10}(M_\star~[{\rm M_\odot}]) = 7.06 \pm 0.04$ and $\log_{10}({\rm SFR}_{10{\rm Myr}}~[{\rm M_\odot~yr^{-1}}]) = -0.48 \pm 0.03$.
The resulting sSFR is $\log_{10}({\rm sSFR}~[{\rm yr^{-1}}]) \simeq -7.54 \pm 0.05$, also in this case suggesting stellar mass doubling time of $\approx 30 - 40$ Myr at the estimated SFR.
Also for \namejades\ the SFH is bursty with $\log_{10}({\rm SFR}_{10{\rm Myr}}/{\rm SFR}_{100{\rm Myr}}) = 0.83 \pm 0.04$.
The best-fit solution requires low line reddening ($E(B-V)_{\rm neb}\simeq 0.10$) and a prominent obscured AGN component with $f_{\rm AGN}\simeq 0.73$ viewed edge-on.
The derived accretion power is $\log_{10}(L_{\rm bol}~[{\rm erg~s^{-1}}]) = 43.94 \pm 0.05$.

Overall, under the updated {\sc cigale} configuration, both targets appear as $\sim 10^7~\rm M_\odot$ systems undergoing intense, bursty star formation while simultaneously hosting an obscured AGN that dominates the rest-frame NIR-to-MIR emission.

\begin{table}
\centering
\caption{{\sc cigale} Bayesian estimates from the final configuration described in \S~\ref{sec:sed_fitting}.}
\label{tab:cigale_bestfit}
\begin{tabular}{lcc}
\hline
\hline
Parameter & \namemiric\ & \namejades\ \\
\hline
$\log_{10}(M_\star~[{\rm M_\odot}])$ & $7.07 \pm 0.09^\dagger$ & $7.06 \pm 0.04$ \\
$\log_{10}({\rm SFR}_{10{\rm Myr}}~[{\rm M_\odot~yr^{-1}}])$ & $-0.52 \pm 0.06^\dagger$ & $-0.48 \pm 0.03$ \\
$\log_{10}({\rm SFR}_{100{\rm Myr}}~[{\rm M_\odot~yr^{-1}}])$ & $-1.27 \pm 0.13^\dagger$ & $-1.31 \pm 0.03$ \\
$\log_{10}({\rm sSFR}_{10{\rm Myr}}~[{\rm yr^{-1}}])$ & $-7.59 \pm 0.11$ & $-7.54 \pm 0.05$ \\
$\log_{10}({\rm SFR}_{10{\rm Myr}}/{\rm SFR}_{100{\rm Myr}})$ & $0.75\pm 0.14$ & $0.83\pm 0.04$ \\
$E(B-V)_{\rm lines}$ & $0.10 \pm 0.02$ & $0.10 \pm 0.02$ \\
$f_{\rm AGN}$ (1.5 - 7 $\mu$m) & $0.70 \pm 0.02$ & $0.73 \pm 0.07$ \\
$i_{\rm AGN}$ [deg] & $89.96 \pm 1.09$ & $90.00 \pm 0.13$ \\
$\log_{10}(L_{\rm bol}~[{\rm erg~s^{-1}}])$ & $43.77 \pm 0.04^\dagger$ & $43.94 \pm 0.05$ \\
\hline
\hline
\end{tabular}
\tablefoot{For \namemiric, quantities highlighted with $^\dagger$ are not corrected for lensing magnification $\mu = 1.18$. Uncertainties are 1$\sigma$ as derived from {\sc cigale}.}
\end{table}

\begin{figure*}
    \centering
    \includegraphics[width=0.497\linewidth]{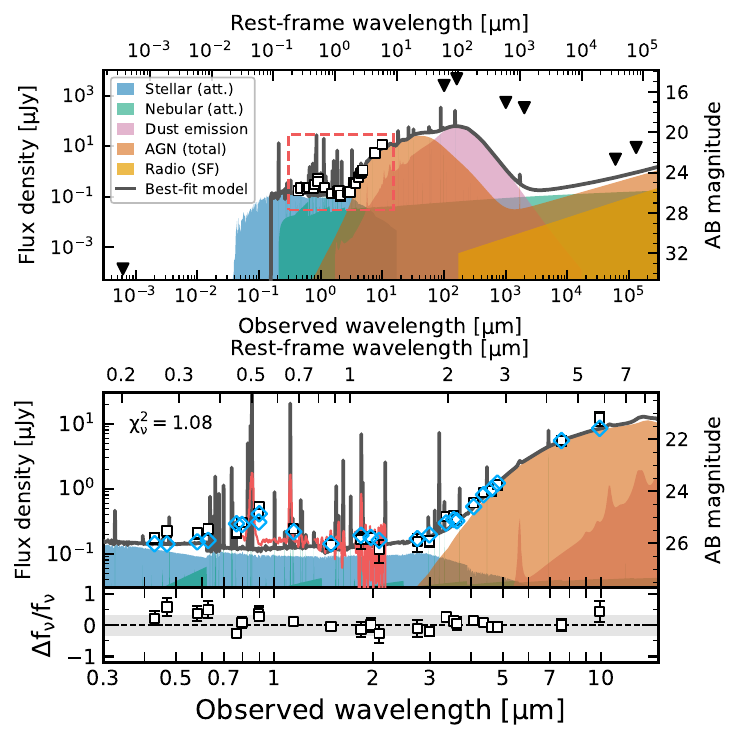}
    \includegraphics[width=0.497\linewidth]{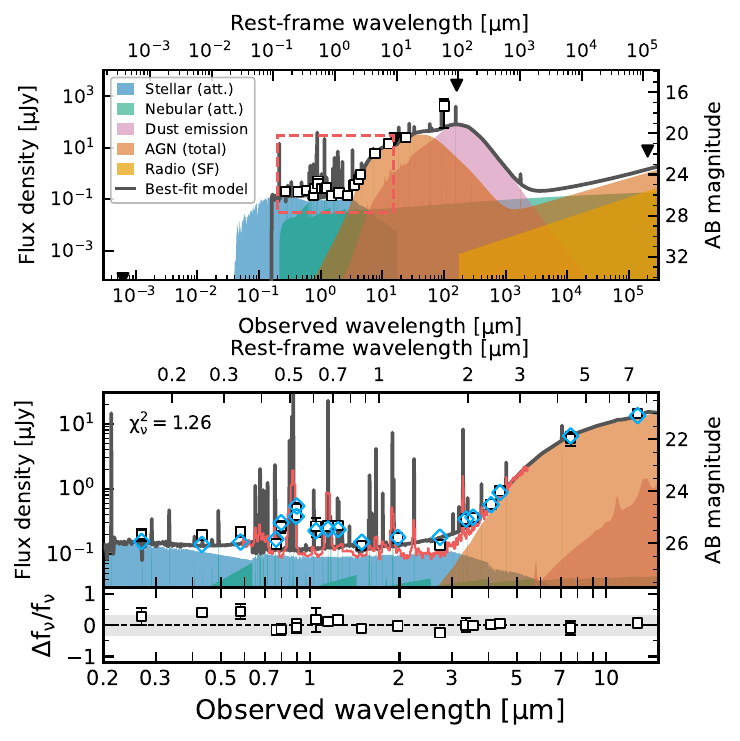}
    \caption{Spectral decomposition of the SED of \namemiric\ (left panels) and \namejades\ (right panels) as obtained by the \textsc{cigale} best-fit models (in grey) highlighting: the attenuated stellar component (light blue), the nebular emission (green), the AGN contribution (orange), the dust emission (purple) and radio emission (yellow). The photometry of our targets is reported with squares while the light blue open diamonds (in the bottom panels) are indicative of the expected (synthetic) photometry predicted by \textsc{cigale} best-fit models. We also show (in red) the observed spectra of our targets. In the case of \namemiric\, to the NIRISS spectrum (only emission lines) we added a continuum emission following our estimates of the optical $\beta$-slope (see \S~\ref{sec:slopes}), i.e. $\beta_{\rm opt} = -2.18$. The bottom inset shows the residuals of the best-fit models. The grey shaded area is representative of residuals $\leq 30\%$.}
    \label{fig:sed_cigale}
\end{figure*}

\begin{figure*}
    \centering
    \includegraphics[width=0.84\textwidth]{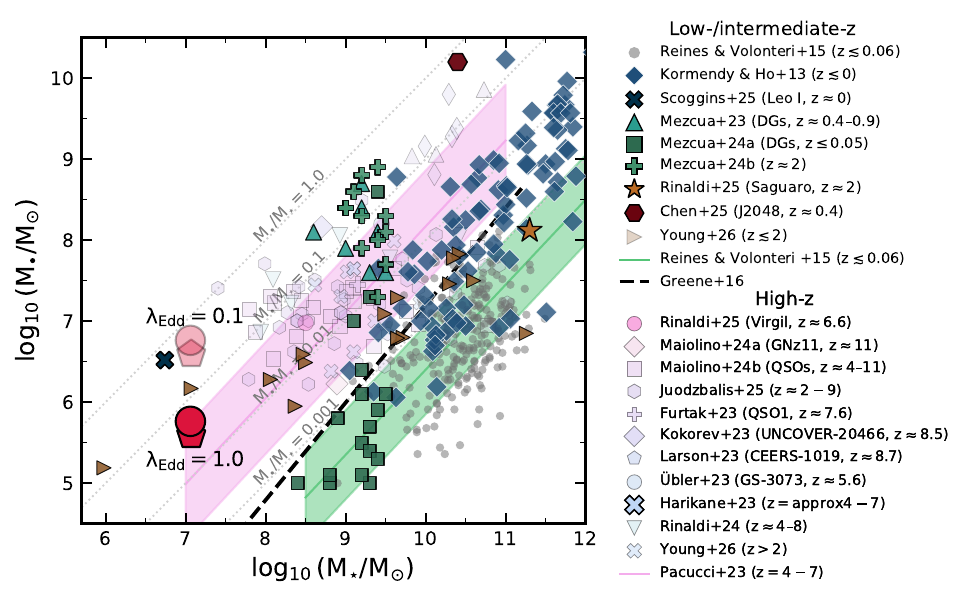}
    \caption{Galaxy $M_\bullet - M_\star$ relation for our targets \namemiric\ (red pentagon) and \namejades\ (red circle) assuming an Eddington ratio $\lambda_{\rm Edd} = 0.1$ (shaded color) and $\lambda_{\rm Edd} = 1$ (full color). We add available literature data from low-redshift galaxies \citep{Kormendy_coevolution_2013, Reines+15, Chen+25, Rinaldi+25b, Mezcua+23, Mezcua+24a, Mezcua+24, Young+26, Scoggins+25} and from high-$z$ studies \citep{Furtak+23, Kokorev+23, Larson+23, Harikane+23, Ubler+23, Maiolino+24, Maiolino_jades_2024, Rinaldi+25c, Rinaldi+25, Juodzbalis+25}. We also report the estimated $M_\star - M_\bullet$ relation from \cite{Reines+15, Greene+16, Pacucci+23}.}
    \label{fig:mbh-mstar_rel}
\end{figure*}

\begin{figure}
    \centering
    \includegraphics[width=\linewidth]{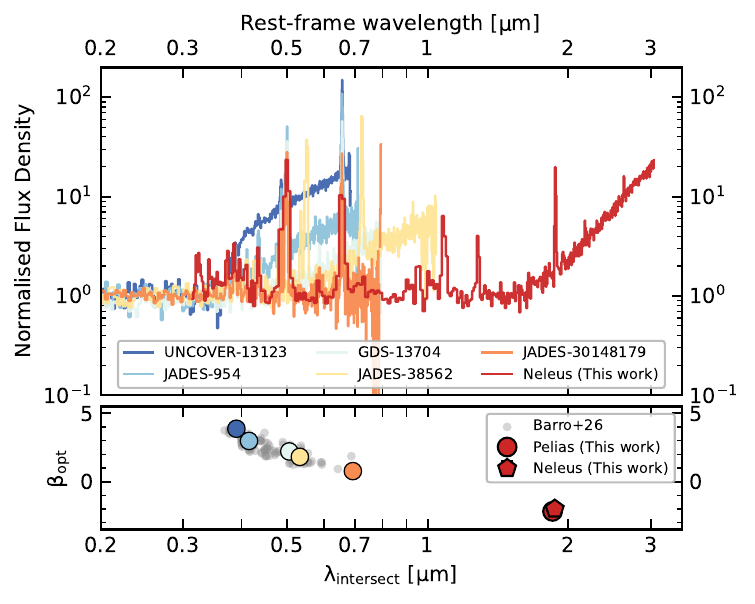}
    \caption{The top panel shows the spectra of five LRDs (UNCOVER-13123, JADES-954, GDS-13704, JADES-38562, JADES-30148179) and the NIRSpec PRISM spectrum of our target \namejades\ (red). All the LRD spectra were normalised at 0.36~$\mu$m. The bottom panel shows the relation between the optical slope $\beta_{\rm opt}$ and the intersection wavelength $\lambda_{\rm intersect}$ where the AGN contribution is comparable to that of the host galaxy. The gray scatter points are LRDs taken from \cite{Barro+26}, while the colored marker are representative of the LRDs whose spectra are shown in the top panel. With red markers we highlight the position of our two targets: \namemiric\ (circle) and \namejades\ (pentagon).}
    \label{fig:lrds_comp_barro}
\end{figure}

\section{Discussion and Conclusions}
\label{sec:discussion}
Our multi-wavelength analysis of \namemiric\ and \namejades\ reveals that both systems are low-mass and compact galaxies hosting a dust-obscured AGN, naturally explaining their steep rest-frame NIR-to-MIR colours. 
This interpretation is further supported by the comparison of the observed photometry with empirical templates (see \S~\ref{sec:known_templates}), which highlights the similarity of their SEDs to composite AGN–starburst systems.

\subsection{Host galaxy properties}
The inferred stellar masses place our targets among the least massive confirmed AGN hosts ($\log_{10}(M_\star~[\rm M_\odot])\approx7.8$; \citealt{Mezcua+18}) and within the extreme low-mass tail of candidate AGN-host DGs compiled in recent studies \citep{Wasleske+24,Messick+25}. 
MIR investigations have demonstrated that black-hole accretion can occur even in galaxies with $\log_{10}(M_\star~[\rm M_\odot])\sim7{-}8$ \citep[e.g.][]{Marleau+17}. 
The stellar masses derived via {\sc cigale} for our targets should nevertheless be interpreted in light of the known limitations of SED fitting in compact, burst-dominated DGs. 
Systems below $\log_{10}(M_\star~[\rm M_\odot])\sim 9$ commonly experience highly stochastic star-formation histories, in which short-lived bursts dominate the luminosity while contributing only with modest stellar mass \citep[e.g.][]{Weisz+12,Amorin+14}. 
This {\it overshining} effect can bias stellar masses toward lower values when young stellar populations dominate the UV–to-optical emission. 
Studies adopting flexible or non-parametric star-formation histories have shown that inferred stellar masses may vary by $\sim 0.3 - 0.5$~dex depending on the assumed SFH parametrisation and the recovery of older stellar components \citep{Pacifici+12, Pacifici+16, Carnall+19, Leja+19}. 
Compact EELGs analysed by \citet{Calabro+17} already highlighted this effect, showing that systems dominated by recent bursts can appear unusually young and low-mass. 
In this context, the use of bursty non-parametric SFHs (see \S~\ref{sec:sed_fitting}) represents a physically motivated modelling choice, and the consistency between independent constraints (e.g. compact exponential morphologies, large emission-line equivalent widths, and extreme sSFRs) supports the interpretation that both galaxies are indeed low-mass systems undergoing rapid recent growth.
We also highlight how a variation of $\sim 0.3 - 0.5$~dex in stellar mass would not qualitatively change our conclusions regarding the nature of these systems as compact starbursting DGs hosting AGN, and it would only marginally affect their precise placement within the low-mass regime. 

The structural properties derived from the JWST imaging (see \S~\ref{sec:compactness}) closely resemble those reported for compact star-forming DGs at similar redshift ($0.1<z<0.9$) studied by \citet{Calabro+17}. 
Circularised effective radii of $r_{e,\rm circ}\sim0.4 - 1$~pkpc and low S\'ersic indices ($n\sim1$) indicate disk-dominated systems with centrally concentrated star formation rather than classical bulge growth. 
At the highest angular resolution, the NIRCam data further reveal that these compact systems can decompose into multiple exponential components. 

From the morphology measured in the F115W filter, which includes the \ha\ emission, we derived star-formation rate surface densities $\Sigma_{\rm SFR}\equiv {\rm SFR}/(2 \pi r_e^2)\approx0.5{-}0.6~{\rm M_\odot~yr^{-1}~kpc^{-2}}$, placing both systems within the regime of intense starbursts \citep{Kennicutt+98}. 
Such values are also characteristic of EELGs at similar redshifts \citep{Amorin+14,Calabro+17}. 
Comparable $\Sigma_{\rm SFR}$ values are observed in massive star-forming clumps at $z\sim1{-}3$, reaching $\sim0.03{-}1~{\rm M_\odot~yr^{-1}~kpc^{-2}}$ \cite[e.g., ][]{Genzel+11, Swinbank+12}, and occasionally even higher for the densest clumps and compact starbursts \cite[e.g., ][]{Fisher+17, Gillman+20}.

The combination of compact structure, low stellar mass, high equivalent widths, and extremely elevated ionisation conditions (O32$>25$ for \namejades) places both galaxies in a regime similar to local {\it Green Pea} (GP) systems, which host many confirmed LyC emitters \citep{Izotov+11, Flury+22}.  
However, the presence of an AGN introduces ambiguity regarding the origin of the ionising radiation, as AGN emission or feedback may either mimic LyC-leaker diagnostics or help create low-column-density escape channels. 
Confirmation of their LyC-leakage therefore requires additional constraints, such as low-ionisation absorption-line covering fractions, Ly$\alpha$ radiative-transfer diagnostics, or direct measurements of escaping radiation below 912~\AA\;that cannot be investigated with the available dataset.

Assuming that the ionised gas traces the gravitational potential and adopting a virial estimator $M_{\rm dyn}(<r_e)=k~\sigma^2 r_e/G$ ($k=3 - 5$), the observed nebular line widths (FWHM~$=70 - 130$~km~s$^{-1}$) imply $\log_{10}(M_{\rm dyn}(<r_e)~[\rm M_\odot]) \simeq 8.3 - 9.6$. 
Compared to the inferred stellar masses ($\log_{10}(M_\star~[\rm M_\odot]) \sim 7$), stars would therefore contribute only a minor fraction ($\lesssim 10\%$) of the enclosed mass, consistent with the high gas fractions and large dynamical-to-stellar mass ratios commonly observed in low-mass, gas-rich DGs \cite[e.g.][]{Geha+06, Courteau+14, Bradford+15}. 
We stress, however, that this estimate relies on nebular emission lines rather than stellar absorption features. 
Dynamical measurements based on ionised gas can differ systematically from those derived from stellar tracers, as different components probe distinct dynamical structures within galaxies \cite[][]{Courteau+14}. 
In compact starbursts, ionised-gas velocity dispersions are frequently dominated by turbulence and stellar-feedback–driven motions rather than purely gravitational broadening \cite[][]{Moiseev+15}, while unresolved rotation and multi-component gas structures can further increase the observed line widths \cite[][]{Hogarth+20}. 
In addition, in the presence of an AGN, narrow-line or extended emission-line region kinematics and outflows may significantly broaden nebular emission lines and bias dynamical mass estimates \citep{Holt+08}.
The derived $M_{\rm dyn}$ should therefore be regarded as an order-of-magnitude constraint and likely an upper limit. 
Indeed, for the measured stellar masses and continuum sizes ($r_e \sim 0.3 - 1$~pkpc), the velocity dispersion expected from the stellar potential alone is only $\sigma_\star\lesssim 10$~km~s$^{-1}$, significantly lower than the observed nebular dispersions. 
If interpreted as due to pure gravitational origin, the measured line widths would require substantially more enclosed mass than provided by stars alone, a smaller effective radius, or additional non-gravitational broadening mechanisms such as turbulence, outflows, or AGN-driven motions.

\subsection{Super-Eddington accretion and/or overmassive BH?}
The best-fit obtained with {\sc cigale} suggest that our targets AGN have bolometric luminosities of $\log_{10}(L_{\rm bol}~[\rm erg~s^{-1}]) \approx 43.7 - 43.9$.
If we extrapolate to the DGs regime the local $M_\bullet$ - $M_\star$ (or $M_\bullet$ - $M_{\rm bulge}$) relations \citep[e.g.][]{Reines+15,Greene+16}, for galaxies with $\log_{10}(M_\star~[\rm M_\odot]) \approx 7$ the extrapolated $M_\bullet/M_\star$ values are $\approx 10^{-3} - 10^{-2}$ thus implying $\log_{10}(M_\bullet ~[\rm M_\odot])\approx 4 - 5$ (depending on the relation adopted).
Assuming such values and considering the values of $L_{\rm bol}$ derived via {\sc cigale}, we would expect an Eddington ratio $\lambda_{\rm Edd} \equiv L_{\rm bol}/L_{\rm Edd} \approx 3 - 70 $, formally implying super-Eddington accretion. 
In this regime the standard thin SMBH accretion disk is expected to transition into a geometrically thick (“slim”) disk in which photon trapping reduces the radiative efficiency, allowing mass accretion rates to exceed the classical Eddington limit \cite[e.g.][]{Abramowicz+88, Ohsuga+02, Massonneau+23}. 
Such flows are predicted to drive strong radiation-pressure–driven winds and produce anisotropic emission, so that the observed luminosity may surpass the isotropic Eddington limit without violating local force balance \cite[e.g.][]{Dotan+11, Wang+14}. 
Super-Eddington phases are thought to enable rapid early black-hole growth, particularly in low-mass galaxies; however, similarly high inferred $\lambda_{\rm Edd}$ values may also arise from systematic uncertainties in black-hole mass estimates when local scaling relations are extrapolated into the dwarf regime \cite[e.g.][]{Reines+15, Maiolino+24}.
In fact, if we assume $\lambda_{\rm Edd} = 0.1 -1$, i.e. the classical AGN/quasar regime of efficient, radiatively-thin accretion,  our targets AGN would have SMBH masses on the order of $\log_{10}(M_\bullet~[\rm M_\odot]) = 5.8 - 6.8$ (see \S~\ref{sec:sed_fitting}), assuming $\lambda = 1$ and 0.1, respectively.
When compared to the total stellar mass of the host galaxies, such SMBH masses result in constituting from 6\% to 60\% of it.
Even allowing for substantial systematic uncertainties in the SED-based $M_\bullet$ estimates, such ratios lie well above the extrapolation of the local $M_\bullet$ - $M_\star$ relation into the dwarf regime, see Figure~\ref{fig:mbh-mstar_rel}. 
This suggests that either the black holes in these systems are already over-massive relative to their hosts, or that we are catching the galaxies at an early phase in which the stellar component has not yet “caught up” with the rapid black hole growth.
Interestingly, similarly extreme black-hole-to-stellar mass ratios have been proposed in other low-$z$ dwarf galaxies. 
The Local Group dwarf Leo I \cite[$M_\star\approx 10^{6.7} ~\rm M_\odot$][]{McConnachie+12} may host a black hole with mass comparable to that of its stellar component, implying $M_\bullet / M_\star \approx 1$, far above the local scaling relations and potentially reflecting the relic signature of an early heavy-seed black hole \cite[e.g.,][]{Bustamanate-Rossel+21, Scoggins+25}.
Recently, also \cite{Young+26} reported the identification of two galaxies at $z \lesssim 1$ with candidate intermediate mass BHs that also lie above the local $M_\bullet - M_\star$ relation, one of which at similar $M_\star$ to our targets (ID 3976, $z_{\rm phot} = 1.03$).

The absence of X-ray detections in the available \textit{Chandra} observations provides an additional constraint on the nature of the accretion. 
Both sources remain undetected in the 0.5 - 7 keV band, with $3\sigma$ flux upper limits of $1.22\times10^{-15}$ and $6.96\times10^{-16}\rm ~erg~s^{-1}~cm^{-2}$ for \namemiric\ and \namejades, respectively. 
Assuming a power-law spectrum with photon index $\Gamma=1.7$, these limits correspond to $\log_{10}(L_{2-10~\rm keV}~[\rm erg~s^{-1}]) \lesssim 42.2$ and $\lesssim 42.0$. 
For comparison, the bolometric luminosities inferred from {\sc cigale} ($\log_{10}( L_{\rm bol})\simeq 43.6 - 43.9$) would predict $\log_{10}(L_{2-10~\rm keV}~ [\rm erg~s^{-1}]) \sim 42.3-43.1$ (taking into account intrinsic scatter of the bolometric correction relation) for AGN accreting in the classical radiatively efficient regime, suggesting that both objects are X-ray weak by at least a factor of a few relative to typical AGN type II expectations \cite[e.g.][]{Lusso+10, Duras+20}. 
Such X-ray weakness is qualitatively consistent with super-Eddington accretion scenarios, in which thick inner disks and radiation-driven winds can suppress or reprocess coronal emission, as found for some high-$\lambda_{\rm Edd}$ AGN and Broad Absorption Line (BAL) quasars \citep[e.g.][]{Luo+14,Teng+14,Laurenti+22,Pacucci+24,Ishibashi+25, Ighina+25}.
We underline, however, that the current X-ray limits support, but do not uniquely require, a super-Eddington interpretation.
In fact, X-ray weakness can arise from heavy obscuration, including Compton-thick or near–Compton-thick columns that strongly attenuate the intrinsic X-ray emission \citep[e.g.][]{Maiolino+98,Comastri04}. 

Independent insights on the possible disk–corona connection, i.e. how efficiently the accretion energy is transferred from an AGN disk into the corona, can be derived from the $\alpha_{OX}$ parameter ($\equiv 0.3839\cdot \log_{10}(f({\rm 2keV})/f({\rm 2500\AA}))$; e.g., \citealt{Tananbaum+79, Lusso+10}) and can provide an important diagnostic: super-Eddington accretion is expected to produce comparatively X-ray–weak sources with steeper $\alpha_{OX}$, whereas systems accreting in the classical thin-disk regime should follow the standard $\alpha_{OX}$ – UV luminosity relation.
For both our targets, we can derive $f({\rm 2500\AA})$ directly from the \hst/F435W filter that covers such rest-frame wavelengths at our targets redshifts.
However, because the rest-frame UV emission traced by the F435W photometry in our target is likely originated by a combination of stellar, nebular, and AGN light, meaningful constraints on the $\alpha_{\rm OX}$ parameter cannot be derived without isolating the AGN contribution to the UV continuum.
In fact, assuming that all the F435W flux is due to the AGN component, we would derive $\alpha_{OX} \lesssim -1.1$ for both targets, thus formally placing our targets in the X-ray-weak regime. 
Nonetheless, in the case of strong dilution of the AGN continuum by a compact, intense starburst that dominates the soft X-ray and UV output would mimic the $\alpha_{OX}$ values obtained for an intrinsically X-ray-weak source \citep[e.g.][]{Guainazzi+09,Brandt+15}. 
In fact, if we assume the best-fit values from {\sc cigale}, we would expect $\alpha_{OX} \lesssim -0.7$ implying that, in this case, the current depth of the X-ray observations is not enough to properly constrain the AGN X-ray emission.

In the context of DGs, such MIR-bright but X-ray-faint AGN are particularly intriguing, as they may trace short-lived episodes of rapid, dust-enshrouded black hole growth in low-mass systems undergoing bursty star formation and stochastic accretion. 
The combination of extremely low host-galaxy masses, high $\Sigma_{\rm SFR}$, and X-ray weak, MIR-luminous AGN therefore suggests that \namemiric\ and \namejades\ may be experiencing a phase of rapid, heavily obscured black hole growth that remains largely invisible even in deep X-ray observations \cite[e.g.][]{Pacucci+26}.

\subsection{Are \namemiric\ and \namejades\ possible low-mass analogues of blue Hot DOGs?}
\label{sec:hot_dogs}
The MIR-bright and X-ray–weak nature of \namemiric\ and \namejades\ invites comparison with dust-obscured AGN populations identified at higher stellar masses, in particular blue DOGs \cite[e.g,][]{Noboriguchi+19} and blue Hot DOGs \citep[e.g.,][]{Assef+16,Assef+20}. 
Blue DOGs retain strong mid-infrared emission produced by AGN-heated dust while exhibiting relatively blue rest-frame UV/optical colours, consistent with composite starburst–AGN systems \citep{Dey+08,Melbourne+12,Riguccini+15}. 
Many objects in these populations are X-ray weak compared to their MIR luminosities, suggesting either heavy nuclear obscuration or intrinsically suppressed coronal emission.

An even closer analogue is provided by the blue Hot DOGs discussed by \citet{Assef+20}, which combine AGN-dominated MIR SEDs, blue UV/optical continua, and pronounced X-ray weakness. 
These authors argue that such properties may arise from very high, possibly super-Eddington accretion rates capable of disrupting or weakening the X-ray corona. 
This interpretation closely mirrors that inferred for \namemiric\ and \namejades, whose SED shapes constraints point toward rapid, dust-enshrouded black-hole growth.

The comparison with the best-fit templates of three Hot Blue DOGs from \cite{Assef+20} (see \S~\ref{sec:known_templates}), however, showed that while the longest wavelengths are well-reproduced, the templates systematically under-predict the flux at the shortest wavelengths thus suggesting that in our target the role of the host galaxy and/or the scattered light from the dust-obscured AGN could play an even more prominent role. 
Besides, while blue DOGs and Hot DOGs are typically hosted by massive galaxies ($\log_{10}(M_\star~[\rm M_\odot]) = 10 - 11$), our targets reside in extremely low-mass, compact systems with high star-formation-rate surface densities. 

They may therefore represent low-mass analogues of this phenomenon: short-lived phases of buried, X-ray–weak AGN activity in which rapid black-hole growth dominates the MIR emission even in DGs. 
We note that if part of the rest-frame UV–optical continuum originates from the accretion disk, as observed in blue Hot DOGs, the stellar masses derived from SED fitting could be overestimated, further increasing the tension with the local $M_\bullet$–$M_\star$ relation.

\subsection{A possible connection to LRDs?}
\label{sec:lrds}
The shape of the SEDs observed in our targets also motivates a comparison with the recently identified class of LRDs, which typically exhibit the characteristic “V-shaped” SED, with flat UV continuum and a rising optical component, often with the transition occurring near the Balmer limit \cite[$\approx3600$~\AA; e.g.,][]{Wang+24a, Furtak+23, Labbe+24, Setton_little_2025}.
In the observed frame, our targets show a similar SED to the LRDs' SED: adopting the photometric selection criteria proposed for LRDs (e.g., \citealt{Barro+24}), both galaxies would formally qualify as photometrically selected LRD candidates, with colours (F150W$-$F200W)~$\approx0.3$, (F277W$-$F444W)~$\approx2$, and compact morphologies in F444W ($c_{\rm F444W}<2$). 
In our case, however, when taking into account redshift, the rise of the SED occurs at significantly longer wavelengths, well beyond the one observed in more canonical LRDs. 
Within the LRD population, however, recent studies have reported clear diversity in SED shapes \cite[e.g.,][]{Barro+26, Perez-Gonzalez+26}. 
While many LRDs exhibit pronounced rest-frame optical continuum curvature and SED turnovers near strong Balmer breaks \cite[e.g.,][]{ Wang+24a, Furtak+23, Labbe+24, Setton_little_2025}, others display smoother, power-law–like continua with turnover taking place at longer wavelengths \cite[e.g.,][]{Perez-Gonzalez+24, Iani+25, Rinaldi+25}. 
Indeed, using a sample of 118 LRDs at $z \gtrsim 3$ and assuming a two-components model made of a star forming host galaxy and dense hot gas envelopes around accreting SMBH \cite[the so-called BH$^\star$ model,][]{Naidu+25}, \cite{Barro+26} showed that these two behaviours form a continuous sequence: systems with power-law optical slopes tend to exhibit SED turnovers at progressively longer wavelengths, establishing a correlation between the rest-frame optical slope ($\beta_{\rm opt}$) and the turnover wavelength ($\lambda_{\rm intersect}$). 
In particular, LRDs with bluer optical slopes show increasingly redshifted turnover wavelengths, as observed in sources such as {\it Virgil} and JADES-30148179 ($\beta_{\rm opt}\approx0.8$, $\lambda_{\rm intersect}\approx7000$~\AA, \citealt{Barro+26}). 
In this framework, our targets, with $\beta_{\rm opt}\approx -2.3$ and $\lambda_{\rm intersect}\approx1.8~\mu$m, would extend this relation toward more extreme values while remaining broadly consistent with its extrapolation (Fig.~\ref{fig:lrds_comp_barro}).

Beyond the location of the SED turnover, our objects also appear consistent with other several trends reported by \cite{Barro+26}. 
In particular, systems with bluer optical slopes were shown to preferentially exhibit lower stellar masses ($\log_{10}(M_\star~[\rm M_\odot]) = 7 - 8$), lower stellar dust attenuation ($A_V \lesssim 0.5$~mag), reduced H$\alpha$/H$\beta$ ratios, narrower emission lines (no broad component), and higher [\ion{O}{iii}]/H$\beta$ ratios. 
The observed properties of our sources qualitatively follow this behaviour. 
We caution, however, that $\beta_{\rm opt}$ is not measured here using the exact same definition adopted by \cite{Barro+26}, and therefore the comparison should be regarded as indicative rather than strictly quantitative. 

Differently from the \cite{Barro+26} sample, which probes higher redshifts, our sources are located at substantially lower redshift. 
While LRD analogues have been reported even down to $z\sim0.1$, these objects generally show clear continuum curvature and prominent Balmer breaks \cite[e.g.][]{Juodzbalis_JADES_2024, Chen+25, Ji_lord_2025, Ma_counting_2025, Lin_discovery_2025, Loiacono_big_2025, Rinaldi+25b}.
If confirmed, our targets may represent the first low-redshift examples consistent with the power-law subset of the LRD population.

\subsection{The unique role of MIRI in detecting these sources}
\label{sec:miri_impact}
\begin{figure}
    \centering
    \includegraphics[width=\linewidth]{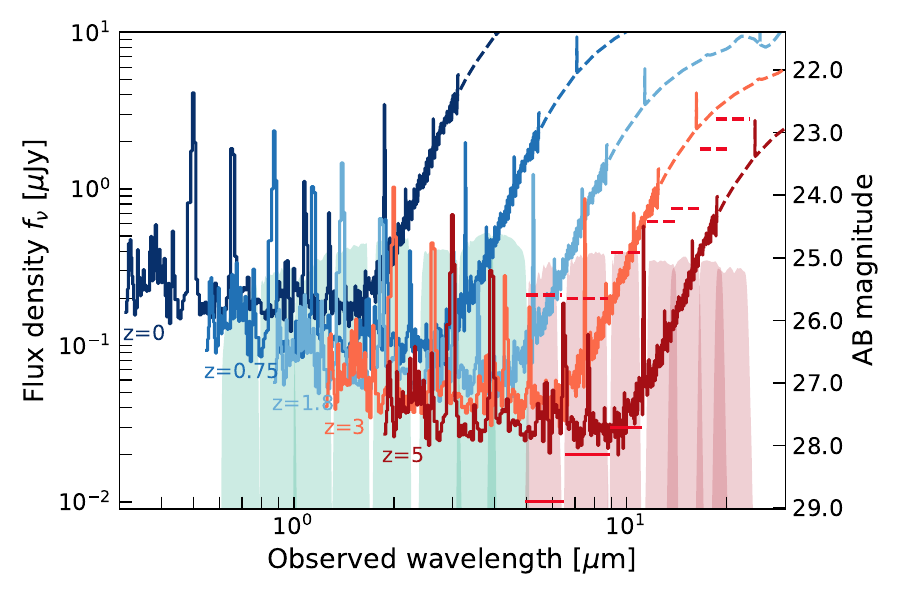}
    \caption{Example of cosmological redshift impact on the detection of objects similar to our targets. We present from blue to red the spectrum of \namejades\ at redshifts $z = 0, 0.75, 1.5, 3, 5$. With only NIRCam coverage (green filters) and no MIRI bands (red filters), the red NIR-to-MIR rest-frame slope of our target could not be observed at $z \geq 1.8$. For the MIRI filters, we also show the $5\sigma$ depth that has been reached in each filter by the SMILES \cite[red dashed lines;][]{Rieke+24, Alberts+24} and MIDIS + MIDIS-RED \cite[red solid lines; e.g.][]{Oestlin+25}.}
    \label{fig:miri_detectability}
\end{figure}

An important implication of our results concerns the detectability of analogous systems at earlier cosmic epochs. 
The distinctive photometric signature that enabled the identification of \namemiric\ and \namejades\ (the sharp SED upturn produced by hot dust emission from an obscured AGN) is strongly wavelength dependent. 
At the redshifts of our targets, this feature appears at observed wavelengths $\gtrsim2.7~\mu$m and is therefore accessible to \jwst\ NIRCam and MIRI imaging. 
For otherwise similar systems at higher redshift, however, the hot-dust excess rapidly shifts beyond the wavelength range probed by NIRCam alone: the NIR-to-MIR rise moves to $\lambda_{\rm obs}\gtrsim5~\mu$m at $z\gtrsim1.8$ and beyond $10~\mu$m by $z\gtrsim5$, leaving MIRI as the only instrument capable of tracing this component, see Figure~\ref{fig:miri_detectability}.
As a consequence, galaxies analogous to our targets observed during the Epoch of Reionisation would likely appear in rest-frame UV–optical data as compact, blue star-forming systems, with little or no photometric indication of AGN activity (similarly to {\it Virgil}; \citealt{Iani+25, Rinaldi+25}). 
In the absence of long-wavelength MIR observations, the dominant hot-dust emission would remain undetected, potentially biasing classifications toward purely stellar interpretations. 
This selection effect suggests that obscured accretion in low-mass galaxies may be systematically under-represented in current $z > 2$ samples.
Besides, to detect such objects at $z > 2$ in the MIRI bands, deep MIRI imaging is required: at $z \approx 3$ depths comparable to those of SMILES \citep{Rieke+24, Alberts+24} would suffice to identify similar targets but at $z \geq 5$ depths of more than 25 mag would be necessary at the longest MIRI wavelength channels (i.e. $\geq 15~\mu$m).

If such compact, dust-enshrouded AGN phases are common, their contribution to the ionising photon budget during reionisation (either directly through AGN emission or indirectly via feedback-regulated escape of ionising radiation) may be underestimated. 
Following \cite{Leitherer+95}, we can estimate the ionizing photon production efficiency (assuming $f_{\rm esc, LyC} = 0$) to be $\xi_{\rm ion,0} = 25.7 - 25.8~\rm Hz~erg^{-1}$.
These $\xi_{\rm ion,0}$ estimates are also in agreement with the $\xi_{\rm ion,0} - {\rm EW_0(H\alpha)}$ and $\xi_{\rm ion,0} - M_{UV}$  reported in recent literature \cite[e.g.][]{Prieto-Lyon+23, Simmonds+23, Rinaldi+24}. 
We note that in case of $f_{\rm esc, LyC} > 0$ (as possibly suggested by their compact structure, low stellar mass, high O32), the ionizing photon production efficiency would be even higher than the one predicted above.
Our results therefore emphasise the critical importance of long-wavelength \jwst\ observations, and in particular deep and wide MIRI surveys, for uncovering a population of rapidly growing black holes in low-mass galaxies that would otherwise remain hidden. 
In this context, \namemiric\ and \namejades\ may represent nearby analogues of systems whose higher-redshift counterparts currently evade detection, offering a direct window into an obscured phase of early galaxy – SMBH co-evolution.

\section{Summary}
\label{sec:summary}

Using deep \hst\ and \jwst\ imaging (NIRCam and MIRI) and spectroscopic (NIRSpec and NIRISS) observations of the lensing cluster M0416 and the GOODS-N field, we reported the discovery and detailed characterisation of two compact galaxies, \namemiric\ and \namejades. 
Our main results can be summarised as follows:

\begin{itemize}

\item Both sources were previously classified as compact blue galaxies in pre-\jwst\ imaging. 
\jwst\ observations revealed an unexpected steep upturn in their spectral energy distributions at observed wavelengths $\gtrsim 2.7\,\mu$m. 
NIRSpec and NIRISS spectroscopy placed them at $z\approx 0.71$ and $z\approx 0.75$ through the detection of strong optical emission lines, including \ha\ and \hb+\oiii.

\item Their SEDs show an unusual combination of a blue optical continuum ($\beta_{\rm opt}\approx-2.1$) and very red NIR-to-MIR colours ($\beta_{\rm NIR}>2$), which cannot be reproduced by standard galaxy templates. 
SED modelling with \textsc{cigale} favours dwarf galaxies with $\log_{10}(M_\star~[\rm M_\odot])\approx 7$ hosting obscured AGN with bolometric luminosities $\log_{10}(L_{\rm bol}\,[{\rm erg\,s^{-1}}])\approx43.7 - 44$.

\item Morphological modelling of the rest-frame UV--optical emission reveals compact disk-like structures with S\'ersic indices $n\approx1$ and circularised effective radii $r_{e,\rm circ}\sim 0.4 - 1$~pkpc. 
These systems exhibit high star-formation rate surface densities ($\Sigma_{\rm SFR}\approx0.5 - 0.6~{\rm M_\odot~yr^{-1}~kpc^{-2}}$), consistent with compact starbursts and EELGs at similar redshift.

\item Although both galaxies are X-ray undetected, multiple diagnostics supported the presence of AGN activity. 
Pure stellar or dust models failed to reproduce the observed MIR luminosities without violating energy-balance constraints, while nebular line ratios indicated only modest attenuation ($A_V \approx 0.2$~mag), ruling out heavily obscured star formation as the dominant power source. 
Excitation diagnostics further revealed extremely high ionisation conditions and elevated [\ion{O}{iii}]/\hb\ ratios, consistent with a hard ionising spectrum.

\item The bolometric luminosities inferred from SED modelling implied black-hole masses of $\log_{10}(M_\bullet~[\rm M_\odot])\approx 5.8 $ assuming Eddington-limited accretion ($\lambda_{\rm Edd} \leq 1$). 
These correspond to unusually large $M_\bullet/M_\star$ ratios relative to local scaling relations, suggesting either over-massive black holes or systems observed during a phase in which black-hole growth temporarily outpaces stellar assembly.

\item The SED shape and photometric properties connect these systems to the recently identified population of Little Red Dots (LRDs). 
Their stellar masses and colours suggest that they may represent lower-redshift analogues of the same physical phase: a compact, dust-embedded accretion episode in which AGN emission dominates while the host galaxy is still assembling.

\item These galaxies lie at the intersection of several populations often studied separately (e.g. EELGs, candidate LyC-leakers, dust-obscured AGNs, LRDs) suggesting that they may trace a common evolutionary stage driven by compact gas inflows, bursty star formation, and rapid black-hole growth.

\end{itemize}

Overall, \namemiric\ and \namejades\ demonstrate that rapid, dust-enshrouded black-hole growth can occur in galaxies with stellar masses of only $\sim10^7\rm ~M_\odot$. 
Their discovery highlights the unique role of long-wavelength \jwst\ observations in uncovering compact, low-mass AGN hosts that would remain hidden in UV-to-NIR surveys alone, and suggests that similar systems at higher redshift may currently evade identification without deep MIRI coverage.

Future observations will be essential to further constrain this interpretation. 
High-resolution rest-frame optical spectroscopy with \jwst/NIRSpec\ IFU could disentangle AGN and star-forming components through spatially resolved kinematics and excitation diagnostics. 
Deep X-ray observations with \textit{Chandra} or future missions such as \textit{Athena} will test whether the nuclei are intrinsically X-ray weak or heavily obscured. 
Mid-infrared spectroscopy with \jwst/MIRI\ will directly probe hot-dust continua and high-ionisation lines, while ALMA observations of cold gas tracers could constrain gas fractions and dynamical masses. 
At longer timescales, facilities such as the \textit{Roman Space Telescope} and ELT-class observatories will enable systematic searches for similar low-mass obscured AGN and resolve their internal structure, providing crucial tests of whether the embedded accretion phase inferred here represents a common pathway in DG evolution.

\begin{acknowledgements}
We thank Carys J. E. Gilbert and Luigi Barchiesi for useful feedback on the estimate of the X-rays upper-limits.
This research has made use of data obtained from the Chandra Source Catalog, provided by the Chandra X-ray Center (CXC).
Some of the data products presented herein were retrieved from the \textsc{Dawn JWST Archive} (DJA). DJA is an initiative of the Cosmic Dawn Center (DAWN), which is funded by the Danish National Research Foundation under grant DNRF140.
J.M., A.T., I.K, C.D.C. acknowledge funding by the European Union (ERC, AGENTS,  101076224).
GN acknowledges support by the Canadian Space Agency under a contract with NRC Herzberg Astronomy and Astrophysics.
RNC, KIC and GD acknowledge funding from the Dutch Research Council (NWO) through the award of the Vici Grant
VI.C.212.036.
AAH acknowledges support from grant PID2021-124665NB-I00  funded by MCIN/AEI/10.13039/501100011033 and by ERDF A way of making Europe. 
PB acknowledges financial support through grant PRIN-MIUR 2020SKSTHZ and support from the Italian Space Agency (ASI) through contract ``Euclid - Phase E'', INAF Grants ``The Big-Data era of cluster lensing'' and ``Probing Dark Matter and Galaxy Formation in Galaxy Clusters through Strong Gravitational Lensing''.
\end{acknowledgements}



\bibliographystyle{aa}
\bibliography{paper}

@ARTICLE{Barro+26,
       author = {{Barro}, Guillermo and {Perez-Gonzalez}, Pablo G. and {Kocevski}, Dale and {Trump}, Jonathan R. and {Dickinson}, Mark and {Arrabal Haro}, Pablo and {Brooks}, Madisyn and {Donnan}, Callum T. and {Dunlop}, James S. and {Finkelstein}, Steven L. and {Franco}, Maximilien and {Gandolfi}, Giovanni and {Giavalisco}, Mauro and {Grogin}, Norman A. and {Hirschmann}, Michaela and {Kartaltepe}, Jeyhan S. and {Koekemoer}, Anton M. and {Larson}, Rebecca L. and {Leung}, Gene C.~K. and {Lucas}, Ray A. and {McGrath}, Elizabeth J. and {Papovich}, Casey and {Perez-Diaz}, Borja and {Somerville}, Rachel S. and {Taylor}, Elizabeth and {Taylor}, Anthony J. and {Tripodi}, Roberta and {Yung}, L.~Y. Aaron and {Wang}, Xin},
        title = "{From ``The Cliff'' to ``Virgil'': Mapping the Spectral Diversity of Little Red Dots with JWST/NIRSpec}",
      journal = {arXiv e-prints},
         year = 2025,
        month = dec,
          eid = {arXiv:2512.15853},
        pages = {arXiv:2512.15853},
          doi = {10.48550/arXiv.2512.15853},
archivePrefix = {arXiv},
       eprint = {2512.15853},
 primaryClass = {astro-ph.GA},
       adsurl = {https://ui.adsabs.harvard.edu/abs/2025arXiv251215853B}
}

@ARTICLE{Backhaus+22,
       author = {{Backhaus}, Bren E. and {Trump}, Jonathan R. and {Cleri}, Nikko J. and {Simons}, Raymond and {Momcheva}, Ivelina and {Papovich}, Casey and {Estrada-Carpenter}, Vicente and {Finkelstein}, Steven L. and {Matharu}, Jasleen and {Ji}, Zhiyuan and {Weiner}, Benjamin and {Giavalisco}, Mauro and {Jung}, Intae},
        title = "{CLEAR: Emission-line Ratios at Cosmic High Noon}",
      journal = {\apj},
         year = 2022,
        month = feb,
       volume = {926},
       number = {2},
          eid = {161},
        pages = {161},
          doi = {10.3847/1538-4357/ac3919},
archivePrefix = {arXiv},
       eprint = {2109.08147},
 primaryClass = {astro-ph.GA},
       adsurl = {https://ui.adsabs.harvard.edu/abs/2022ApJ...926..161B}
}

@ARTICLE{Tang+25,
       author = {{Tang}, Mengtao and {Stark}, Daniel P. and {Mason}, Charlotte A. and {Gelli}, Viola and {Chen}, Zuyi and {Topping}, Michael W.},
        title = "{The JWST Spectroscopic Properties of Galaxies at $z=9-14$}",
      journal = {arXiv e-prints},
         year = 2025,
        month = jul,
          eid = {arXiv:2507.08245},
        pages = {arXiv:2507.08245},
          doi = {10.48550/arXiv.2507.08245},
archivePrefix = {arXiv},
       eprint = {2507.08245},
 primaryClass = {astro-ph.GA},
       adsurl = {https://ui.adsabs.harvard.edu/abs/2025arXiv250708245T}
}

@ARTICLE{Cairos+15,
       author = {{Cair{\'o}s}, L.~M. and {Caon}, N. and {Weilbacher}, P.~M.},
        title = "{VIMOS integral field spectroscopy of blue compact galaxies. I. Morphological properties, diagnostic emission-line ratios, and kinematics}",
      journal = {\aap},
         year = 2015,
        month = may,
       volume = {577},
          eid = {A21},
        pages = {A21},
          doi = {10.1051/0004-6361/201322518},
       adsurl = {https://ui.adsabs.harvard.edu/abs/2015A&A...577A..21C}
}

@ARTICLE{Backhaus+23,
       author = {{Backhaus}, Bren E. and {Bridge}, Joanna S. and {Trump}, Jonathan R. and {Cleri}, Nikko J. and {Papovich}, Casey and {Simons}, Raymond C. and {Momcheva}, Ivelina and {Holwerda}, Benne W. and {Ji}, Zhiyuan and {Jung}, Intae and {Matharu}, Jasleen},
        title = "{CLEAR: Spatially Resolved Emission Lines and Active Galactic Nuclei at 0.6 < z < 1.3}",
      journal = {\apj},
         year = 2023,
        month = jan,
       volume = {943},
       number = {1},
          eid = {37},
        pages = {37},
          doi = {10.3847/1538-4357/aca668},
archivePrefix = {arXiv},
       eprint = {2207.11265},
 primaryClass = {astro-ph.GA},
       adsurl = {https://ui.adsabs.harvard.edu/abs/2023ApJ...943...37B}
}

@ARTICLE{Feuillet+24,
       author = {{Feuillet}, L{\'e}a M. and {Mel{\'e}ndez}, Marcio and {Kraemer}, Steve and {Schmitt}, Henrique R. and {Fischer}, Travis C. and {Reeves}, James N.},
        title = "{Classifying Intermediate-redshift Galaxies in SDSS: Alternative Diagnostic Diagrams}",
      journal = {\apj},
         year = 2024,
        month = feb,
       volume = {962},
       number = {2},
          eid = {104},
        pages = {104},
          doi = {10.3847/1538-4357/ad1a09},
archivePrefix = {arXiv},
       eprint = {2312.17381},
 primaryClass = {astro-ph.GA},
       adsurl = {https://ui.adsabs.harvard.edu/abs/2024ApJ...962..104F}
}

@ARTICLE{Tripodi+24,
       author = {{Tripodi}, Roberta and {D'Eugenio}, Francesco and {Maiolino}, Roberto and {Curti}, Mirko and {Scholtz}, Jan and {Tacchella}, Sandro and {Marconcini}, Cosimo and {Bunker}, Andrew J. and {Trussler}, James A.~A. and {Cameron}, Alex J. and {Arribas}, Santiago and {Baker}, William M. and {Brada{\v{c}}}, Maru{\v{s}}a and {Carniani}, Stefano and {Charlot}, St{\'e}phane and {Ji}, Xihan and {Ji}, Zhiyuan and {Robertson}, Brant and {{\"U}bler}, Hannah and {Venturi}, Giacomo and {Willmer}, Christopher N.~A. and {Witstok}, Joris},
        title = "{Spatially resolved emission lines in galaxies at 4 {\ensuremath{\leq}} z < 10 from the JADES survey: Evidence for enhanced central star formation}",
      journal = {\aap},
         year = 2024,
        month = dec,
       volume = {692},
          eid = {A184},
        pages = {A184},
          doi = {10.1051/0004-6361/202449980},
archivePrefix = {arXiv},
       eprint = {2403.08431},
 primaryClass = {astro-ph.GA},
       adsurl = {https://ui.adsabs.harvard.edu/abs/2024A&A...692A.184T}
}

@ARTICLE{Trump+23,
       author = {{Trump}, Jonathan R. and {Arrabal Haro}, Pablo and {Simons}, Raymond C. and {Backhaus}, Bren E. and {Amor{\'\i}n}, Ricardo O. and {Dickinson}, Mark and {Fern{\'a}ndez}, Vital and {Papovich}, Casey and {Nicholls}, David C. and {Kewley}, Lisa J. and {Brunker}, Samantha W. and {Salzer}, John J. and {Wilkins}, Stephen M. and {Almaini}, Omar and {Bagley}, Micaela B. and {Berg}, Danielle A. and {Bhatawdekar}, Rachana and {Bisigello}, Laura and {Buat}, V{\'e}ronique and {Burgarella}, Denis and {Calabr{\`o}}, Antonello and {Casey}, Caitlin M. and {Ciesla}, Laure and {Cleri}, Nikko J. and {Cole}, Justin W. and {Cooper}, M.~C. and {Cooray}, Asantha R. and {Costantin}, Luca and {Croton}, Darren and {Ferguson}, Henry C. and {Finkelstein}, Steven L. and {Fujimoto}, Seiji and {Gardner}, Jonathan P. and {Gawiser}, Eric and {Giavalisco}, Mauro and {Grazian}, Andrea and {Grogin}, Norman A. and {Hathi}, Nimish P. and {Hirschmann}, Michaela and {Holwerda}, Benne W. and {Huertas-Company}, Marc and {Hutchison}, Taylor A. and {Jogee}, Shardha and {Juneau}, St{\'e}phanie and {Jung}, Intae and {Kartaltepe}, Jeyhan S. and {Kirkpatrick}, Allison and {Kocevski}, Dale D. and {Koekemoer}, Anton M. and {Lotz}, Jennifer M. and {Lucas}, Ray A. and {Magnelli}, Benjamin and {Matharu}, Jasleen and {P{\'e}rez-Gonz{\'a}lez}, Pablo G. and {Pirzkal}, Nor and {Rafelski}, Marc and {Rose}, Caitlin and {Seill{\'e}}, Lise-Marie and {Somerville}, Rachel S. and {Straughn}, Amber N. and {Tacchella}, Sandro and {Vanderhoof}, Brittany N. and {Weiner}, Benjamin J. and {Wuyts}, Stijn and {Yung}, L.~Y. Aaron and {Zavala}, Jorge A.},
        title = "{The Physical Conditions of Emission-line Galaxies at Cosmic Dawn from JWST/NIRSpec Spectroscopy in the SMACS 0723 Early Release Observations}",
      journal = {\apj},
         year = 2023,
        month = mar,
       volume = {945},
       number = {1},
          eid = {35},
        pages = {35},
          doi = {10.3847/1538-4357/acba8a},
archivePrefix = {arXiv},
       eprint = {2207.12388},
 primaryClass = {astro-ph.GA},
       adsurl = {https://ui.adsabs.harvard.edu/abs/2023ApJ...945...35T}
}

@ARTICLE{Zeimann+15,
       author = {{Zeimann}, Gregory R. and {Ciardullo}, Robin and {Gebhardt}, Henry and {Gronwall}, Caryl and {Hagen}, Alex and {Trump}, Jonathan R. and {Bridge}, Joanna S. and {Luo}, Bin and {Schneider}, Donald P.},
        title = "{Hubble Space Telescope Emission-line Galaxies at z \raisebox{-0.5ex}\textasciitilde 2: The Mystery of Neon}",
      journal = {\apj},
         year = 2015,
        month = jan,
       volume = {798},
       number = {1},
          eid = {29},
        pages = {29},
          doi = {10.1088/0004-637X/798/1/29},
archivePrefix = {arXiv},
       eprint = {1410.6159},
 primaryClass = {astro-ph.GA},
       adsurl = {https://ui.adsabs.harvard.edu/abs/2015ApJ...798...29Z}
}

@ARTICLE{Gordon+03,
       author = {{Gordon}, Karl D. and {Clayton}, Geoffrey C. and {Misselt}, K.~A. and {Landolt}, Arlo U. and {Wolff}, Michael J.},
        title = "{A Quantitative Comparison of the Small Magellanic Cloud, Large Magellanic Cloud, and Milky Way Ultraviolet to Near-Infrared Extinction Curves}",
      journal = {\apj},
         year = 2003,
        month = sep,
       volume = {594},
       number = {1},
        pages = {279-293},
          doi = {10.1086/376774},
archivePrefix = {arXiv},
       eprint = {astro-ph/0305257},
 primaryClass = {astro-ph},
       adsurl = {https://ui.adsabs.harvard.edu/abs/2003ApJ...594..279G}
}

@ARTICLE{Luridiana+15,
       author = {{Luridiana}, V. and {Morisset}, C. and {Shaw}, R.~A.},
        title = "{PyNeb: a new tool for analyzing emission lines. I. Code description and validation of results}",
      journal = {\aap},
         year = 2015,
        month = jan,
       volume = {573},
          eid = {A42},
        pages = {A42},
          doi = {10.1051/0004-6361/201323152},
archivePrefix = {arXiv},
       eprint = {1410.6662},
 primaryClass = {astro-ph.IM},
       adsurl = {https://ui.adsabs.harvard.edu/abs/2015A&A...573A..42L}
}

@ARTICLE{Scholtz+25,
       author = {{Scholtz}, Jan and {Maiolino}, Roberto and {D'Eugenio}, Francesco and {Curtis-Lake}, Emma and {Carniani}, Stefano and {Charlot}, Stephane and {Curti}, Mirko and {Silcock}, Maddie S. and {Arribas}, Santiago and {Baker}, William and {Bhatawdekar}, Rachana and {Boyett}, Kristan and {Bunker}, Andrew J. and {Chevallard}, Jacopo and {Circosta}, Chiara and {Eisenstein}, Daniel J. and {Hainline}, Kevin and {Hausen}, Ryan and {Ji}, Xihan and {Ji}, Zhiyuan and {Johnson}, Benjamin D. and {Kumari}, Nimisha and {Looser}, Tobias J. and {Lyu}, Jianwei and {Maseda}, Michael V. and {Parlanti}, Eleonora and {Perna}, Michele and {Rieke}, Marcia and {Robertson}, Brant and {Del Pino}, Bruno Rodr{\'\i}guez and {Sun}, Fengwu and {Tacchella}, Sandro and {{\"U}bler}, Hannah and {Venturi}, Giacomo and {Williams}, Christina C. and {Willmer}, Christopher N.~A. and {Willott}, Chris and {Witstok}, Joris},
        title = "{JADES: A large population of obscured, narrow-line active galactic nuclei at high redshift}",
      journal = {\aap},
         year = 2025,
        month = may,
       volume = {697},
          eid = {A175},
        pages = {A175},
          doi = {10.1051/0004-6361/202348804},
archivePrefix = {arXiv},
       eprint = {2311.18731},
 primaryClass = {astro-ph.GA},
       adsurl = {https://ui.adsabs.harvard.edu/abs/2025A&A...697A.175S}
}

@ARTICLE{Leitherer+02,
       author = {{Leitherer}, Claus and {Li}, I.-Hui and {Calzetti}, Daniela and {Heckman}, Timothy M.},
        title = "{Global Far-Ultraviolet (912-1800 {\r{A}}) Properties of Star-forming Galaxies}",
      journal = {\apjs},
         year = 2002,
        month = jun,
       volume = {140},
       number = {2},
        pages = {303-329},
          doi = {10.1086/342486},
       adsurl = {https://ui.adsabs.harvard.edu/abs/2002ApJS..140..303L}
}

@ARTICLE{Lyu+24,
       author = {{Lyu}, Jianwei and {Alberts}, Stacey and {Rieke}, George H. and {Shivaei}, Irene and {P{\'e}rez-Gonz{\'a}lez}, Pablo G. and {Sun}, Fengwu and {Hainline}, Kevin N. and {Baum}, Stefi and {Bonaventura}, Nina and {Bunker}, Andrew J. and {Egami}, Eiichi and {Eisenstein}, Daniel J. and {Florian}, Michael and {Ji}, Zhiyuan and {Johnson}, Benjamin D. and {Morrison}, Jane and {Rieke}, Marcia and {Robertson}, Brant and {Rujopakarn}, Wiphu and {Tacchella}, Sandro and {Scholtz}, Jan and {Willmer}, Christopher N.~A.},
        title = "{Active Galactic Nuclei Selection and Demographics: A New Age with JWST/MIRI}",
      journal = {\apj},
         year = 2024,
        month = may,
       volume = {966},
       number = {2},
          eid = {229},
        pages = {229},
          doi = {10.3847/1538-4357/ad3643},
archivePrefix = {arXiv},
       eprint = {2310.12330},
 primaryClass = {astro-ph.GA},
       adsurl = {https://ui.adsabs.harvard.edu/abs/2024ApJ...966..229L}
}

@ARTICLE{Draine+07,
       author = {{Draine}, B.~T. and {Dale}, D.~A. and {Bendo}, G. and {Gordon}, K.~D. and {Smith}, J.~D.~T. and {Armus}, L. and {Engelbracht}, C.~W. and {Helou}, G. and {Kennicutt}, Jr., R.~C. and {Li}, A. and {Roussel}, H. and {Walter}, F. and {Calzetti}, D. and {Moustakas}, J. and {Murphy}, E.~J. and {Rieke}, G.~H. and {Bot}, C. and {Hollenbach}, D.~J. and {Sheth}, K. and {Teplitz}, H.~I.},
        title = "{Dust Masses, PAH Abundances, and Starlight Intensities in the SINGS Galaxy Sample}",
      journal = {\apj},
         year = 2007,
        month = jul,
       volume = {663},
       number = {2},
        pages = {866-894},
          doi = {10.1086/518306},
archivePrefix = {arXiv},
       eprint = {astro-ph/0703213},
 primaryClass = {astro-ph},
       adsurl = {https://ui.adsabs.harvard.edu/abs/2007ApJ...663..866D}
}

@ARTICLE{Sturm+25,
       author = {{Sturm}, Megan R. and {Hayes}, Bayli and {Reines}, Amy E.},
        title = "{Star-forming Nuclear Clusters in Dwarf Galaxies Mimicking Active Galactic Nucleus Signatures in the Mid-infrared}",
      journal = {\apj},
         year = 2025,
        month = jan,
       volume = {979},
       number = {1},
          eid = {36},
        pages = {36},
          doi = {10.3847/1538-4357/ada02f},
       adsurl = {https://ui.adsabs.harvard.edu/abs/2025ApJ...979...36S}
}

@ARTICLE{Leja+19,
       author = {{Leja}, Joel and {Carnall}, Adam C. and {Johnson}, Benjamin D. and {Conroy}, Charlie and {Speagle}, Joshua S.},
        title = "{How to Measure Galaxy Star Formation Histories. II. Nonparametric Models}",
      journal = {\apj},
         year = 2019,
        month = may,
       volume = {876},
       number = {1},
          eid = {3},
        pages = {3},
          doi = {10.3847/1538-4357/ab133c},
archivePrefix = {arXiv},
       eprint = {1811.03637},
 primaryClass = {astro-ph.GA},
       adsurl = {https://ui.adsabs.harvard.edu/abs/2019ApJ...876....3L}
}

@ARTICLE{Izotov+11,
       author = {{Izotov}, Yuri I. and {Guseva}, Natalia G. and {Thuan}, Trinh X.},
        title = "{Green Pea Galaxies and Cohorts: Luminous Compact Emission-line Galaxies in the Sloan Digital Sky Survey}",
      journal = {\apj},
         year = 2011,
        month = feb,
       volume = {728},
       number = {2},
          eid = {161},
        pages = {161},
          doi = {10.1088/0004-637X/728/2/161},
archivePrefix = {arXiv},
       eprint = {1012.5639},
 primaryClass = {astro-ph.CO},
       adsurl = {https://ui.adsabs.harvard.edu/abs/2011ApJ...728..161I}
}

@ARTICLE{Wang+14,
       author = {{Wang}, Jian-Min and {Qiu}, Jie and {Du}, Pu and {Ho}, Luis C.},
        title = "{Self-shadowing Effects of Slim Accretion Disks in Active Galactic Nuclei: The Diverse Appearance of the Broad-line Region}",
      journal = {\apj},
         year = 2014,
        month = dec,
       volume = {797},
       number = {1},
          eid = {65},
        pages = {65},
          doi = {10.1088/0004-637X/797/1/65},
archivePrefix = {arXiv},
       eprint = {1410.5285},
 primaryClass = {astro-ph.GA},
       adsurl = {https://ui.adsabs.harvard.edu/abs/2014ApJ...797...65W}
}

@ARTICLE{Dotan+11,
       author = {{Dotan}, Calanit and {Shaviv}, Nir J.},
        title = "{Super-Eddington slim accretion discs with winds}",
      journal = {\mnras},
         year = 2011,
        month = may,
       volume = {413},
       number = {3},
        pages = {1623-1632},
          doi = {10.1111/j.1365-2966.2011.18235.x},
archivePrefix = {arXiv},
       eprint = {1004.1797},
 primaryClass = {astro-ph.HE},
       adsurl = {https://ui.adsabs.harvard.edu/abs/2011MNRAS.413.1623D}
}

@ARTICLE{Genin+25,
       author = {{Genin}, Aur{\'e}lien and {Shuntov}, Marko and {Brammer}, Gabe and {Allen}, Natalie and {Ito}, Kei and {Magdis}, Georgios and {Matharu}, Jasleen and {Oesch}, Pascal A. and {Toft}, Sune and {Valentino}, Francesco},
        title = "{DAWN JWST Archive: Morphology from profile fitting of over 340 000 galaxies in major JWST fields: Morphology evolution with redshift and galaxy type}",
      journal = {\aap},
         year = 2025,
        month = jul,
       volume = {699},
          eid = {A343},
        pages = {A343},
          doi = {10.1051/0004-6361/202555504},
archivePrefix = {arXiv},
       eprint = {2505.21622},
 primaryClass = {astro-ph.GA},
       adsurl = {https://ui.adsabs.harvard.edu/abs/2025A&A...699A.343G}
}

@ARTICLE{Steidel+14,
       author = {{Steidel}, Charles C. and {Rudie}, Gwen C. and {Strom}, Allison L. and {Pettini}, Max and {Reddy}, Naveen A. and {Shapley}, Alice E. and {Trainor}, Ryan F. and {Erb}, Dawn K. and {Turner}, Monica L. and {Konidaris}, Nicholas P. and {Kulas}, Kristin R. and {Mace}, Gregory and {Matthews}, Keith and {McLean}, Ian S.},
        title = "{Strong Nebular Line Ratios in the Spectra of z \raisebox{-0.5ex}\textasciitilde 2-3 Star Forming Galaxies: First Results from KBSS-MOSFIRE}",
      journal = {\apj},
         year = 2014,
        month = nov,
       volume = {795},
       number = {2},
          eid = {165},
        pages = {165},
          doi = {10.1088/0004-637X/795/2/165},
archivePrefix = {arXiv},
       eprint = {1405.5473},
 primaryClass = {astro-ph.GA},
       adsurl = {https://ui.adsabs.harvard.edu/abs/2014ApJ...795..165S}
}

@ARTICLE{Liddle+07,
       author = {{Liddle}, Andrew R.},
        title = "{Information criteria for astrophysical model selection}",
      journal = {\mnras},
         year = 2007,
        month = may,
       volume = {377},
       number = {1},
        pages = {L74-L78},
          doi = {10.1111/j.1745-3933.2007.00306.x},
archivePrefix = {arXiv},
       eprint = {astro-ph/0701113},
 primaryClass = {astro-ph},
       adsurl = {https://ui.adsabs.harvard.edu/abs/2007MNRAS.377L..74L}
}

@ARTICLE{Schwarz+78,
       author = {{Schwarz}, Gideon},
        title = "{Estimating the Dimension of a Model}",
      journal = {Annals of Statistics},
         year = 1978,
        month = jul,
       volume = {6},
       number = {2},
        pages = {461-464},
       adsurl = {https://ui.adsabs.harvard.edu/abs/1978AnSta...6..461S}
}

@ARTICLE{Mannucci+10,
       author = {{Mannucci}, F. and {Cresci}, G. and {Maiolino}, R. and {Marconi}, A. and {Gnerucci}, A.},
        title = "{A fundamental relation between mass, star formation rate and metallicity in local and high-redshift galaxies}",
      journal = {\mnras},
         year = 2010,
        month = nov,
       volume = {408},
       number = {4},
        pages = {2115-2127},
          doi = {10.1111/j.1365-2966.2010.17291.x},
archivePrefix = {arXiv},
       eprint = {1005.0006},
 primaryClass = {astro-ph.CO},
       adsurl = {https://ui.adsabs.harvard.edu/abs/2010MNRAS.408.2115M}
}

@ARTICLE{Bian+18,
       author = {{Bian}, Fuyan and {Kewley}, Lisa J. and {Dopita}, Michael A.},
        title = "{{\textquotedblleft}Direct{\textquotedblright} Gas-phase Metallicity in Local Analogs of High-redshift Galaxies: Empirical Metallicity Calibrations for High-redshift Star-forming Galaxies}",
      journal = {\apj},
         year = 2018,
        month = jun,
       volume = {859},
       number = {2},
          eid = {175},
        pages = {175},
          doi = {10.3847/1538-4357/aabd74},
archivePrefix = {arXiv},
       eprint = {1805.08224},
 primaryClass = {astro-ph.GA},
       adsurl = {https://ui.adsabs.harvard.edu/abs/2018ApJ...859..175B}
}

@ARTICLE{Rodriguez-Munoz+22,
       author = {{Rodr{\'\i}guez-Mu{\~n}oz}, L. and {Rodighiero}, G. and {P{\'e}rez-Gonz{\'a}lez}, P.~G. and {Talia}, M. and {Baronchelli}, I. and {Morselli}, L. and {Renzini}, A. and {Puglisi}, A. and {Grazian}, A. and {Zanella}, A. and {Mancini}, C. and {Feltre}, A. and {Romano}, M. and {Vidal Garc{\'\i}a}, A. and {Franceschini}, A. and {Alcalde Pampliega}, B. and {Cassata}, P. and {Costantin}, L. and {Dom{\'\i}nguez S{\'a}nchez}, H. and {Espino-Briones}, N. and {Iani}, E. and {Koekemoer}, A. and {Lumbreras-Calle}, A. and {Rodr{\'\i}guez-Espinosa}, J.~M.},
        title = "{Differential attenuation in star-forming galaxies at 0.3 {\ensuremath{\lesssim}} z {\ensuremath{\lesssim}} 1.5 in the SHARDS/CANDELS field}",
      journal = {\mnras},
         year = 2022,
        month = feb,
       volume = {510},
       number = {2},
        pages = {2061-2083},
          doi = {10.1093/mnras/stab3558},
archivePrefix = {arXiv},
       eprint = {2112.01885},
 primaryClass = {astro-ph.GA},
       adsurl = {https://ui.adsabs.harvard.edu/abs/2022MNRAS.510.2061R}
}

@ARTICLE{Momcheva+13,
       author = {{Momcheva}, Ivelina G. and {Lee}, Janice C. and {Ly}, Chun and {Salim}, Samir and {Dale}, Daniel A. and {Ouchi}, Masami and {Finn}, Rose and {Ono}, Yoshiaki},
        title = "{Nebular Attenuation in H{\ensuremath{\alpha}}-selected Star-forming Galaxies at z = 0.8 from the NewH{\ensuremath{\alpha}} Survey}",
      journal = {\aj},
         year = 2013,
        month = feb,
       volume = {145},
       number = {2},
          eid = {47},
        pages = {47},
          doi = {10.1088/0004-6256/145/2/47},
archivePrefix = {arXiv},
       eprint = {1207.5479},
 primaryClass = {astro-ph.CO},
       adsurl = {https://ui.adsabs.harvard.edu/abs/2013AJ....145...47M}
}

@ARTICLE{Oestlin+25,
       author = {{{\"O}stlin}, G{\"o}ran and {P{\'e}rez-Gonz{\'a}lez}, Pablo G. and {Melinder}, Jens and {Gillman}, Steven and {Iani}, Edoardo and {Costantin}, Luca and {Boogaard}, Leindert A. and {Rinaldi}, Pierluigi and {Colina}, Luis and {Ulrik N{\o}rgaard-Nielsen}, Hans and {Dicken}, Daniel and {Greve}, Thomas R. and {Wright}, Gillian and {Alonso-Herrero}, Almudena and {{\'A}lvarez-M{\'a}rquez}, Javier and {Annunziatella}, Marianna and {Bik}, Arjan and {Bosman}, Sarah E.~I. and {Caputi}, Karina I. and {Crespo Gomez}, Alejandro and {Eckart}, Andreas and {Garcia-Marin}, Macarena and {Hjorth}, Jens and {Ilbert}, Olivier and {Jermann}, Iris and {Kendrew}, Sarah and {Labiano}, Alvaro and {Langeroodi}, Danial and {Le Fevre}, Olivier and {Libralato}, Mattia and {Meyer}, Romain A. and {Moutard}, Thibaud and {Peissker}, Florian and {Pye}, John P. and {Tikkanen}, Tuomo V. and {Topinka}, Martin and {Walter}, Fabian and {Ward}, Martin and {van der Werf}, Paul and {van Dishoeck}, Ewine F. and {G{\"u}del}, Manuel and {Henning}, Thomas and {Lagage}, Pierre-Olivier and {Ray}, Tom P. and {Vandenbussche}, Bart},
        title = "{MIRI Deep Imaging Survey (MIDIS) of the Hubble Ultra Deep Field: Survey description and early results for the galaxy population detected at 5.6 {\ensuremath{\mu}}m}",
      journal = {\aap},
         year = 2025,
        month = apr,
       volume = {696},
          eid = {A57},
        pages = {A57},
          doi = {10.1051/0004-6361/202451723},
archivePrefix = {arXiv},
       eprint = {2411.19686},
 primaryClass = {astro-ph.GA},
       adsurl = {https://ui.adsabs.harvard.edu/abs/2025A&A...696A..57O}
}

@ARTICLE{Iani+24,
       author = {{Iani}, Edoardo and {Caputi}, Karina I. and {Rinaldi}, Pierluigi and {Annunziatella}, Marianna and {Boogaard}, Leindert A. and {{\"O}stlin}, G{\"o}ran and {Costantin}, Luca and {Gillman}, Steven and {P{\'e}rez-Gonz{\'a}lez}, Pablo G. and {Colina}, Luis and {Greve}, Thomas R. and {Wright}, Gillian and {Alonso-Herrero}, Almudena and {{\'A}lvarez-M{\'a}rquez}, Javier and {Bik}, Arjan and {Bosman}, Sarah E.~I. and {Crespo G{\'o}mez}, Alejandro and {Eckart}, Andreas and {Hjorth}, Jens and {Jermann}, Iris and {Labiano}, Alvaro and {Langeroodi}, Danial and {Melinder}, Jens and {Moutard}, Thibaud and {Pei{\ss}ker}, Florian and {Pye}, John P. and {Tikkanen}, Tuomo V. and {van der Werf}, Paul P. and {Walter}, Fabian and {Henning}, Thomas K. and {Lagage}, Pierre-Olivier and {van Dishoeck}, Ewine F.},
        title = "{MIDIS: JWST NIRCam and MIRI Unveil the Stellar Population Properties of Ly{\ensuremath{\alpha}} Emitters and Lyman-break Galaxies at z ≃ 3{\textendash}7}",
      journal = {\apj},
         year = 2024,
        month = mar,
       volume = {963},
       number = {2},
          eid = {97},
        pages = {97},
          doi = {10.3847/1538-4357/ad15f6},
archivePrefix = {arXiv},
       eprint = {2309.08515},
 primaryClass = {astro-ph.GA},
       adsurl = {https://ui.adsabs.harvard.edu/abs/2024ApJ...963...97I}
}

@ARTICLE{Rinaldi+23,
       author = {{Rinaldi}, P. and {Caputi}, K.~I. and {Costantin}, L. and {Gillman}, S. and {Iani}, E. and {P{\'e}rez-Gonz{\'a}lez}, P.~G. and {{\"O}stlin}, G. and {Colina}, L. and {Greve}, T.~R. and {Noorgard-Nielsen}, H.~U. and {Wright}, G.~S. and {Alonso-Herrero}, A. and {{\'A}lvarez-M{\'a}rquez}, J. and {Eckart}, A. and {Garc{\'\i}a-Mar{\'\i}n}, M. and {Hjorth}, J. and {Ilbert}, O. and {Kendrew}, S. and {Labiano}, A. and {Le F{\`e}vre}, O. and {Pye}, J. and {Tikkanen}, T. and {Walter}, F. and {van der Werf}, P. and {Ward}, M. and {Annunziatella}, M. and {Azzollini}, R. and {Bik}, A. and {Boogaard}, L. and {Bosman}, S.~E.~I. and {Crespo G{\'o}mez}, A. and {Jermann}, I. and {Langeroodi}, D. and {Melinder}, J. and {Meyer}, R.~A. and {Moutard}, T. and {Peissker}, F. and {Topinka}, M. and {van Dishoeck}, E. and {G{\"u}del}, M. and {Henning}, Th. and {Lagage}, P.-O. and {Ray}, T. and {Vandenbussche}, B. and {Waelkens}, C. and {Navarro-Carrera}, R. and {Kokorev}, V.},
        title = "{MIDIS: Strong (H{\ensuremath{\beta}}+[O III]) and H{\ensuremath{\alpha}} Emitters at Redshift z ≃ 7-8 Unveiled with JWST NIRCam and MIRI Imaging in the Hubble eXtreme Deep Field}",
      journal = {\apj},
         year = 2023,
        month = aug,
       volume = {952},
       number = {2},
          eid = {143},
        pages = {143},
          doi = {10.3847/1538-4357/acdc27},
archivePrefix = {arXiv},
       eprint = {2301.10717},
 primaryClass = {astro-ph.GA},
       adsurl = {https://ui.adsabs.harvard.edu/abs/2023ApJ...952..143R}
}

@ARTICLE{Weisz+12,
       author = {{Weisz}, Daniel R. and {Johnson}, Benjamin D. and {Johnson}, L. Clifton and {Skillman}, Evan D. and {Lee}, Janice C. and {Kennicutt}, Robert C. and {Calzetti}, Daniela and {van Zee}, Liese and {Bothwell}, Matthew S. and {Dalcanton}, Julianne J. and {Dale}, Daniel A. and {Williams}, Benjamin F.},
        title = "{Modeling the Effects of Star Formation Histories on H{\ensuremath{\alpha}} and Ultraviolet Fluxes in nearby Dwarf Galaxies}",
      journal = {\apj},
         year = 2012,
        month = jan,
       volume = {744},
       number = {1},
          eid = {44},
        pages = {44},
          doi = {10.1088/0004-637X/744/1/44},
archivePrefix = {arXiv},
       eprint = {1109.2905},
 primaryClass = {astro-ph.CO},
       adsurl = {https://ui.adsabs.harvard.edu/abs/2012ApJ...744...44W}
}

@ARTICLE{Ohsuga+02,
       author = {{Ohsuga}, Ken and {Mineshige}, Shin and {Mori}, Masao and {Umemura}, Masayuki},
        title = "{Does the Slim-Disk Model Correctly Consider Photon-trapping Effects?}",
      journal = {\apj},
         year = 2002,
        month = jul,
       volume = {574},
       number = {1},
        pages = {315-324},
          doi = {10.1086/340798},
archivePrefix = {arXiv},
       eprint = {astro-ph/0203425},
 primaryClass = {astro-ph},
       adsurl = {https://ui.adsabs.harvard.edu/abs/2002ApJ...574..315O}
}

@ARTICLE{Massonneau+23,
       author = {{Massonneau}, Warren and {Volonteri}, Marta and {Dubois}, Yohan and {Beckmann}, Ricarda S.},
        title = "{How the super-Eddington regime regulates black hole growth in high-redshift galaxies}",
      journal = {\aap},
         year = 2023,
        month = feb,
       volume = {670},
          eid = {A180},
        pages = {A180},
          doi = {10.1051/0004-6361/202243170},
archivePrefix = {arXiv},
       eprint = {2201.08766},
 primaryClass = {astro-ph.GA},
       adsurl = {https://ui.adsabs.harvard.edu/abs/2023A&A...670A.180M}
}

@ARTICLE{Abramowicz+88,
       author = {{Abramowicz}, M.~A. and {Czerny}, B. and {Lasota}, J.~P. and {Szuszkiewicz}, E.},
        title = "{Slim Accretion Disks}",
      journal = {\apj},
         year = 1988,
        month = sep,
       volume = {332},
        pages = {646},
          doi = {10.1086/166683},
       adsurl = {https://ui.adsabs.harvard.edu/abs/1988ApJ...332..646A}
}

@ARTICLE{Duras+20,
       author = {{Duras}, F. and {Bongiorno}, A. and {Ricci}, F. and {Piconcelli}, E. and {Shankar}, F. and {Lusso}, E. and {Bianchi}, S. and {Fiore}, F. and {Maiolino}, R. and {Marconi}, A. and {Onori}, F. and {Sani}, E. and {Schneider}, R. and {Vignali}, C. and {La Franca}, F.},
        title = "{Universal bolometric corrections for active galactic nuclei over seven luminosity decades}",
      journal = {\aap},
         year = 2020,
        month = apr,
       volume = {636},
          eid = {A73},
        pages = {A73},
          doi = {10.1051/0004-6361/201936817},
archivePrefix = {arXiv},
       eprint = {2001.09984},
 primaryClass = {astro-ph.GA},
       adsurl = {https://ui.adsabs.harvard.edu/abs/2020A&A...636A..73D}
}

@ARTICLE{Flury+22,
       author = {{Flury}, Sophia R. and {Jaskot}, Anne E. and {Ferguson}, Harry C. and {Worseck}, G{\'a}bor and {Makan}, Kirill and {Chisholm}, John and {Saldana-Lopez}, Alberto and {Schaerer}, Daniel and {McCandliss}, Stephan and {Wang}, Bingjie and {Ford}, N.~M. and {Heckman}, Timothy and {Ji}, Zhiyuan and {Giavalisco}, Mauro and {Amorin}, Ricardo and {Atek}, Hakim and {Blaizot}, Jeremy and {Borthakur}, Sanchayeeta and {Carr}, Cody and {Castellano}, Marco and {Cristiani}, Stefano and {De Barros}, Stephane and {Dickinson}, Mark and {Finkelstein}, Steven L. and {Fleming}, Brian and {Fontanot}, Fabio and {Garel}, Thibault and {Grazian}, Andrea and {Hayes}, Matthew and {Henry}, Alaina and {Mauerhofer}, Valentin and {Micheva}, Genoveva and {Oey}, M.~S. and {Ostlin}, Goran and {Papovich}, Casey and {Pentericci}, Laura and {Ravindranath}, Swara and {Rosdahl}, Joakim and {Rutkowski}, Michael and {Santini}, Paola and {Scarlata}, Claudia and {Teplitz}, Harry and {Thuan}, Trinh and {Trebitsch}, Maxime and {Vanzella}, Eros and {Verhamme}, Anne and {Xu}, Xinfeng},
        title = "{The Low-redshift Lyman Continuum Survey. I. New, Diverse Local Lyman Continuum Emitters}",
      journal = {\apjs},
         year = 2022,
        month = may,
       volume = {260},
       number = {1},
          eid = {1},
        pages = {1},
          doi = {10.3847/1538-4365/ac5331},
archivePrefix = {arXiv},
       eprint = {2201.11716},
 primaryClass = {astro-ph.GA},
       adsurl = {https://ui.adsabs.harvard.edu/abs/2022ApJS..260....1F}
}

@ARTICLE{Carnall+19,
       author = {{Carnall}, Adam C. and {Leja}, Joel and {Johnson}, Benjamin D. and {McLure}, Ross J. and {Dunlop}, James S. and {Conroy}, Charlie},
        title = "{How to Measure Galaxy Star Formation Histories. I. Parametric Models}",
      journal = {\apj},
         year = 2019,
        month = mar,
       volume = {873},
       number = {1},
          eid = {44},
        pages = {44},
          doi = {10.3847/1538-4357/ab04a2},
archivePrefix = {arXiv},
       eprint = {1811.03635},
 primaryClass = {astro-ph.GA},
       adsurl = {https://ui.adsabs.harvard.edu/abs/2019ApJ...873...44C}
}

@ARTICLE{Pacifici+16,
       author = {{Pacifici}, Camilla and {Oh}, Sree and {Oh}, Kyuseok and {Lee}, Jaehyun and {Yi}, Sukyoung K.},
        title = "{Timing the Evolution of Quiescent and Star-forming Local Galaxies}",
      journal = {\apj},
         year = 2016,
        month = jun,
       volume = {824},
       number = {1},
          eid = {45},
        pages = {45},
          doi = {10.3847/0004-637X/824/1/45},
archivePrefix = {arXiv},
       eprint = {1604.02460},
 primaryClass = {astro-ph.GA},
       adsurl = {https://ui.adsabs.harvard.edu/abs/2016ApJ...824...45P}
}

@ARTICLE{Pacifici+12,
       author = {{Pacifici}, Camilla and {Charlot}, St{\'e}phane and {Blaizot}, J{\'e}r{\'e}my and {Brinchmann}, Jarle},
        title = "{Relative merits of different types of rest-frame optical observations to constrain galaxy physical parameters}",
      journal = {\mnras},
         year = 2012,
        month = apr,
       volume = {421},
       number = {3},
        pages = {2002-2024},
          doi = {10.1111/j.1365-2966.2012.20431.x},
archivePrefix = {arXiv},
       eprint = {1201.0780},
 primaryClass = {astro-ph.CO},
       adsurl = {https://ui.adsabs.harvard.edu/abs/2012MNRAS.421.2002P}
}

@ARTICLE{Larson+23,
       author = {{Larson}, Rebecca L. and {Finkelstein}, Steven L. and {Kocevski}, Dale D. and {Hutchison}, Taylor A. and {Trump}, Jonathan R. and {Arrabal Haro}, Pablo and {Bromm}, Volker and {Cleri}, Nikko J. and {Dickinson}, Mark and {Fujimoto}, Seiji and {Kartaltepe}, Jeyhan S. and {Koekemoer}, Anton M. and {Papovich}, Casey and {Pirzkal}, Nor and {Tacchella}, Sandro and {Zavala}, Jorge A. and {Bagley}, Micaela and {Behroozi}, Peter and {Champagne}, Jaclyn B. and {Cole}, Justin W. and {Jung}, Intae and {Morales}, Alexa M. and {Yang}, Guang and {Zhang}, Haowen and {Zitrin}, Adi and {Amor{\'\i}n}, Ricardo O. and {Burgarella}, Denis and {Casey}, Caitlin M. and {Ch{\'a}vez Ortiz}, {\'O}scar A. and {Cox}, Isabella G. and {Chworowsky}, Katherine and {Fontana}, Adriano and {Gawiser}, Eric and {Grazian}, Andrea and {Grogin}, Norman A. and {Harish}, Santosh and {Hathi}, Nimish P. and {Hirschmann}, Michaela and {Holwerda}, Benne W. and {Juneau}, St{\'e}phanie and {Leung}, Gene C.~K. and {Lucas}, Ray A. and {McGrath}, Elizabeth J. and {P{\'e}rez-Gonz{\'a}lez}, Pablo G. and {Rigby}, Jane R. and {Seill{\'e}}, Lise-Marie and {Simons}, Raymond C. and {de La Vega}, Alexander and {Weiner}, Benjamin J. and {Wilkins}, Stephen M. and {Yung}, L.~Y. Aaron and {Ceers Team}},
        title = "{A CEERS Discovery of an Accreting Supermassive Black Hole 570 Myr after the Big Bang: Identifying a Progenitor of Massive z > 6 Quasars}",
      journal = {\apjl},
         year = 2023,
        month = aug,
       volume = {953},
       number = {2},
          eid = {L29},
        pages = {L29},
          doi = {10.3847/2041-8213/ace619},
archivePrefix = {arXiv},
       eprint = {2303.08918},
 primaryClass = {astro-ph.GA},
       adsurl = {https://ui.adsabs.harvard.edu/abs/2023ApJ...953L..29L}
}

@ARTICLE{Hainline+16,
       author = {{Hainline}, Kevin N. and {Reines}, Amy E. and {Greene}, Jenny E. and {Stern}, Daniel},
        title = "{Mid-infrared Colors of Dwarf Galaxies: Young Starbursts Mimicking Active Galactic Nuclei}",
      journal = {\apj},
         year = 2016,
        month = dec,
       volume = {832},
       number = {2},
          eid = {119},
        pages = {119},
          doi = {10.3847/0004-637X/832/2/119},
archivePrefix = {arXiv},
       eprint = {1609.06721},
 primaryClass = {astro-ph.GA},
       adsurl = {https://ui.adsabs.harvard.edu/abs/2016ApJ...832..119H}
}

@ARTICLE{Rieke+25,
       author = {{Rieke}, George H. and {Sun}, Yang and {Lyu}, Jianwei and {Willmer}, Christopher N.~A. and {Zhu}, Yongda and {Rinaldi}, Pierluigi and {Stone}, Meredith A. and {Hainline}, Kevin N. and {P{\'e}rez-Gonz{\'a}lez}, Pablo G.},
        title = "{Confirming Near- to Mid-infrared Photometrically Identified Obscured AGNs in the JWST Era}",
      journal = {\apj},
         year = 2025,
        month = nov,
       volume = {994},
       number = {1},
          eid = {35},
        pages = {35},
          doi = {10.3847/1538-4357/adff79},
archivePrefix = {arXiv},
       eprint = {2510.07303},
 primaryClass = {astro-ph.GA},
       adsurl = {https://ui.adsabs.harvard.edu/abs/2025ApJ...994...35R}
}

@ARTICLE{Labbe+24,
       author = {{Labbe}, Ivo and {Greene}, Jenny E. and {Matthee}, Jorryt and {Treiber}, Helena and {Kokorev}, Vasily and {Miller}, Tim B. and {Kramarenko}, Ivan and {Setton}, David J. and {Ma}, Yilun and {Goulding}, Andy D. and {Bezanson}, Rachel and {Naidu}, Rohan P. and {Williams}, Christina C. and {Atek}, Hakim and {Brammer}, Gabriel and {Cutler}, Sam E. and {Chemerynska}, Iryna and {Cloonan}, Aidan P. and {Dayal}, Pratika and {de Graaff}, Anna and {Fudamoto}, Yoshinobu and {Fujimoto}, Seiji and {Furtak}, Lukas J. and {Glazebrook}, Karl and {Heintz}, Kasper E. and {Leja}, Joel and {Marchesini}, Danilo and {Nanayakkara}, Themiya and {Nelson}, Erica J. and {Oesch}, Pascal A. and {Pan}, Richard and {Price}, Sedona H. and {Shivaei}, Irene and {Sobral}, David and {Suess}, Katherine A. and {van Dokkum}, Pieter and {Wang}, Bingjie and {Weaver}, John R. and {Whitaker}, Katherine E. and {Zitrin}, Adi},
        title = "{An unambiguous AGN and a Balmer break in an Ultraluminous Little Red Dot at z=4.47 from Ultradeep UNCOVER and All the Little Things Spectroscopy}",
      journal = {arXiv e-prints},
         year = 2024,
        month = dec,
          eid = {arXiv:2412.04557},
        pages = {arXiv:2412.04557},
          doi = {10.48550/arXiv.2412.04557},
archivePrefix = {arXiv},
       eprint = {2412.04557},
 primaryClass = {astro-ph.GA},
       adsurl = {https://ui.adsabs.harvard.edu/abs/2024arXiv241204557L}
}

@ARTICLE{Wang+24a,
       author = {{Wang}, Bingjie and {de Graaff}, Anna and {Davies}, Rebecca L. and {Greene}, Jenny E. and {Leja}, Joel and {Brammer}, Gabriel B. and {Goulding}, Andy D. and {Miller}, Tim B. and {Suess}, Katherine A. and {Weibel}, Andrea and {Williams}, Christina C. and {Bezanson}, Rachel and {Boogaard}, Leindert A. and {Cleri}, Nikko J. and {Hirschmann}, Michaela and {Katz}, Harley and {Labb{\'e}}, Ivo and {Maseda}, Michael V. and {Matthee}, Jorryt and {McConachie}, Ian and {Naidu}, Rohan P. and {Oesch}, Pascal A. and {Rix}, Hans-Walter and {Setton}, David J. and {Whitaker}, Katherine E.},
        title = "{RUBIES: JWST/NIRSpec Confirmation of an Infrared-luminous, Broad-line Little Red Dot with an Ionized Outflow}",
      journal = {\apj},
         year = 2025,
        month = may,
       volume = {984},
       number = {2},
          eid = {121},
        pages = {121},
          doi = {10.3847/1538-4357/adc1ca},
archivePrefix = {arXiv},
       eprint = {2403.02304},
 primaryClass = {astro-ph.GA},
       adsurl = {https://ui.adsabs.harvard.edu/abs/2025ApJ...984..121W}
}

@ARTICLE{Naidu+25,
       author = {{Naidu}, Rohan P. and {Matthee}, Jorryt and {Katz}, Harley and {de Graaff}, Anna and {Oesch}, Pascal and {Smith}, Aaron and {Greene}, Jenny E. and {Brammer}, Gabriel and {Weibel}, Andrea and {Hviding}, Raphael and {Chisholm}, John and {Labb\textbackslash'e}, Ivo and {Simcoe}, Robert A. and {Witten}, Callum and {Atek}, Hakim and {Baggen}, Josephine F.~W. and {Belli}, Sirio and {Bezanson}, Rachel and {Boogaard}, Leindert A. and {Bose}, Sownak and {Covelo-Paz}, Alba and {Dayal}, Pratika and {Fudamoto}, Yoshinobu and {Furtak}, Lukas J. and {Giovinazzo}, Emma and {Goulding}, Andy and {Gronke}, Max and {Heintz}, Kasper E. and {Hirschmann}, Michaela and {Illingworth}, Garth and {Inoue}, Akio K. and {Johnson}, Benjamin D. and {Leja}, Joel and {Leonova}, Ecaterina and {McConachie}, Ian and {Maseda}, Michael V. and {Natarajan}, Priyamvada and {Nelson}, Erica and {Setton}, David J. and {Shivaei}, Irene and {Sobral}, David and {Stefanon}, Mauro and {Tacchella}, Sandro and {Toft}, Sune and {Torralba}, Alberto and {van Dokkum}, Pieter and {van der Wel}, Arjen and {Volonteri}, Marta and {Walter}, Fabian and {Wang}, Bingjie and {Watson}, Darach},
        title = "{A ``Black Hole Star'' Reveals the Remarkable Gas-Enshrouded Hearts of the Little Red Dots}",
      journal = {arXiv e-prints},
         year = 2025,
        month = mar,
          eid = {arXiv:2503.16596},
        pages = {arXiv:2503.16596},
          doi = {10.48550/arXiv.2503.16596},
archivePrefix = {arXiv},
       eprint = {2503.16596},
 primaryClass = {astro-ph.GA},
       adsurl = {https://ui.adsabs.harvard.edu/abs/2025arXiv250316596N}
}

@ARTICLE{Perez-Gonzalez+24,
       author = {{P{\'e}rez-Gonz{\'a}lez}, Pablo G. and {Barro}, Guillermo and {Rieke}, George H. and {Lyu}, Jianwei and {Rieke}, Marcia and {Alberts}, Stacey and {Williams}, Christina C. and {Hainline}, Kevin and {Sun}, Fengwu and {Pusk{\'a}s}, D{\'a}vid and {Annunziatella}, Marianna and {Baker}, William M. and {Bunker}, Andrew J. and {Egami}, Eiichi and {Ji}, Zhiyuan and {Johnson}, Benjamin D. and {Robertson}, Brant and {Rodr{\'\i}guez Del Pino}, Bruno and {Rujopakarn}, Wiphu and {Shivaei}, Irene and {Tacchella}, Sandro and {Willmer}, Christopher N.~A. and {Willott}, Chris},
        title = "{What Is the Nature of Little Red Dots and what Is Not, MIRI SMILES Edition}",
      journal = {\apj},
         year = 2024,
        month = jun,
       volume = {968},
       number = {1},
          eid = {4},
        pages = {4},
          doi = {10.3847/1538-4357/ad38bb},
archivePrefix = {arXiv},
       eprint = {2401.08782},
 primaryClass = {astro-ph.GA},
       adsurl = {https://ui.adsabs.harvard.edu/abs/2024ApJ...968....4P}
}

@ARTICLE{Pasha+23,
       author = {{Pasha}, Imad and {Miller}, Tim B.},
        title = "{pysersic: A Python package for determining galaxy structural properties via Bayesian inference, accelerated with jax}",
      journal = {The Journal of Open Source Software},
         year = 2023,
        month = sep,
       volume = {8},
       number = {89},
          eid = {5703},
        pages = {5703},
          doi = {10.21105/joss.05703},
archivePrefix = {arXiv},
       eprint = {2306.05454},
 primaryClass = {astro-ph.GA},
       adsurl = {https://ui.adsabs.harvard.edu/abs/2023JOSS....8.5703P}
}

@ARTICLE{Theios+19,
       author = {{Theios}, Rachel L. and {Steidel}, Charles C. and {Strom}, Allison L. and {Rudie}, Gwen C. and {Trainor}, Ryan F. and {Reddy}, Naveen A.},
        title = "{Dust Attenuation, Star Formation, and Metallicity in z {\ensuremath{\sim}} 2-3 Galaxies from KBSS-MOSFIRE}",
      journal = {\apj},
         year = 2019,
        month = jan,
       volume = {871},
       number = {1},
          eid = {128},
        pages = {128},
          doi = {10.3847/1538-4357/aaf386},
archivePrefix = {arXiv},
       eprint = {1805.00016},
 primaryClass = {astro-ph.GA},
       adsurl = {https://ui.adsabs.harvard.edu/abs/2019ApJ...871..128T}
}

@ARTICLE{Willott+22,
       author = {{Willott}, Chris J. and {Doyon}, Ren{\'e} and {Albert}, Loic and {Brammer}, Gabriel B. and {Dixon}, William V. and {Muzic}, Koraljka and {Ravindranath}, Swara and {Scholz}, Aleks and {Abraham}, Roberto and {Artigau}, {\'E}tienne and {Brada{\v{c}}}, Maru{\v{s}}a and {Goudfrooij}, Paul and {Hutchings}, John B. and {Iyer}, Kartheik G. and {Jayawardhana}, Ray and {LaMassa}, Stephanie and {Martis}, Nicholas and {Meyer}, Michael R. and {Morishita}, Takahiro and {Mowla}, Lamiya and {Muzzin}, Adam and {Noirot}, Ga{\"e}l and {Pacifici}, Camilla and {Rowlands}, Neil and {Sarrouh}, Ghassan and {Sawicki}, Marcin and {Taylor}, Joanna M. and {Volk}, Kevin and {Zabl}, Johannes},
        title = "{The Near-infrared Imager and Slitless Spectrograph for the James Webb Space Telescope. II. Wide Field Slitless Spectroscopy}",
      journal = {\pasp},
         year = 2022,
        month = feb,
       volume = {134},
       number = {1032},
          eid = {025002},
        pages = {025002},
          doi = {10.1088/1538-3873/ac5158},
archivePrefix = {arXiv},
       eprint = {2202.01714},
 primaryClass = {astro-ph.IM},
       adsurl = {https://ui.adsabs.harvard.edu/abs/2022PASP..134b5002W}
}

@ARTICLE{Heywood+21,
       author = {{Heywood}, I. and {Murphy}, E.~J. and {Jim{\'e}nez-Andrade}, E.~F. and {Armus}, L. and {Cotton}, W.~D. and {DeCoursey}, C. and {Dickinson}, M. and {Lazio}, T.~J.~W. and {Momjian}, E. and {Penner}, K. and {Smail}, I. and {Smirnov}, O.~M.},
        title = "{The VLA Frontier Fields Survey: Deep, High-resolution Radio Imaging of the MACS Lensing Clusters at 3 and 6 GHz}",
      journal = {\apj},
         year = 2021,
        month = apr,
       volume = {910},
       number = {2},
          eid = {105},
        pages = {105},
          doi = {10.3847/1538-4357/abdf61},
archivePrefix = {arXiv},
       eprint = {2103.07806},
 primaryClass = {astro-ph.GA},
       adsurl = {https://ui.adsabs.harvard.edu/abs/2021ApJ...910..105H}
}

@ARTICLE{Xue+16,
       author = {{Xue}, Y.~Q. and {Luo}, B. and {Brandt}, W.~N. and {Alexander}, D.~M. and {Bauer}, F.~E. and {Lehmer}, B.~D. and {Yang}, G.},
        title = "{The 2 Ms Chandra Deep Field-North Survey and the 250 ks Extended Chandra Deep Field-South Survey: Improved Point-source Catalogs}",
      journal = {\apjs},
         year = 2016,
        month = jun,
       volume = {224},
       number = {2},
          eid = {15},
        pages = {15},
          doi = {10.3847/0067-0049/224/2/15},
archivePrefix = {arXiv},
       eprint = {1602.06299},
 primaryClass = {astro-ph.GA},
       adsurl = {https://ui.adsabs.harvard.edu/abs/2016ApJS..224...15X}
}

@ARTICLE{Magnelli+13,
       author = {{Magnelli}, B. and {Popesso}, P. and {Berta}, S. and {Pozzi}, F. and {Elbaz}, D. and {Lutz}, D. and {Dickinson}, M. and {Altieri}, B. and {Andreani}, P. and {Aussel}, H. and {B{\'e}thermin}, M. and {Bongiovanni}, A. and {Cepa}, J. and {Charmandaris}, V. and {Chary}, R.-R. and {Cimatti}, A. and {Daddi}, E. and {F{\"o}rster Schreiber}, N.~M. and {Genzel}, R. and {Gruppioni}, C. and {Harwit}, M. and {Hwang}, H.~S. and {Ivison}, R.~J. and {Magdis}, G. and {Maiolino}, R. and {Murphy}, E. and {Nordon}, R. and {Pannella}, M. and {P{\'e}rez Garc{\'\i}a}, A. and {Poglitsch}, A. and {Rosario}, D. and {Sanchez-Portal}, M. and {Santini}, P. and {Scott}, D. and {Sturm}, E. and {Tacconi}, L.~J. and {Valtchanov}, I.},
        title = "{The deepest Herschel-PACS far-infrared survey: number counts and infrared luminosity functions from combined PEP/GOODS-H observations}",
      journal = {\aap},
         year = 2013,
        month = may,
       volume = {553},
          eid = {A132},
        pages = {A132},
          doi = {10.1051/0004-6361/201321371},
archivePrefix = {arXiv},
       eprint = {1303.4436},
 primaryClass = {astro-ph.CO},
       adsurl = {https://ui.adsabs.harvard.edu/abs/2013A&A...553A.132M}
}

@INPROCEEDINGS{Teplitz+04,
       author = {{Teplitz}, H.~I. and {Appleton}, P.~N. and {Armus}, L. and {Grillmair}, C.~J. and {Wu}, X. and {Charmandaris}, V. and {Uchida}, K. and {van Cleve}, J. and {Spitzer Science Center Team} and {IRS Instrument Team}},
        title = "{PeakUp Imaging mode for the Spitzer IRS}",
    booktitle = {American Astronomical Society Meeting Abstracts},
         year = 2004,
       series = {American Astronomical Society Meeting Abstracts},
       volume = {205},
        month = dec,
          eid = {05.05},
        pages = {05.05},
       adsurl = {https://ui.adsabs.harvard.edu/abs/2004AAS...205.0505T}
}

@INPROCEEDINGS{Rieke+04,
       author = {{Rieke}, George H. and {Young}, Erick T. and {Cadien}, James and {Engelbracht}, Charles W. and {Gordon}, Karl D. and {Kelly}, Douglas M. and {Low}, Frank J. and {Misselt}, Karl A. and {Morrison}, Jane E. and {Muzerolle}, James and {Rivlis}, G. and {Stansberry}, John A. and {Beeman}, Jeffrey W. and {Haller}, Eugene E. and {Frayer}, David T. and {Latter}, William B. and {Noriega-Crespo}, Alberto and {Padgett}, Deborah L. and {Hines}, Dean C. and {Bean}, J. Douglas and {Burmester}, William and {Heim}, Gerald B. and {Glenn}, Thomas and {Ordonez}, R. and {Schwenker}, John P. and {Siewert}, S. and {Strecker}, Donald W. and {Tennant}, S. and {Troeltzsch}, John R. and {Unruh}, Bryce and {Warden}, R.~M. and {Ade}, Peter A. and {Alonso-Herrero}, Almudena and {Blaylock}, Myra and {Dole}, Herve and {Egami}, Eiichi and {Hinz}, Joannah L. and {Le Floc'h}, Emeric and {Papovich}, Casey and {Perez-Gonzalez}, Pablo G. and {Rieke}, Marcia J. and {Smith}, Paul S. and {Su}, Kate Y.~L. and {Bennett}, Lee and {Henderson}, David and {Lu}, Nanyao and {Masci}, Frank J. and {Pesenson}, Misha and {Rebull}, Luisa and {Rho}, Jeonghee and {Keene}, Jocelyn and {Stolovy}, Susan and {Wachter}, Stefanie and {Wheaton}, William and {Richards}, Paul L. and {Garner}, Harry W. and {Hegge}, M. and {Henderson}, Monte L. and {MacFeely}, Kim I. and {Michika}, David and {Miller}, Chris D. and {Neitenbach}, Mark and {Winghart}, Jeremiah and {Woodruff}, R. and {Arens}, E. and {Beichman}, Charles A. and {Gaalema}, Stephen D. and {Gautier}, III, Thomas N. and {Lada}, Charles J. and {Mould}, Jeremy and {Neugebauer}, Gerry X. and {Stapelfeldt}, Karl R.},
        title = "{On-orbit performance of the MIPS instrument}",
    booktitle = {Optical, Infrared, and Millimeter Space Telescopes},
         year = 2004,
       editor = {{Mather}, John C.},
       series = {Society of Photo-Optical Instrumentation Engineers (SPIE) Conference Series},
       volume = {5487},
        month = oct,
        pages = {50-61},
          doi = {10.1117/12.551965},
       adsurl = {https://ui.adsabs.harvard.edu/abs/2004SPIE.5487...50R}
}

@INPROCEEDINGS{Dickinson+03,
       author = {{Dickinson}, Mark and {Giavalisco}, Mauro and {GOODS Team}},
        title = "{The Great Observatories Origins Deep Survey}",
    booktitle = {The Mass of Galaxies at Low and High Redshift},
         year = 2003,
       editor = {{Bender}, Ralf and {Renzini}, Alvio},
        month = jan,
        pages = {324},
          doi = {10.1007/10899892_78},
archivePrefix = {arXiv},
       eprint = {astro-ph/0204213},
 primaryClass = {astro-ph},
       adsurl = {https://ui.adsabs.harvard.edu/abs/2003mglh.conf..324D}
}

@ARTICLE{Griffin+10,
       author = {{Griffin}, M.~J. and {Abergel}, A. and {Abreu}, A. and {Ade}, P.~A.~R. and {Andr{\'e}}, P. and {Augueres}, J.-L. and {Babbedge}, T. and {Bae}, Y. and {Baillie}, T. and {Baluteau}, J.-P. and {Barlow}, M.~J. and {Bendo}, G. and {Benielli}, D. and {Bock}, J.~J. and {Bonhomme}, P. and {Brisbin}, D. and {Brockley-Blatt}, C. and {Caldwell}, M. and {Cara}, C. and {Castro-Rodriguez}, N. and {Cerulli}, R. and {Chanial}, P. and {Chen}, S. and {Clark}, E. and {Clements}, D.~L. and {Clerc}, L. and {Coker}, J. and {Communal}, D. and {Conversi}, L. and {Cox}, P. and {Crumb}, D. and {Cunningham}, C. and {Daly}, F. and {Davis}, G.~R. and {de Antoni}, P. and {Delderfield}, J. and {Devin}, N. and {di Giorgio}, A. and {Didschuns}, I. and {Dohlen}, K. and {Donati}, M. and {Dowell}, A. and {Dowell}, C.~D. and {Duband}, L. and {Dumaye}, L. and {Emery}, R.~J. and {Ferlet}, M. and {Ferrand}, D. and {Fontignie}, J. and {Fox}, M. and {Franceschini}, A. and {Frerking}, M. and {Fulton}, T. and {Garcia}, J. and {Gastaud}, R. and {Gear}, W.~K. and {Glenn}, J. and {Goizel}, A. and {Griffin}, D.~K. and {Grundy}, T. and {Guest}, S. and {Guillemet}, L. and {Hargrave}, P.~C. and {Harwit}, M. and {Hastings}, P. and {Hatziminaoglou}, E. and {Herman}, M. and {Hinde}, B. and {Hristov}, V. and {Huang}, M. and {Imhof}, P. and {Isaak}, K.~J. and {Israelsson}, U. and {Ivison}, R.~J. and {Jennings}, D. and {Kiernan}, B. and {King}, K.~J. and {Lange}, A.~E. and {Latter}, W. and {Laurent}, G. and {Laurent}, P. and {Leeks}, S.~J. and {Lellouch}, E. and {Levenson}, L. and {Li}, B. and {Li}, J. and {Lilienthal}, J. and {Lim}, T. and {Liu}, S.~J. and {Lu}, N. and {Madden}, S. and {Mainetti}, G. and {Marliani}, P. and {McKay}, D. and {Mercier}, K. and {Molinari}, S. and {Morris}, H. and {Moseley}, H. and {Mulder}, J. and {Mur}, M. and {Naylor}, D.~A. and {Nguyen}, H. and {O'Halloran}, B. and {Oliver}, S. and {Olofsson}, G. and {Olofsson}, H.-G. and {Orfei}, R. and {Page}, M.~J. and {Pain}, I. and {Panuzzo}, P. and {Papageorgiou}, A. and {Parks}, G. and {Parr-Burman}, P. and {Pearce}, A. and {Pearson}, C. and {P{\'e}rez-Fournon}, I. and {Pinsard}, F. and {Pisano}, G. and {Podosek}, J. and {Pohlen}, M. and {Polehampton}, E.~T. and {Pouliquen}, D. and {Rigopoulou}, D. and {Rizzo}, D. and {Roseboom}, I.~G. and {Roussel}, H. and {Rowan-Robinson}, M. and {Rownd}, B. and {Saraceno}, P. and {Sauvage}, M. and {Savage}, R. and {Savini}, G. and {Sawyer}, E. and {Scharmberg}, C. and {Schmitt}, D. and {Schneider}, N. and {Schulz}, B. and {Schwartz}, A. and {Shafer}, R. and {Shupe}, D.~L. and {Sibthorpe}, B. and {Sidher}, S. and {Smith}, A. and {Smith}, A.~J. and {Smith}, D. and {Spencer}, L. and {Stobie}, B. and {Sudiwala}, R. and {Sukhatme}, K. and {Surace}, C. and {Stevens}, J.~A. and {Swinyard}, B.~M. and {Trichas}, M. and {Tourette}, T. and {Triou}, H. and {Tseng}, S. and {Tucker}, C. and {Turner}, A. and {Vaccari}, M. and {Valtchanov}, I. and {Vigroux}, L. and {Virique}, E. and {Voellmer}, G. and {Walker}, H. and {Ward}, R. and {Waskett}, T. and {Weilert}, M. and {Wesson}, R. and {White}, G.~J. and {Whitehouse}, N. and {Wilson}, C.~D. and {Winter}, B. and {Woodcraft}, A.~L. and {Wright}, G.~S. and {Xu}, C.~K. and {Zavagno}, A. and {Zemcov}, M. and {Zhang}, L. and {Zonca}, E.},
        title = "{The Herschel-SPIRE instrument and its in-flight performance}",
      journal = {\aap},
         year = 2010,
        month = jul,
       volume = {518},
          eid = {L3},
        pages = {L3},
          doi = {10.1051/0004-6361/201014519},
archivePrefix = {arXiv},
       eprint = {1005.5123},
 primaryClass = {astro-ph.IM},
       adsurl = {https://ui.adsabs.harvard.edu/abs/2010A&A...518L...3G}
}

@ARTICLE{Sun+22,
       author = {{Sun}, Fengwu and {Egami}, Eiichi and {Fujimoto}, Seiji and {Rawle}, Timothy and {Bauer}, Franz E. and {Kohno}, Kotaro and {Smail}, Ian and {P{\'e}rez-Gonz{\'a}lez}, Pablo G. and {Ao}, Yiping and {Chapman}, Scott C. and {Combes}, Francoise and {Dessauges-Zavadsky}, Miroslava and {Espada}, Daniel and {Gonz{\'a}lez-L{\'o}pez}, Jorge and {Koekemoer}, Anton M. and {Kokorev}, Vasily and {Lee}, Minju M. and {Morokuma-Matsui}, Kana and {Mu{\~n}oz Arancibia}, Alejandra M. and {Oguri}, Masamune and {Pell{\'o}}, Roser and {Ueda}, Yoshihiro and {Uematsu}, Ryosuke and {Valentino}, Francesco and {Van der Werf}, Paul and {Walth}, Gregory L. and {Zemcov}, Michael and {Zitrin}, Adi},
        title = "{ALMA Lensing Cluster Survey: ALMA-Herschel Joint Study of Lensed Dusty Star-forming Galaxies across z ≃ 0.5 - 6}",
      journal = {\apj},
         year = 2022,
        month = jun,
       volume = {932},
       number = {2},
          eid = {77},
        pages = {77},
          doi = {10.3847/1538-4357/ac6e3f},
archivePrefix = {arXiv},
       eprint = {2204.07187},
 primaryClass = {astro-ph.GA},
       adsurl = {https://ui.adsabs.harvard.edu/abs/2022ApJ...932...77S}
}

@ARTICLE{Kokorev+22,
       author = {{Kokorev}, V. and {Brammer}, G. and {Fujimoto}, S. and {Kohno}, K. and {Magdis}, G.~E. and {Valentino}, F. and {Toft}, S. and {Oesch}, P. and {Davidzon}, I. and {Bauer}, F.~E. and {Coe}, D. and {Egami}, E. and {Oguri}, M. and {Ouchi}, M. and {Postman}, M. and {Richard}, J. and {Jolly}, J.-B. and {Knudsen}, K.~K. and {Sun}, F. and {Weaver}, J.~R. and {Ao}, Y. and {Baker}, A.~J. and {Bradley}, L. and {Caputi}, K.~I. and {Dessauges-Zavadsky}, M. and {Espada}, D. and {Hatsukade}, B. and {Koekemoer}, A.~M. and {Mu{\~n}oz Arancibia}, A.~M. and {Shimasaku}, K. and {Umehata}, H. and {Wang}, T. and {Wang}, W.-H.},
        title = "{ALMA Lensing Cluster Survey: Hubble Space Telescope and Spitzer Photometry of 33 Lensed Fields Built with CHArGE}",
      journal = {\apjs},
         year = 2022,
        month = dec,
       volume = {263},
       number = {2},
          eid = {38},
        pages = {38},
          doi = {10.3847/1538-4365/ac9909},
archivePrefix = {arXiv},
       eprint = {2207.07125},
 primaryClass = {astro-ph.GA},
       adsurl = {https://ui.adsabs.harvard.edu/abs/2022ApJS..263...38K}
}

@ARTICLE{Egami+10,
       author = {{Egami}, E. and {Rex}, M. and {Rawle}, T.~D. and {P{\'e}rez-Gonz{\'a}lez}, P.~G. and {Richard}, J. and {Kneib}, J.-P. and {Schaerer}, D. and {Altieri}, B. and {Valtchanov}, I. and {Blain}, A.~W. and {Fadda}, D. and {Zemcov}, M. and {Bock}, J.~J. and {Boone}, F. and {Bridge}, C.~R. and {Clement}, B. and {Combes}, F. and {Dessauges-Zavadsky}, M. and {Dowell}, C.~D. and {Ilbert}, O. and {Ivison}, R.~J. and {Jauzac}, M. and {Lutz}, D. and {Metcalfe}, L. and {Omont}, A. and {Pell{\'o}}, R. and {Pereira}, M.~J. and {Rieke}, G.~H. and {Rodighiero}, G. and {Smail}, I. and {Smith}, G.~P. and {Tramoy}, G. and {Walth}, G.~L. and {van der Werf}, P. and {Werner}, M.~W.},
        title = "{The Herschel Lensing Survey (HLS): Overview}",
      journal = {\aap},
         year = 2010,
        month = jul,
       volume = {518},
          eid = {L12},
        pages = {L12},
          doi = {10.1051/0004-6361/201014696},
archivePrefix = {arXiv},
       eprint = {1005.3820},
 primaryClass = {astro-ph.CO},
       adsurl = {https://ui.adsabs.harvard.edu/abs/2010A&A...518L..12E}
}

@ARTICLE{Poglitsch+10,
       author = {{Poglitsch}, A. and {Waelkens}, C. and {Geis}, N. and {Feuchtgruber}, H. and {Vandenbussche}, B. and {Rodriguez}, L. and {Krause}, O. and {Renotte}, E. and {van Hoof}, C. and {Saraceno}, P. and {Cepa}, J. and {Kerschbaum}, F. and {Agn{\`e}se}, P. and {Ali}, B. and {Altieri}, B. and {Andreani}, P. and {Augueres}, J.-L. and {Balog}, Z. and {Barl}, L. and {Bauer}, O.~H. and {Belbachir}, N. and {Benedettini}, M. and {Billot}, N. and {Boulade}, O. and {Bischof}, H. and {Blommaert}, J. and {Callut}, E. and {Cara}, C. and {Cerulli}, R. and {Cesarsky}, D. and {Contursi}, A. and {Creten}, Y. and {De Meester}, W. and {Doublier}, V. and {Doumayrou}, E. and {Duband}, L. and {Exter}, K. and {Genzel}, R. and {Gillis}, J.-M. and {Gr{\"o}zinger}, U. and {Henning}, T. and {Herreros}, J. and {Huygen}, R. and {Inguscio}, M. and {Jakob}, G. and {Jamar}, C. and {Jean}, C. and {de Jong}, J. and {Katterloher}, R. and {Kiss}, C. and {Klaas}, U. and {Lemke}, D. and {Lutz}, D. and {Madden}, S. and {Marquet}, B. and {Martignac}, J. and {Mazy}, A. and {Merken}, P. and {Montfort}, F. and {Morbidelli}, L. and {M{\"u}ller}, T. and {Nielbock}, M. and {Okumura}, K. and {Orfei}, R. and {Ottensamer}, R. and {Pezzuto}, S. and {Popesso}, P. and {Putzeys}, J. and {Regibo}, S. and {Reveret}, V. and {Royer}, P. and {Sauvage}, M. and {Schreiber}, J. and {Stegmaier}, J. and {Schmitt}, D. and {Schubert}, J. and {Sturm}, E. and {Thiel}, M. and {Tofani}, G. and {Vavrek}, R. and {Wetzstein}, M. and {Wieprecht}, E. and {Wiezorrek}, E.},
        title = "{The Photodetector Array Camera and Spectrometer (PACS) on the Herschel Space Observatory}",
      journal = {\aap},
         year = 2010,
        month = jul,
       volume = {518},
          eid = {L2},
        pages = {L2},
          doi = {10.1051/0004-6361/201014535},
archivePrefix = {arXiv},
       eprint = {1005.1487},
 primaryClass = {astro-ph.IM},
       adsurl = {https://ui.adsabs.harvard.edu/abs/2010A&A...518L...2P}
}

@ARTICLE{Gonzalez-Lopez+17b,
       author = {{Gonz{\'a}lez-L{\'o}pez}, J. and {Bauer}, F.~E. and {Aravena}, M. and {Laporte}, N. and {Bradley}, L. and {Carrasco}, M. and {Carvajal}, R. and {Demarco}, R. and {Infante}, L. and {Kneissl}, R. and {Koekemoer}, A.~M. and {Mu{\~n}oz Arancibia}, A.~M. and {Troncoso}, P. and {Villard}, E. and {Zitrin}, A.},
        title = "{The ALMA Frontier Fields Survey. III. 1.1 mm emission line identifications in Abell 2744, MACSJ 0416.1-2403, MACSJ 1149.5+2223, Abell 370, and Abell S1063}",
      journal = {\aap},
         year = 2017,
        month = dec,
       volume = {608},
          eid = {A138},
        pages = {A138},
          doi = {10.1051/0004-6361/201730961},
archivePrefix = {arXiv},
       eprint = {1704.03007},
 primaryClass = {astro-ph.GA},
       adsurl = {https://ui.adsabs.harvard.edu/abs/2017A&A...608A.138G}
}

@ARTICLE{Gonzalez-Lopez+17a,
       author = {{Gonz{\'a}lez-L{\'o}pez}, J. and {Bauer}, F.~E. and {Romero-Ca{\~n}izales}, C. and {Kneissl}, R. and {Villard}, E. and {Carvajal}, R. and {Kim}, S. and {Laporte}, N. and {Anguita}, T. and {Aravena}, M. and {Bouwens}, R.~J. and {Bradley}, L. and {Carrasco}, M. and {Demarco}, R. and {Ford}, H. and {Ibar}, E. and {Infante}, L. and {Messias}, H. and {Mu{\~n}oz Arancibia}, A.~M. and {Nagar}, N. and {Padilla}, N. and {Treister}, E. and {Troncoso}, P. and {Zitrin}, A.},
        title = "{The ALMA Frontier Fields Survey. I. 1.1 mm continuum detections in Abell 2744, MACS J0416.1-2403 and MACS J1149.5+2223}",
      journal = {\aap},
         year = 2017,
        month = jan,
       volume = {597},
          eid = {A41},
        pages = {A41},
          doi = {10.1051/0004-6361/201628806},
archivePrefix = {arXiv},
       eprint = {1607.03808},
 primaryClass = {astro-ph.GA},
       adsurl = {https://ui.adsabs.harvard.edu/abs/2017A&A...597A..41G}
}

@ARTICLE{Rawle+16,
       author = {{Rawle}, T.~D. and {Altieri}, B. and {Egami}, E. and {P{\'e}rez-Gonz{\'a}lez}, P.~G. and {Boone}, F. and {Clement}, B. and {Ivison}, R.~J. and {Richard}, J. and {Rujopakarn}, W. and {Valtchanov}, I. and {Walth}, G. and {Weiner}, B.~J. and {Blain}, A.~W. and {Dessauges-Zavadsky}, M. and {Kneib}, J.-P. and {Lutz}, D. and {Rodighiero}, G. and {Schaerer}, D. and {Smail}, I.},
        title = "{A complete census of Herschel-detected infrared sources within the HST Frontier Fields}",
      journal = {\mnras},
         year = 2016,
        month = jun,
       volume = {459},
       number = {2},
        pages = {1626-1645},
          doi = {10.1093/mnras/stw712},
archivePrefix = {arXiv},
       eprint = {1508.00586},
 primaryClass = {astro-ph.GA},
       adsurl = {https://ui.adsabs.harvard.edu/abs/2016MNRAS.459.1626R}
}

@ARTICLE{Liu+18,
       author = {{Liu}, Daizhong and {Daddi}, Emanuele and {Dickinson}, Mark and {Owen}, Frazer and {Pannella}, Maurilio and {Sargent}, Mark and {B{\'e}thermin}, Matthieu and {Magdis}, Georgios and {Gao}, Yu and {Shu}, Xinwen and {Wang}, Tao and {Jin}, Shuowen and {Inami}, Hanae},
        title = "{{\textquotedblleft}Super-deblended{\textquotedblright} Dust Emission in Galaxies. I. The GOODS-North Catalog and the Cosmic Star Formation Rate Density out to Redshift 6}",
      journal = {\apj},
         year = 2018,
        month = feb,
       volume = {853},
       number = {2},
          eid = {172},
        pages = {172},
          doi = {10.3847/1538-4357/aaa600},
archivePrefix = {arXiv},
       eprint = {1703.05281},
 primaryClass = {astro-ph.GA},
       adsurl = {https://ui.adsabs.harvard.edu/abs/2018ApJ...853..172L}
}

@ARTICLE{Owen+18,
       author = {{Owen}, Frazer N.},
        title = "{Deep JVLA Imaging of GOODS-N at 20 cm}",
      journal = {\apjs},
         year = 2018,
        month = apr,
       volume = {235},
       number = {2},
          eid = {34},
        pages = {34},
          doi = {10.3847/1538-4365/aab4a1},
archivePrefix = {arXiv},
       eprint = {1803.05455},
 primaryClass = {astro-ph.GA},
       adsurl = {https://ui.adsabs.harvard.edu/abs/2018ApJS..235...34O}
}

@ARTICLE{Ighina+25,
       author = {{Ighina}, Luca and {Caccianiga}, Alessandro and {Connor}, Thomas and {Moretti}, Alberto and {Pacucci}, Fabio and {Reynolds}, Cormac and {Afonso}, Jos{\'e} and {Arsioli}, Bruno and {Belladitta}, Silvia and {Broderick}, Jess W. and {Dallacasa}, Daniele and {Della Ceca}, Roberto and {Haardt}, Francesco and {Lambrides}, Erini and {Leung}, James K. and {Lupi}, Alessandro and {Matute}, Israel and {Rigamonti}, Fabio and {Severgnini}, Paola and {Seymour}, Nick and {Tavecchio}, Fabrizio and {Vignali}, Cristian},
        title = "{X-Ray Investigation of Possible Super-Eddington Accretion in a Radio-loud Quasar at z = 6.13}",
      journal = {\apjl},
         year = 2025,
        month = sep,
       volume = {990},
       number = {2},
          eid = {L56},
        pages = {L56},
          doi = {10.3847/2041-8213/aded0a},
archivePrefix = {arXiv},
       eprint = {2509.04559},
 primaryClass = {astro-ph.GA},
       adsurl = {https://ui.adsabs.harvard.edu/abs/2025ApJ...990L..56I}
}

@ARTICLE{Yang+22,
       author = {{Yang}, Guang and {Boquien}, M{\'e}d{\'e}ric and {Brandt}, W.~N. and {Buat}, V{\'e}ronique and {Burgarella}, Denis and {Ciesla}, Laure and {Lehmer}, Bret D. and {Ma{\l}ek}, Katarzyna and {Mountrichas}, George and {Papovich}, Casey and {Pons}, Estelle and {Stalevski}, Marko and {Theul{\'e}}, Patrice and {Zhu}, Shifu},
        title = "{Fitting AGN/Galaxy X-Ray-to-radio SEDs with CIGALE and Improvement of the Code}",
      journal = {\apj},
         year = 2022,
        month = mar,
       volume = {927},
       number = {2},
          eid = {192},
        pages = {192},
          doi = {10.3847/1538-4357/ac4971},
archivePrefix = {arXiv},
       eprint = {2201.03718},
 primaryClass = {astro-ph.GA},
       adsurl = {https://ui.adsabs.harvard.edu/abs/2022ApJ...927..192Y}
}

@ARTICLE{Yang+20,
       author = {{Yang}, G. and {Boquien}, M. and {Buat}, V. and {Burgarella}, D. and {Ciesla}, L. and {Duras}, F. and {Stalevski}, M. and {Brandt}, W.~N. and {Papovich}, C.},
        title = "{X-CIGALE: Fitting AGN/galaxy SEDs from X-ray to infrared}",
      journal = {\mnras},
         year = 2020,
        month = jan,
       volume = {491},
       number = {1},
        pages = {740-757},
          doi = {10.1093/mnras/stz3001},
archivePrefix = {arXiv},
       eprint = {2001.08263},
 primaryClass = {astro-ph.GA},
       adsurl = {https://ui.adsabs.harvard.edu/abs/2020MNRAS.491..740Y}
}

@ARTICLE{Noirot+23,
       author = {{Noirot}, Ga{\"e}l and {Desprez}, Guillaume and {Asada}, Yoshihisa and {Sawicki}, Marcin and {Estrada-Carpenter}, Vicente and {Martis}, Nicholas and {Sarrouh}, Ghassan and {Strait}, Victoria and {Abraham}, Roberto and {Brada{\v{c}}}, Maru{\v{s}}a and {Brammer}, Gabriel and {Iyer}, Kartheik and {MacFarland}, Shannon and {Matharu}, Jasleen and {Mowla}, Lamiya and {Muzzin}, Adam and {Pacifici}, Camilla and {Ravindranath}, Swara and {Willott}, Chris J. and {Albert}, Lo{\"\i}c and {Doyon}, Ren{\'e} and {Hutchings}, John B. and {Rowlands}, Neil},
        title = "{The first large catalogue of spectroscopic redshifts in Webb's first deep field, SMACS J0723.3-7327}",
      journal = {\mnras},
         year = 2023,
        month = oct,
       volume = {525},
       number = {2},
        pages = {1867-1884},
          doi = {10.1093/mnras/stad1019},
archivePrefix = {arXiv},
       eprint = {2212.07366},
 primaryClass = {astro-ph.GA},
       adsurl = {https://ui.adsabs.harvard.edu/abs/2023MNRAS.525.1867N}
}

@ARTICLE{Maseda+24,
       author = {{Maseda}, Michael V. and {de Graaff}, Anna and {Franx}, Marijn and {Rix}, Hans-Walter and {Carniani}, Stefano and {Laseter}, Isaac and {Dudzevi{\v{c}}i{\={u}}t{\.{e}}}, Ugn{\.{e}} and {Rawle}, Tim and {Parlanti}, Eleonora and {Arribas}, Santiago and {Bunker}, Andrew J. and {Cameron}, Alex J. and {Charlot}, Stephane and {Curti}, Mirko and {D'Eugenio}, Francesco and {Jones}, Gareth C. and {Kumari}, Nimisha and {Maiolino}, Roberto and {{\"U}bler}, Hannah and {Saxena}, Aayush and {Smit}, Renske and {Willott}, Chris and {Witstok}, Joris},
        title = "{The NIRSpec Wide GTO Survey}",
      journal = {\aap},
         year = 2024,
        month = sep,
       volume = {689},
          eid = {A73},
        pages = {A73},
          doi = {10.1051/0004-6361/202449914},
archivePrefix = {arXiv},
       eprint = {2403.05506},
 primaryClass = {astro-ph.GA},
       adsurl = {https://ui.adsabs.harvard.edu/abs/2024A&A...689A..73M}
}

@ARTICLE{Kormendy_coevolution_2013,
       author = {{Kormendy}, John and {Ho}, Luis C.},
        title = "{Coevolution (Or Not) of Supermassive Black Holes and Host Galaxies}",
      journal = {\araa},
         year = 2013,
        month = aug,
       volume = {51},
       number = {1},
        pages = {511-653},
          doi = {10.1146/annurev-astro-082708-101811},
archivePrefix = {arXiv},
       eprint = {1304.7762},
 primaryClass = {astro-ph.CO},
       adsurl = {https://ui.adsabs.harvard.edu/abs/2013ARA&A..51..511K}
}

@ARTICLE{Mortlock_luminous_2011,
       author = {{Mortlock}, Daniel J. and {Warren}, Stephen J. and {Venemans}, Bram P. and {Patel}, Mitesh and {Hewett}, Paul C. and {McMahon}, Richard G. and {Simpson}, Chris and {Theuns}, Tom and {Gonz{\'a}les-Solares}, Eduardo A. and {Adamson}, Andy and {Dye}, Simon and {Hambly}, Nigel C. and {Hirst}, Paul and {Irwin}, Mike J. and {Kuiper}, Ernst and {Lawrence}, Andy and {R{\"o}ttgering}, Huub J.~A.},
        title = "{A luminous quasar at a redshift of z = 7.085}",
      journal = {\nat},
         year = 2011,
        month = jun,
       volume = {474},
       number = {7353},
        pages = {616-619},
          doi = {10.1038/nature10159},
archivePrefix = {arXiv},
       eprint = {1106.6088},
 primaryClass = {astro-ph.CO},
       adsurl = {https://ui.adsabs.harvard.edu/abs/2011Natur.474..616M}
}

@ARTICLE{Pacucci_growth_2015,
       author = {{Pacucci}, Fabio and {Volonteri}, Marta and {Ferrara}, Andrea},
        title = "{The growth efficiency of high-redshift black holes}",
      journal = {\mnras},
         year = 2015,
        month = sep,
       volume = {452},
       number = {2},
        pages = {1922-1933},
          doi = {10.1093/mnras/stv1465},
archivePrefix = {arXiv},
       eprint = {1506.04750},
 primaryClass = {astro-ph.GA},
       adsurl = {https://ui.adsabs.harvard.edu/abs/2015MNRAS.452.1922P}
}

@ARTICLE{Rinaldi+25c,
       author = {{Rinaldi}, P. and {Bonaventura}, N. and {Rieke}, G.~H. and {Alberts}, S. and {Caputi}, K.~I. and {Baker}, W.~M. and {Baum}, S. and {Bhatawdekar}, R. and {Bunker}, A.~J. and {Carniani}, S. and {Curtis-Lake}, E. and {D'Eugenio}, F. and {Egami}, E. and {Ji}, Z. and {Johnson}, B.~D. and {Hainline}, K. and {Helton}, J.~M. and {Lin}, X. and {Lyu}, J. and {Ma}, Z. and {Maiolino}, R. and {P{\'e}rez-Gonz{\'a}lez}, P.~G. and {Rieke}, M. and {Robertson}, B.~E. and {Shivaei}, I. and {Stone}, M. and {Sun}, Y. and {Tacchella}, S. and {{\"U}bler}, H. and {Williams}, C.~C. and {Willmer}, C.~N.~A. and {Willott}, C. and {Zhang}, J. and {Zhu}, Y.},
        title = "{Not Just a Dot: The Complex UV Morphology and Underlying Properties of Little Red Dots}",
      journal = {\apj},
         year = 2025,
        month = oct,
       volume = {992},
       number = {1},
          eid = {71},
        pages = {71},
          doi = {10.3847/1538-4357/adfa10},
archivePrefix = {arXiv},
       eprint = {2411.14383},
 primaryClass = {astro-ph.GA},
       adsurl = {https://ui.adsabs.harvard.edu/abs/2025ApJ...992...71R}
}

@ARTICLE{Reines_dwarf_2013,
       author = {{Reines}, Amy E. and {Greene}, Jenny E. and {Geha}, Marla},
        title = "{Dwarf Galaxies with Optical Signatures of Active Massive Black Holes}",
      journal = {\apj},
         year = 2013,
        month = oct,
       volume = {775},
       number = {2},
          eid = {116},
        pages = {116},
          doi = {10.1088/0004-637X/775/2/116},
archivePrefix = {arXiv},
       eprint = {1308.0328},
 primaryClass = {astro-ph.CO},
       adsurl = {https://ui.adsabs.harvard.edu/abs/2013ApJ...775..116R}
}

@ARTICLE{Donley_spitzer_2008,
       author = {{Donley}, J.~L. and {Rieke}, G.~H. and {P{\'e}rez-Gonz{\'a}lez}, P.~G. and {Barro}, G.},
        title = "{Spitzer's Contribution to the AGN Population}",
      journal = {\apj},
         year = 2008,
        month = nov,
       volume = {687},
       number = {1},
        pages = {111-132},
          doi = {10.1086/591510},
archivePrefix = {arXiv},
       eprint = {0806.4610},
 primaryClass = {astro-ph},
       adsurl = {https://ui.adsabs.harvard.edu/abs/2008ApJ...687..111D}
}

@ARTICLE{Assef_low-resolution_2010,
       author = {{Assef}, R.~J. and {Kochanek}, C.~S. and {Brodwin}, M. and {Cool}, R. and {Forman}, W. and {Gonzalez}, A.~H. and {Hickox}, R.~C. and {Jones}, C. and {Le Floc'h}, E. and {Moustakas}, J. and {Murray}, S.~S. and {Stern}, D.},
        title = "{Low-Resolution Spectral Templates for Active Galactic Nuclei and Galaxies from 0.03 to 30 {\ensuremath{\mu}}m}",
      journal = {\apj},
         year = 2010,
        month = apr,
       volume = {713},
       number = {2},
        pages = {970-985},
          doi = {10.1088/0004-637X/713/2/970},
archivePrefix = {arXiv},
       eprint = {0909.3849},
 primaryClass = {astro-ph.CO},
       adsurl = {https://ui.adsabs.harvard.edu/abs/2010ApJ...713..970A}
}

@ARTICLE{Hainline_spectroscopic_2014,
       author = {{Hainline}, Kevin N. and {Hickox}, Ryan C. and {Carroll}, Christopher M. and {Myers}, Adam D. and {DiPompeo}, Michael A. and {Trouille}, Laura},
        title = "{A Spectroscopic Survey of WISE-selected Obscured Quasars with the Southern African Large Telescope}",
      journal = {\apj},
         year = 2014,
        month = nov,
       volume = {795},
       number = {2},
          eid = {124},
        pages = {124},
          doi = {10.1088/0004-637X/795/2/124},
archivePrefix = {arXiv},
       eprint = {1409.4773},
 primaryClass = {astro-ph.GA},
       adsurl = {https://ui.adsabs.harvard.edu/abs/2014ApJ...795..124H}
}

@ARTICLE{Izotov_star-forming_2011,
       author = {{Izotov}, Y.~I. and {Guseva}, N.~G. and {Fricke}, K.~J. and {Henkel}, C.},
        title = "{Star-forming galaxies with hot dust emission in the Sloan Digital Sky Survey discovered by the Wide-field Infrared Survey Explorer (WISE)}",
      journal = {\aap},
         year = 2011,
        month = dec,
       volume = {536},
          eid = {L7},
        pages = {L7},
          doi = {10.1051/0004-6361/201118402},
archivePrefix = {arXiv},
       eprint = {1111.5450},
 primaryClass = {astro-ph.CO},
       adsurl = {https://ui.adsabs.harvard.edu/abs/2011A&A...536L...7I}
}

@ARTICLE{Stern_mid-infrared_2005,
       author = {{Stern}, Daniel and {Eisenhardt}, Peter and {Gorjian}, Varoujan and {Kochanek}, Christopher S. and {Caldwell}, Nelson and {Eisenstein}, Daniel and {Brodwin}, Mark and {Brown}, Michael J.~I. and {Cool}, Richard and {Dey}, Arjun and {Green}, Paul and {Jannuzi}, Buell T. and {Murray}, Stephen S. and {Pahre}, Michael A. and {Willner}, S.~P.},
        title = "{Mid-Infrared Selection of Active Galaxies}",
      journal = {\apj},
         year = 2005,
        month = sep,
       volume = {631},
       number = {1},
        pages = {163-168},
          doi = {10.1086/432523},
archivePrefix = {arXiv},
       eprint = {astro-ph/0410523},
 primaryClass = {astro-ph},
       adsurl = {https://ui.adsabs.harvard.edu/abs/2005ApJ...631..163S}
}

@ARTICLE{Alonso-Herrero_infared_2006,
       author = {{Alonso-Herrero}, A. and {P{\'e}rez-Gonz{\'a}lez}, P.~G. and {Alexander}, D.~M. and {Rieke}, G.~H. and {Rigopoulou}, D. and {Le Floc'h}, E. and {Barmby}, P. and {Papovich}, C. and {Rigby}, J.~R. and {Bauer}, F.~E. and {Brandt}, W.~N. and {Egami}, E. and {Willner}, S.~P. and {Dole}, H. and {Huang}, J.-S.},
        title = "{Infrared Power-Law Galaxies in the Chandra Deep Field-South: Active Galactic Nuclei and Ultraluminous Infrared Galaxies}",
      journal = {\apj},
         year = 2006,
        month = mar,
       volume = {640},
       number = {1},
        pages = {167-184},
          doi = {10.1086/499800},
archivePrefix = {arXiv},
       eprint = {astro-ph/0511507},
 primaryClass = {astro-ph},
       adsurl = {https://ui.adsabs.harvard.edu/abs/2006ApJ...640..167A}
}

@ARTICLE{Caputi_generalized_2013,
       author = {{Caputi}, K.~I.},
        title = "{A Generalized Power-law Diagnostic for Infrared Galaxies at z > 1: Active Galactic Nuclei and Hot Interstellar Dust}",
      journal = {\apj},
         year = 2013,
        month = may,
       volume = {768},
       number = {2},
          eid = {103},
        pages = {103},
          doi = {10.1088/0004-637X/768/2/103},
archivePrefix = {arXiv},
       eprint = {1303.1906},
 primaryClass = {astro-ph.CO},
       adsurl = {https://ui.adsabs.harvard.edu/abs/2013ApJ...768..103C}
}

@ARTICLE{Lupi_difficulties_2020,
       author = {{Lupi}, Alessandro and {Sbarrato}, Tullia and {Carniani}, Stefano},
        title = "{Difficulties in mid-infrared selection of AGNs in dwarf galaxies}",
      journal = {\mnras},
         year = 2020,
        month = feb,
       volume = {492},
       number = {2},
        pages = {2528-2534},
          doi = {10.1093/mnras/stz3636},
archivePrefix = {arXiv},
       eprint = {2001.02234},
 primaryClass = {astro-ph.GA},
       adsurl = {https://ui.adsabs.harvard.edu/abs/2020MNRAS.492.2528L}
}

@ARTICLE{Wright_wise_2010,
       author = {{Wright}, Edward L. and {Eisenhardt}, Peter R.~M. and {Mainzer}, Amy K. and {Ressler}, Michael E. and {Cutri}, Roc M. and {Jarrett}, Thomas and {Kirkpatrick}, J. Davy and {Padgett}, Deborah and {McMillan}, Robert S. and {Skrutskie}, Michael and {Stanford}, S.~A. and {Cohen}, Martin and {Walker}, Russell G. and {Mather}, John C. and {Leisawitz}, David and {Gautier}, III, Thomas N. and {McLean}, Ian and {Benford}, Dominic and {Lonsdale}, Carol J. and {Blain}, Andrew and {Mendez}, Bryan and {Irace}, William R. and {Duval}, Valerie and {Liu}, Fengchuan and {Royer}, Don and {Heinrichsen}, Ingolf and {Howard}, Joan and {Shannon}, Mark and {Kendall}, Martha and {Walsh}, Amy L. and {Larsen}, Mark and {Cardon}, Joel G. and {Schick}, Scott and {Schwalm}, Mark and {Abid}, Mohamed and {Fabinsky}, Beth and {Naes}, Larry and {Tsai}, Chao-Wei},
        title = "{The Wide-field Infrared Survey Explorer (WISE): Mission Description and Initial On-orbit Performance}",
      journal = {\aj},
         year = 2010,
        month = dec,
       volume = {140},
       number = {6},
        pages = {1868-1881},
          doi = {10.1088/0004-6256/140/6/1868},
archivePrefix = {arXiv},
       eprint = {1008.0031},
 primaryClass = {astro-ph.IM},
       adsurl = {https://ui.adsabs.harvard.edu/abs/2010AJ....140.1868W}
}

@ARTICLE{Hviding_RUBIES_2025,
       author = {{Hviding}, Raphael E. and {de Graaff}, Anna and {Miller}, Tim B. and {Setton}, David J. and {Greene}, Jenny E. and {Labb{\'e}}, Ivo and {Brammer}, Gabriel and {Bezanson}, Rachel and {Boogaard}, Leindert A. and {Cleri}, Nikko J. and {Leja}, Joel and {Maseda}, Michael V. and {McConachie}, Ian and {Matthee}, Jorryt and {Naidu}, Rohan P. and {Oesch}, Pascal A. and {Wang}, Bingjie and {Whitaker}, Katherine E. and {Williams}, Christina C.},
        title = "{RUBIES: A spectroscopic census of little red dots: All point sources with v-shaped continua have broad lines}",
      journal = {\aap},
         year = 2025,
        month = oct,
       volume = {702},
          eid = {A57},
        pages = {A57},
          doi = {10.1051/0004-6361/202555816},
archivePrefix = {arXiv},
       eprint = {2506.05459},
 primaryClass = {astro-ph.GA},
       adsurl = {https://ui.adsabs.harvard.edu/abs/2025A&A...702A..57H}
}

@ARTICLE{Juodzbalis_JADES_2024,
       author = {{Juod{\v{z}}balis}, Ignas and {Ji}, Xihan and {Maiolino}, Roberto and {D'Eugenio}, Francesco and {Scholtz}, Jan and {Risaliti}, Guido and {Fabian}, Andrew C. and {Mazzolari}, Giovanni and {Gilli}, Roberto and {Prandoni}, Isabella and {Arribas}, Santiago and {Bunker}, Andrew J. and {Carniani}, Stefano and {Charlot}, St{\'e}phane and {Curtis-Lake}, Emma and {de Graaff}, Anna and {Hainline}, Kevin and {Parlanti}, Eleonora and {Perna}, Michele and {P{\'e}rez-Gonz{\'a}lez}, Pablo G. and {Robertson}, Brant and {Tacchella}, Sandro and {{\"U}bler}, Hannah and {Williams}, Christina C. and {Willott}, Chris and {Witstok}, Joris},
        title = "{JADES - the Rosetta stone of JWST-discovered AGN: deciphering the intriguing nature of early AGN}",
      journal = {\mnras},
         year = 2024,
        month = nov,
       volume = {535},
       number = {1},
        pages = {853-873},
          doi = {10.1093/mnras/stae2367},
archivePrefix = {arXiv},
       eprint = {2407.08643},
 primaryClass = {astro-ph.GA},
       adsurl = {https://ui.adsabs.harvard.edu/abs/2024MNRAS.535..853J}
}

@ARTICLE{Ji_lord_2025,
       author = {{Ji}, Xihan and {D'Eugenio}, Francesco and {Juod{\v{z}}balis}, Ignas and {Walton}, Dominic J. and {Fabian}, Andrew C. and {Maiolino}, Roberto and {Ramos Almeida}, Cristina and {Acosta Pulido}, Jose A. and {Belokurov}, Vasily A. and {Isobe}, Yuki and {Jones}, Gareth and {Maraston}, Claudia and {Scholtz}, Jan and {Simmonds}, Charlotte and {Tacchella}, Sandro and {Terlevich}, Elena and {Terlevich}, Roberto},
        title = "{Lord of LRDs: Insights into a ``Little Red Dot'' with a low-ionization spectrum at z = 0.1}",
      journal = {arXiv e-prints},
         year = 2025,
        month = jul,
          eid = {arXiv:2507.23774},
        pages = {arXiv:2507.23774},
          doi = {10.48550/arXiv.2507.23774},
archivePrefix = {arXiv},
       eprint = {2507.23774},
 primaryClass = {astro-ph.GA},
       adsurl = {https://ui.adsabs.harvard.edu/abs/2025arXiv250723774J}
}

@ARTICLE{Lin_discovery_2025,
       author = {{Lin}, Xiaojing and {Fan}, Xiaohui and {Cai}, Zheng and {Bian}, Fuyan and {Liu}, Hanpu and {Sun}, Fengwu and {Ma}, Yilun and {Greene}, Jenny E. and {Strauss}, Michael A. and {Green}, Richard and {Lyu}, Jianwei and {Champagne}, Jaclyn B. and {Goulding}, Andy D. and {Inayoshi}, Kohei and {Jin}, Xiangyu and {Leung}, Gene C.~K. and {Li}, Mingyu and {Liu}, Weizhe and {Liu}, Yichen and {Mao}, Junjie and {Pudoka}, Maria Anne and {Tee}, Wei Leong and {Wang}, Ben and {Wang}, Feige and {Wu}, Yunjing and {Yang}, Jinyi and {Zhang}, Haowen and {Zhu}, Yongda},
        title = "{The Discovery of Little Red Dots in the Local Universe: Signatures of Cool Gas Envelopes}",
      journal = {arXiv e-prints},
         year = 2025,
        month = jul,
          eid = {arXiv:2507.10659},
        pages = {arXiv:2507.10659},
          doi = {10.48550/arXiv.2507.10659},
archivePrefix = {arXiv},
       eprint = {2507.10659},
 primaryClass = {astro-ph.GA},
       adsurl = {https://ui.adsabs.harvard.edu/abs/2025arXiv250710659L}
}

@ARTICLE{Ma_counting_2025,
       author = {{Ma}, Yilun and {Greene}, Jenny E. and {Setton}, David J. and {Goulding}, Andy D. and {Annunziatella}, Marianna and {Fan}, Xiaohui and {Kokorev}, Vasily and {Labbe}, Ivo and {Li}, Jiaxuan and {Lin}, Xiaojing and {Marchesini}, Danilo and {Matthee}, Jorryt and {Robbins}, Luke and {Sajina}, Anna and {Sawicki}, Marcin and {Telford}, O. Grace},
        title = "{Counting Little Red Dots at $z<4$ with Ground-based Surveys and Spectroscopic Follow-up}",
      journal = {arXiv e-prints},
         year = 2025,
        month = apr,
          eid = {arXiv:2504.08032},
        pages = {arXiv:2504.08032},
          doi = {10.48550/arXiv.2504.08032},
archivePrefix = {arXiv},
       eprint = {2504.08032},
 primaryClass = {astro-ph.GA},
       adsurl = {https://ui.adsabs.harvard.edu/abs/2025arXiv250408032M}
}

@ARTICLE{Loiacono_big_2025,
       author = {{Loiacono}, Federica and {Gilli}, Roberto and {Mignoli}, Marco and {Mazzolari}, Giovanni and {Decarli}, Roberto and {Brusa}, Marcella and {Calura}, Francesco and {Chiaberge}, Marco and {Comastri}, Andrea and {D'Amato}, Quirino and {Iwasawa}, Kazushi and {Juod{\v{z}}balis}, Ignas and {Lanzuisi}, Giorgio and {Maiolino}, Roberto and {Marchesi}, Stefano and {Norman}, Colin and {Peca}, Alessandro and {Prandoni}, Isabella and {Sapori}, Matteo and {Signorini}, Matilde and {Tozzi}, Paolo and {Vanzella}, Eros and {Vignali}, Cristian and {Vito}, Fabio and {Zamorani}, Gianni},
        title = "{A big red dot at cosmic noon}",
      journal = {\aap},
         year = 2025,
        month = oct,
       volume = {703},
          eid = {A36},
        pages = {A36},
          doi = {10.1051/0004-6361/202555946},
archivePrefix = {arXiv},
       eprint = {2506.12141},
 primaryClass = {astro-ph.GA},
       adsurl = {https://ui.adsabs.harvard.edu/abs/2025A&A...703A..36L}
}

@ARTICLE{Setton_little_2025,
       author = {{Setton}, David J. and {Greene}, Jenny E. and {de Graaff}, Anna and {Ma}, Yilun and {Leja}, Joel and {Matthee}, Jorryt and {Bezanson}, Rachel and {Boogaard}, Leindert A. and {Cleri}, Nikko J. and {Katz}, Harley and {Labbe}, Ivo and {Maseda}, Michael V. and {McConachie}, Ian and {Miller}, Tim B. and {Price}, Sedona H. and {Suess}, Katherine A. and {van Dokkum}, Pieter and {Wang}, Bingjie and {Weibel}, Andrea and {Whitaker}, Katherine E. and {Williams}, Christina C.},
        title = "{Little Red Dots at an Inflection Point: Ubiquitous V-shaped Turnover Consistently Occurs at the Balmer Limit}",
      journal = {\apj},
         year = 2025,
        month = dec,
       volume = {995},
       number = {1},
          eid = {118},
        pages = {118},
          doi = {10.3847/1538-4357/ae1500},
archivePrefix = {arXiv},
       eprint = {2411.03424},
 primaryClass = {astro-ph.GA},
       adsurl = {https://ui.adsabs.harvard.edu/abs/2025ApJ...995..118S}
}

@ARTICLE{Izotov_dust_2014,
       author = {{Izotov}, Y.~I. and {Guseva}, N.~G. and {Fricke}, K.~J. and {Kr{\"u}gel}, E. and {Henkel}, C.},
        title = "{Dust emission in star-forming dwarf galaxies: General properties and the nature of the submm excess}",
      journal = {\aap},
         year = 2014,
        month = oct,
       volume = {570},
          eid = {A97},
        pages = {A97},
          doi = {10.1051/0004-6361/201423539},
archivePrefix = {arXiv},
       eprint = {1408.4623},
 primaryClass = {astro-ph.GA},
       adsurl = {https://ui.adsabs.harvard.edu/abs/2014A&A...570A..97I}
}

@ARTICLE{Remy-Ruyer_linking_2015,
       author = {{R{\'e}my-Ruyer}, A. and {Madden}, S.~C. and {Galliano}, F. and {Lebouteiller}, V. and {Baes}, M. and {Bendo}, G.~J. and {Boselli}, A. and {Ciesla}, L. and {Cormier}, D. and {Cooray}, A. and {Cortese}, L. and {De Looze}, I. and {Doublier-Pritchard}, V. and {Galametz}, M. and {Jones}, A.~P. and {Karczewski}, O. {\L}. and {Lu}, N. and {Spinoglio}, L.},
        title = "{Linking dust emission to fundamental properties in galaxies: the low-metallicity picture}",
      journal = {\aap},
         year = 2015,
        month = oct,
       volume = {582},
          eid = {A121},
        pages = {A121},
          doi = {10.1051/0004-6361/201526067},
archivePrefix = {arXiv},
       eprint = {1507.05432},
 primaryClass = {astro-ph.GA},
       adsurl = {https://ui.adsabs.harvard.edu/abs/2015A&A...582A.121R}
}

@ARTICLE{Sturm_star-forming_2025,
       author = {{Sturm}, Megan R. and {Hayes}, Bayli and {Reines}, Amy E.},
        title = "{Star-forming Nuclear Clusters in Dwarf Galaxies Mimicking Active Galactic Nucleus Signatures in the Mid-infrared}",
      journal = {\apj},
         year = 2025,
        month = jan,
       volume = {979},
       number = {1},
          eid = {36},
        pages = {36},
          doi = {10.3847/1538-4357/ada02f},
       adsurl = {https://ui.adsabs.harvard.edu/abs/2025ApJ...979...36S}
}

@ARTICLE{Gardner_jwst_2023,
       author = {{Gardner}, Jonathan P. and {Mather}, John C. and {Abbott}, Randy and {Abell}, James S. and {Abernathy}, Mark and {Abney}, Faith E. and {Abraham}, John G. and {Abraham}, Roberto and {Abul-Huda}, Yasin M. and {Acton}, Scott and {Adams}, Cynthia K. and {Adams}, Evan and {Adler}, David S. and {Adriaensen}, Maarten and {Aguilar}, Jonathan Albert and {Ahmed}, Mansoor and {Ahmed}, Nasif S. and {Ahmed}, Tanjira and {Albat}, R{\"u}deger and {Albert}, Lo{\"\i}c and {Alberts}, Stacey and {Aldridge}, David and {Allen}, Mary Marsha and {Allen}, Shaune S. and {Altenburg}, Martin and {Altunc}, Serhat and {Alvarez}, Jose Lorenzo and {{\'A}lvarez-M{\'a}rquez}, Javier and {Alves de Oliveira}, Catarina and {Ambrose}, Leslie L. and {Anandakrishnan}, Satya M. and {Andersen}, Gregory C. and {Anderson}, Harry James and {Anderson}, Jay and {Anderson}, Kristen and {Anderson}, Sara M. and {Aprea}, Julio and {Archer}, Benita J. and {Arenberg}, Jonathan W. and {Argyriou}, Ioannis and {Arribas}, Santiago and {Artigau}, {\'E}tienne and {Arvai}, Amanda Rose and {Atcheson}, Paul and {Atkinson}, Charles B. and {Averbukh}, Jesse and {Aymergen}, Cagatay and {Bacinski}, John J. and {Baggett}, Wayne E. and {Bagnasco}, Giorgio and {Baker}, Lynn L. and {Balzano}, Vicki Ann and {Banks}, Kimberly A. and {Baran}, David A. and {Barker}, Elizabeth A. and {Barrett}, Larry K. and {Barringer}, Bruce O. and {Barto}, Allison and {Bast}, William and {Baudoz}, Pierre and {Baum}, Stefi and {Beatty}, Thomas G. and {Beaulieu}, Mathilde and {Bechtold}, Kathryn and {Beck}, Tracy and {Beddard}, Megan M. and {Beichman}, Charles and {Bellagama}, Larry and {Bely}, Pierre and {Berger}, Timothy W. and {Bergeron}, Louis E. and {Bernier}, Antoine-Darveau and {Bertch}, Maria D. and {Beskow}, Charlotte and {Betz}, Laura E. and {Biagetti}, Carl P. and {Birkmann}, Stephan and {Bjorklund}, Kurt F. and {Blackwood}, James D. and {Blazek}, Ronald Paul and {Blossfeld}, Stephen and {Bluth}, Marcel and {Boccaletti}, Anthony and {Boegner}, Jr., Martin E. and {Bohlin}, Ralph C. and {Boia}, John Joseph and {B{\"o}ker}, Torsten and {Bonaventura}, N. and {Bond}, Nicholas A. and {Bosley}, Kari Ann and {Boucarut}, Rene A. and {Bouchet}, Patrice and {Bouwman}, Jeroen and {Bower}, Gary and {Bowers}, Ariel S. and {Bowers}, Charles W. and {Boyce}, Leslye A. and {Boyer}, Christine T. and {Boyer}, Martha L. and {Boyer}, Michael and {Boyer}, Robert and {Bradley}, Larry D. and {Brady}, Gregory R. and {Brandl}, Bernhard R. and {Brannen}, Judith L. and {Breda}, David and {Bremmer}, Harold G. and {Brennan}, David and {Bresnahan}, Pamela A. and {Bright}, Stacey N. and {Broiles}, Brian J. and {Bromenschenkel}, Asa and {Brooks}, Brian H. and {Brooks}, Keira J. and {Brown}, Bob and {Brown}, Bruce and {Brown}, Thomas M. and {Bruce}, Barry W. and {Bryson}, Jonathan G. and {Bujanda}, Edwin D. and {Bullock}, Blake M. and {Bunker}, A.~J. and {Bureo}, Rafael and {Burt}, Irving J. and {Bush}, James Aaron and {Bushouse}, Howard A. and {Bussman}, Marie C. and {Cabaud}, Olivier and {Cale}, Steven and {Calhoon}, Charles D. and {Calvani}, Humberto and {Canipe}, Alicia M. and {Caputo}, Francis M. and {Cara}, Mihai and {Carey}, Larkin and {Case}, Michael Eli and {Cesari}, Thaddeus and {Cetorelli}, Lee D. and {Chance}, Don R. and {Chandler}, Lynn and {Chaney}, Dave and {Chapman}, George N. and {Charlot}, S. and {Chayer}, Pierre and {Cheezum}, Jeffrey I. and {Chen}, Bin and {Chen}, Christine H. and {Cherinka}, Brian and {Chichester}, Sarah C. and {Chilton}, Zachary S. and {Chittiraibalan}, Dharini and {Clampin}, Mark and {Clark}, Charles R. and {Clark}, Kerry W. and {Clark}, Stephanie M. and {Claybrooks}, Edward E. and {Cleveland}, Keith A. and {Cohen}, Andrew L. and {Cohen}, Lester M. and {Col{\'o}n}, Knicole D. and {Coleman}, Benee L. and {Colina}, Luis and {Comber}, Brian J. and {Comeau}, Thomas M. and {Comer}, Thomas and {Conde Reis}, Alain and {Connolly}, Dennis C. and {Conroy}, Kyle E. and {Contos}, Adam R. and {Contreras}, James and {Cook}, Neil J. and {Cooper}, James L. and {Cooper}, Rachel Aviva and {Correia}, Michael F. and {Correnti}, Matteo and {Cossou}, Christophe and {Costanza}, Brian F. and {Coulais}, Alain and {Cox}, Colin R. and {Coyle}, Ray T. and {Cracraft}, Misty M. and {Crew}, Keith A. and {Curtis}, Gary J. and {Cusveller}, Bianca and {Da Costa Maciel}, Cleyciane and {Dailey}, Christopher T. and {Daugeron}, Fr{\'e}d{\'e}ric and {Davidson}, Greg S. and {Davies}, James E. and {Davis}, Katherine Anne and {Davis}, Michael S. and {Day}, Ratna and {de Chambure}, Daniel and {de Jong}, Pauline and {De Marchi}, Guido and {Dean}, Bruce H. and {Decker}, John E. and {Delisa}, Amy S. and {Dell}, Lawrence C. and {Dellagatta}, Gail},
        title = "{The James Webb Space Telescope Mission}",
      journal = {\pasp},
         year = 2023,
        month = jun,
       volume = {135},
       number = {1048},
          eid = {068001},
        pages = {068001},
          doi = {10.1088/1538-3873/acd1b5},
archivePrefix = {arXiv},
       eprint = {2304.04869},
 primaryClass = {astro-ph.IM},
       adsurl = {https://ui.adsabs.harvard.edu/abs/2023PASP..135f8001G}
}

@ARTICLE{Maiolino_jades_2024,
       author = {{Maiolino}, Roberto and {Scholtz}, Jan and {Curtis-Lake}, Emma and {Carniani}, Stefano and {Baker}, William and {de Graaff}, Anna and {Tacchella}, Sandro and {{\"U}bler}, Hannah and {D'Eugenio}, Francesco and {Witstok}, Joris and {Curti}, Mirko and {Arribas}, Santiago and {Bunker}, Andrew J. and {Charlot}, St{\'e}phane and {Chevallard}, Jacopo and {Eisenstein}, Daniel J. and {Egami}, Eiichi and {Ji}, Zhiyuan and {Jones}, Gareth C. and {Lyu}, Jianwei and {Rawle}, Tim and {Robertson}, Brant and {Rujopakarn}, Wiphu and {Perna}, Michele and {Sun}, Fengwu and {Venturi}, Giacomo and {Williams}, Christina C. and {Willott}, Chris},
        title = "{JADES: The diverse population of infant black holes at 4 < z < 11: Merging, tiny, poor, but mighty}",
      journal = {\aap},
         year = 2024,
        month = nov,
       volume = {691},
          eid = {A145},
        pages = {A145},
          doi = {10.1051/0004-6361/202347640},
archivePrefix = {arXiv},
       eprint = {2308.01230},
 primaryClass = {astro-ph.GA},
       adsurl = {https://ui.adsabs.harvard.edu/abs/2024A&A...691A.145M}
}

@ARTICLE{Jakobsen_NIRSpec_2022,
       author = {{Jakobsen}, P. and {Ferruit}, P. and {Alves de Oliveira}, C. and {Arribas}, S. and {Bagnasco}, G. and {Barho}, R. and {Beck}, T.~L. and {Birkmann}, S. and {B{\"o}ker}, T. and {Bunker}, A.~J. and {Charlot}, S. and {de Jong}, P. and {de Marchi}, G. and {Ehrenwinkler}, R. and {Falcolini}, M. and {Fels}, R. and {Franx}, M. and {Franz}, D. and {Funke}, M. and {Giardino}, G. and {Gnata}, X. and {Holota}, W. and {Honnen}, K. and {Jensen}, P.~L. and {Jentsch}, M. and {Johnson}, T. and {Jollet}, D. and {Karl}, H. and {Kling}, G. and {K{\"o}hler}, J. and {Kolm}, M.-G. and {Kumari}, N. and {Lander}, M.~E. and {Lemke}, R. and {L{\'o}pez-Caniego}, M. and {L{\"u}tzgendorf}, N. and {Maiolino}, R. and {Manjavacas}, E. and {Marston}, A. and {Maschmann}, M. and {Maurer}, R. and {Messerschmidt}, B. and {Moseley}, S.~H. and {Mosner}, P. and {Mott}, D.~B. and {Muzerolle}, J. and {Pirzkal}, N. and {Pittet}, J.-F. and {Plitzke}, A. and {Posselt}, W. and {Rapp}, B. and {Rauscher}, B.~J. and {Rawle}, T. and {Rix}, H.-W. and {R{\"o}del}, A. and {Rumler}, P. and {Sabbi}, E. and {Salvignol}, J.-C. and {Schmid}, T. and {Sirianni}, M. and {Smith}, C. and {Strada}, P. and {te Plate}, M. and {Valenti}, J. and {Wettemann}, T. and {Wiehe}, T. and {Wiesmayer}, M. and {Willott}, C.~J. and {Wright}, R. and {Zeidler}, P. and {Zincke}, C.},
        title = "{The Near-Infrared Spectrograph (NIRSpec) on the James Webb Space Telescope. I. Overview of the instrument and its capabilities}",
      journal = {\aap},
         year = 2022,
        month = may,
       volume = {661},
          eid = {A80},
        pages = {A80},
          doi = {10.1051/0004-6361/202142663},
archivePrefix = {arXiv},
       eprint = {2202.03305},
 primaryClass = {astro-ph.IM},
       adsurl = {https://ui.adsabs.harvard.edu/abs/2022A&A...661A..80J}
}

@ARTICLE{Mezcua+24,
       author = {{Mezcua}, Mar and {Pacucci}, Fabio and {Suh}, Hyewon and {Siudek}, Malgorzata and {Natarajan}, Priyamvada},
        title = "{Overmassive Black Holes at Cosmic Noon: Linking the Local and the High-redshift Universe}",
      journal = {\apjl},
         year = 2024,
        month = may,
       volume = {966},
       number = {2},
          eid = {L30},
        pages = {L30},
          doi = {10.3847/2041-8213/ad3c2a},
archivePrefix = {arXiv},
       eprint = {2404.05793},
 primaryClass = {astro-ph.GA},
       adsurl = {https://ui.adsabs.harvard.edu/abs/2024ApJ...966L..30M}
}

@MISC{2017jwst.prop.1211F,
       author = {{Ferruit}, Pierre},
        title = "{NIRSpec WIDE MOS Survey - GOODS-N}",
 howpublished = {JWST Proposal. Cycle 1, ID. \#1211},
         year = 2017,
        month = jul,
        pages = {1211},
       adsurl = {https://ui.adsabs.harvard.edu/abs/2017jwst.prop.1211F}
}

@ARTICLE{Doyon+23,
       author = {{Doyon}, Ren{\'e} and {Willott}, Chris J. and {Hutchings}, John B. and {Sivaramakrishnan}, Anand and {Albert}, Lo{\"\i}c and {Lafreni{\`e}re}, David and {Rowlands}, Neil and {Bego{\~n}a Vila}, M. and {Martel}, Andr{\'e} R. and {LaMassa}, Stephanie and {Aldridge}, David and {Artigau}, {\'E}tienne and {Cameron}, Peter and {Chayer}, Pierre and {Cook}, Neil J. and {Cooper}, Rachel A. and {Darveau-Bernier}, Antoine and {Dupuis}, Jean and {Earnshaw}, Colin and {Espinoza}, N{\'e}stor and {Filippazzo}, Joseph C. and {Fullerton}, Alexander W. and {Gaudreau}, Daniel and {Gawlik}, Roman and {Goudfrooij}, Paul and {Haley}, Craig and {Kammerer}, Jens and {Kendall}, David and {Lambros}, Scott D. and {Ignat}, Luminita Ilinca and {Maszkiewicz}, Michael and {McColgan}, Ashley and {Morishita}, Takahiro and {Ouellette}, Nathalie N.-Q. and {Pacifici}, Camilla and {Philippi}, Natasha and {Radica}, Michael and {Ravindranath}, Swara and {Rowe}, Jason and {Roy}, Arpita and {Roy}, Niladri and {Saad}, Karl and {Sohn}, Sangmo Tony and {Talens}, Geert Jan and {Touahri}, Driss and {Thatte}, Deepashri and {Taylor}, Joanna M. and {Vandal}, Thomas and {Volk}, Kevin and {Wander}, Michel and {Warner}, Gerald and {Zheng}, Sheng-Hai and {Zhou}, Julia and {Abraham}, Roberto and {Beaulieu}, Mathilde and {Benneke}, Bj{\"o}rn and {Ferrarese}, Laura and {Jayawardhana}, Ray and {Johnstone}, Doug and {Kaltenegger}, Lisa and {Meyer}, Michael R. and {Pipher}, Judy L. and {Rameau}, Julien and {Rieke}, Marcia and {Salhi}, Salma and {Sawicki}, Marcin},
        title = "{The Near Infrared Imager and Slitless Spectrograph for the James Webb Space Telescope. I. Instrument Overview and In-flight Performance}",
      journal = {\pasp},
         year = 2023,
        month = sep,
       volume = {135},
       number = {1051},
          eid = {098001},
        pages = {098001},
          doi = {10.1088/1538-3873/acd41b},
archivePrefix = {arXiv},
       eprint = {2306.03277},
 primaryClass = {astro-ph.IM},
       adsurl = {https://ui.adsabs.harvard.edu/abs/2023PASP..135i8001D}
}

@ARTICLE{Giavalisco+04,
       author = {{Giavalisco}, M. and {Ferguson}, H.~C. and {Koekemoer}, A.~M. and {Dickinson}, M. and {Alexander}, D.~M. and {Bauer}, F.~E. and {Bergeron}, J. and {Biagetti}, C. and {Brandt}, W.~N. and {Casertano}, S. and {Cesarsky}, C. and {Chatzichristou}, E. and {Conselice}, C. and {Cristiani}, S. and {Da Costa}, L. and {Dahlen}, T. and {de Mello}, D. and {Eisenhardt}, P. and {Erben}, T. and {Fall}, S.~M. and {Fassnacht}, C. and {Fosbury}, R. and {Fruchter}, A. and {Gardner}, J.~P. and {Grogin}, N. and {Hook}, R.~N. and {Hornschemeier}, A.~E. and {Idzi}, R. and {Jogee}, S. and {Kretchmer}, C. and {Laidler}, V. and {Lee}, K.~S. and {Livio}, M. and {Lucas}, R. and {Madau}, P. and {Mobasher}, B. and {Moustakas}, L.~A. and {Nonino}, M. and {Padovani}, P. and {Papovich}, C. and {Park}, Y. and {Ravindranath}, S. and {Renzini}, A. and {Richardson}, M. and {Riess}, A. and {Rosati}, P. and {Schirmer}, M. and {Schreier}, E. and {Somerville}, R.~S. and {Spinrad}, H. and {Stern}, D. and {Stiavelli}, M. and {Strolger}, L. and {Urry}, C.~M. and {Vandame}, B. and {Williams}, R. and {Wolf}, C.},
        title = "{The Great Observatories Origins Deep Survey: Initial Results from Optical and Near-Infrared Imaging}",
      journal = {\apjl},
         year = 2004,
        month = jan,
       volume = {600},
       number = {2},
        pages = {L93-L98},
          doi = {10.1086/379232},
archivePrefix = {arXiv},
       eprint = {astro-ph/0309105},
 primaryClass = {astro-ph},
       adsurl = {https://ui.adsabs.harvard.edu/abs/2004ApJ...600L..93G}
}

@ARTICLE{Ebeling+01,
       author = {{Ebeling}, H. and {Edge}, A.~C. and {Henry}, J.~P.},
        title = "{MACS: A Quest for the Most Massive Galaxy Clusters in the Universe}",
      journal = {\apj},
         year = 2001,
        month = jun,
       volume = {553},
       number = {2},
        pages = {668-676},
          doi = {10.1086/320958},
archivePrefix = {arXiv},
       eprint = {astro-ph/0009101},
 primaryClass = {astro-ph},
       adsurl = {https://ui.adsabs.harvard.edu/abs/2001ApJ...553..668E}
}

@ARTICLE{Perez-Gonzalez+26,
       author = {{P{\'e}rez-Gonz{\'a}lez}, Pablo G. and {Barro}, Guillermo and {Carniani}, Stefano and {D'Eugenio}, Francesco and {Rieke}, George H. and {Tripodi}, Roberta and {Bunker}, Andrew J. and {Ji}, Xihan and {Marques-Chaves}, Rui and {Schaerer}, Daniel and {Venturi}, Giacomo and {Ar{\'e}valo-Gonz{\'a}lez}, Flor and {Arribas}, Santiago and {Rinaldi}, Pierluigi and {Rodr{\'\i}guez Del Pino}, Bruno and {Witstok}, Joris and {Bhatawdekar}, Rachana and {Boogaard}, Leindert A. and {Charlot}, Stephane and {Chevallard}, Jacopo and {Costantin}, Luca and {Curti}, Mirko and {Curtis-Lake}, Emma and {Daddi}, Emanuele and {Davis}, Kelcey and {Dickinson}, Mark and {Donnan}, Callum T. and {Donnan}, Fergus R. and {Dunlop}, James S. and {Eisenstein}, Daniel J. and {Ferguson}, Henry C. and {Fern{\'a}ndez Aranda}, Rom{\'a}n and {Finkelstein}, Steven L. and {Fujimoto}, Seiji and {Gandolfi}, Giovanni and {Giavalisco}, Mauro and {Grogin}, Norman A. and {Hamed}, Mahmoud and {Hirschmann}, Michaela and {Kartaltepe}, Jeyhan S. and {Kocevski}, Dale D. and {Koekemoer}, Anton M. and {Leung}, Gene C.~K. and {Lofaro}, Cristina M. and {Lucas}, Ray A. and {McLeod}, Derek J. and {Melinder}, Jens and {{\"O}stlin}, Goran and {Papovich}, Casey and {Pentericci}, Laura and {P{\'e}rez-D{\'\i}az}, Borja and {Rieke}, Marcia and {Scholtz}, Jan and {Somerville}, Rachel S. and {Stanton}, Thomas M. and {Stevenson}, Struan D. and {Shivaei}, Irene and {Tacchella}, Sandro and {Trump}, Jonathan R. and {{\"U}bler}, Hannah and {Wang}, Xin and {Williams}, Christina C. and {Willmer}, Christopher N.~A. and {Yung}, L.~Y. Aaron and {Zhu}, Yongda},
        title = "{Little Red Dots: One Photometric Tag Concealing Diverse Spectroscopic Flavors of Massive Star Formation and Black Hole Activity}",
      journal = {arXiv e-prints},
         year = 2026,
        month = feb,
          eid = {arXiv:2602.20247},
        pages = {arXiv:2602.20247},
          doi = {10.48550/arXiv.2602.20247},
archivePrefix = {arXiv},
       eprint = {2602.20247},
 primaryClass = {astro-ph.GA},
       adsurl = {https://ui.adsabs.harvard.edu/abs/2026arXiv260220247P}
}

@ARTICLE{Pacucci+26,
       author = {{Pacucci}, Fabio and {Ferrara}, Andrea and {Kocevski}, Dale D.},
        title = "{The Little Red Dots Are Direct Collapse Black Holes}",
      journal = {arXiv e-prints},
         year = 2026,
        month = jan,
          eid = {arXiv:2601.14368},
        pages = {arXiv:2601.14368},
          doi = {10.48550/arXiv.2601.14368},
archivePrefix = {arXiv},
       eprint = {2601.14368},
 primaryClass = {astro-ph.GA},
       adsurl = {https://ui.adsabs.harvard.edu/abs/2026arXiv260114368P}
}

@ARTICLE{Pacucci+23,
       author = {{Pacucci}, Fabio and {Nguyen}, Bao and {Carniani}, Stefano and {Maiolino}, Roberto and {Fan}, Xiaohui},
        title = "{JWST CEERS and JADES Active Galaxies at z = 4-7 Violate the Local M $_{{\textbullet}}$-M $_{{\ensuremath{\star}}}$ Relation at >3{\ensuremath{\sigma}}: Implications for Low-mass Black Holes and Seeding Models}",
      journal = {\apjl},
         year = 2023,
        month = nov,
       volume = {957},
       number = {1},
          eid = {L3},
        pages = {L3},
          doi = {10.3847/2041-8213/ad0158},
archivePrefix = {arXiv},
       eprint = {2308.12331},
 primaryClass = {astro-ph.GA},
       adsurl = {https://ui.adsabs.harvard.edu/abs/2023ApJ...957L...3P}
}

@ARTICLE{Kokorev+23,
       author = {{Kokorev}, Vasily and {Fujimoto}, Seiji and {Labbe}, Ivo and {Greene}, Jenny E. and {Bezanson}, Rachel and {Dayal}, Pratika and {Nelson}, Erica J. and {Atek}, Hakim and {Brammer}, Gabriel and {Caputi}, Karina I. and {Chemerynska}, Iryna and {Cutler}, Sam E. and {Feldmann}, Robert and {Fudamoto}, Yoshinobu and {Furtak}, Lukas J. and {Goulding}, Andy D. and {de Graaff}, Anna and {Leja}, Joel and {Marchesini}, Danilo and {Miller}, Tim B. and {Nanayakkara}, Themiya and {Oesch}, Pascal A. and {Pan}, Richard and {Price}, Sedona H. and {Setton}, David J. and {Smit}, Renske and {Stefanon}, Mauro and {Wang}, Bingjie and {Weaver}, John R. and {Whitaker}, Katherine E. and {Williams}, Christina C. and {Zitrin}, Adi},
        title = "{UNCOVER: A NIRSpec Identification of a Broad-line AGN at z = 8.50}",
      journal = {\apjl},
         year = 2023,
        month = nov,
       volume = {957},
       number = {1},
          eid = {L7},
        pages = {L7},
          doi = {10.3847/2041-8213/ad037a},
archivePrefix = {arXiv},
       eprint = {2308.11610},
 primaryClass = {astro-ph.GA},
       adsurl = {https://ui.adsabs.harvard.edu/abs/2023ApJ...957L...7K}
}

@ARTICLE{Maiolino+24,
       author = {{Maiolino}, Roberto and {Scholtz}, Jan and {Witstok}, Joris and {Carniani}, Stefano and {D'Eugenio}, Francesco and {de Graaff}, Anna and {{\"U}bler}, Hannah and {Tacchella}, Sandro and {Curtis-Lake}, Emma and {Arribas}, Santiago and {Bunker}, Andrew and {Charlot}, St{\'e}phane and {Chevallard}, Jacopo and {Curti}, Mirko and {Looser}, Tobias J. and {Maseda}, Michael V. and {Rawle}, Timothy D. and {Rodr{\'\i}guez del Pino}, Bruno and {Willott}, Chris J. and {Egami}, Eiichi and {Eisenstein}, Daniel J. and {Hainline}, Kevin N. and {Robertson}, Brant and {Williams}, Christina C. and {Willmer}, Christopher N.~A. and {Baker}, William M. and {Boyett}, Kristan and {DeCoursey}, Christa and {Fabian}, Andrew C. and {Helton}, Jakob M. and {Ji}, Zhiyuan and {Jones}, Gareth C. and {Kumari}, Nimisha and {Laporte}, Nicolas and {Nelson}, Erica J. and {Perna}, Michele and {Sandles}, Lester and {Shivaei}, Irene and {Sun}, Fengwu},
        title = "{A small and vigorous black hole in the early Universe}",
      journal = {\nat},
         year = 2024,
        month = mar,
       volume = {627},
       number = {8002},
        pages = {59-63},
          doi = {10.1038/s41586-024-07052-5},
archivePrefix = {arXiv},
       eprint = {2305.12492},
 primaryClass = {astro-ph.GA},
       adsurl = {https://ui.adsabs.harvard.edu/abs/2024Natur.627...59M}
}

@ARTICLE{Juodzbalis+25,
       author = {{Juod{\v{z}}balis}, Ignas and {Marconcini}, Cosimo and {D'Eugenio}, Francesco and {Maiolino}, Roberto and {Marconi}, Alessandro and {{\"U}bler}, Hannah and {Scholtz}, Jan and {Ji}, Xihan and {Arribas}, Santiago and {Bennett}, Jake S. and {Bromm}, Volker and {Bunker}, Andrew J. and {Carniani}, Stefano and {Charlot}, St{\'e}phane and {Cresci}, Giovanni and {Dayal}, Pratika and {Egami}, Eiichi and {Fabian}, Andrew and {Inayoshi}, Kohei and {Isobe}, Yuki and {Ivey}, Lucy and {Jones}, Gareth C. and {Koudmani}, Sophie and {Laporte}, Nicolas and {Liu}, Boyuan and {Lyu}, Jianwei and {Mazzolari}, Giovanni and {Monty}, Stephanie and {Parlanti}, Eleonora and {P{\'e}rez-Gonz{\'a}lez}, Pablo G. and {Perna}, Michele and {Robertson}, Brant and {Schneider}, Raffaella and {Sijacki}, Debora and {Tacchella}, Sandro and {Trinca}, Alessandro and {Valiante}, Rosa and {Volonteri}, Marta and {Witstok}, Joris and {Zhang}, Saiyang},
        title = "{A direct black hole mass measurement in a Little Red Dot at the Epoch of Reionization}",
      journal = {arXiv e-prints},
         year = 2025,
        month = aug,
          eid = {arXiv:2508.21748},
        pages = {arXiv:2508.21748},
          doi = {10.48550/arXiv.2508.21748},
archivePrefix = {arXiv},
       eprint = {2508.21748},
 primaryClass = {astro-ph.GA},
       adsurl = {https://ui.adsabs.harvard.edu/abs/2025arXiv250821748J}
}

@ARTICLE{Rinaldi+25b,
       author = {{Rinaldi}, Pierluigi and {Rieke}, George H. and {Wu}, Zihao and {Gilbert}, Carys J.~E. and {Pacucci}, Fabio and {Barchiesi}, Luigi and {Alberts}, Stacey and {Carniani}, Stefano and {Bunker}, Andrew J. and {Bhatawdekar}, Rachana and {D'Eugenio}, Francesco and {Ji}, Zhiyuan and {Johnson}, Benjamin D. and {Hainline}, Kevin and {Kokorev}, Vasily and {Kumari}, Nimisha and {Iani}, Edoardo and {Lyu}, Jianwei and {Maiolino}, Roberto and {Parlanti}, Eleonora and {Robertson}, Brant E. and {Sun}, Yang and {Vignali}, Cristian and {Williams}, Christina C. and {Willmer}, Christopher N.~A. and {Zhu}, Yongda},
        title = "{Beyond the Dot: an LRD-like nucleus at the Heart of an IR-Bright Galaxy and its implications for high-redshift LRDs}",
      journal = {arXiv e-prints},
         year = 2025,
        month = jul,
          eid = {arXiv:2507.17738},
        pages = {arXiv:2507.17738},
          doi = {10.48550/arXiv.2507.17738},
archivePrefix = {arXiv},
       eprint = {2507.17738},
 primaryClass = {astro-ph.GA},
       adsurl = {https://ui.adsabs.harvard.edu/abs/2025arXiv250717738R}
}

@ARTICLE{Chen+25,
       author = {{Chen}, Xiaoyang and {Ichikawa}, Kohei and {Akiyama}, Masayuki and {Inayoshi}, Kohei and {Inoue}, Akio K. and {Onoue}, Masafusa and {Toba}, Yoshiki and {Zavala}, Jorge A. and {Bakx}, Tom J. and {Kawaguchi}, Toshihiro and {Lee}, Kianhong and {Matsumoto}, Naoki and {Vijarnwannaluk}, Bovornpratch},
        title = "{A z \raisebox{-0.5ex}\textasciitilde 0.4 Galaxy Reflecting the high-redshift Little Red Dots: An Extended Starburst with an Overmassive Black Hole}",
      journal = {arXiv e-prints},
         year = 2025,
        month = oct,
          eid = {arXiv:2510.02801},
        pages = {arXiv:2510.02801},
          doi = {10.48550/arXiv.2510.02801},
archivePrefix = {arXiv},
       eprint = {2510.02801},
 primaryClass = {astro-ph.GA},
       adsurl = {https://ui.adsabs.harvard.edu/abs/2025arXiv251002801C}
}

@ARTICLE{Volonteri_formation_2010,
       author = {{Volonteri}, Marta},
        title = "{Formation of supermassive black holes}",
      journal = {\aapr},
         year = 2010,
        month = jul,
       volume = {18},
       number = {3},
        pages = {279-315},
          doi = {10.1007/s00159-010-0029-x},
archivePrefix = {arXiv},
       eprint = {1003.4404},
 primaryClass = {astro-ph.CO},
       adsurl = {https://ui.adsabs.harvard.edu/abs/2010A&ARv..18..279V}
}

@ARTICLE{Greene_low-mass_2012,
       author = {{Greene}, Jenny E.},
        title = "{Low-mass black holes as the remnants of primordial black hole formation}",
      journal = {Nature Communications},
         year = 2012,
        month = dec,
       volume = {3},
          eid = {1304},
        pages = {1304},
          doi = {10.1038/ncomms2314},
archivePrefix = {arXiv},
       eprint = {1211.7082},
 primaryClass = {astro-ph.CO},
       adsurl = {https://ui.adsabs.harvard.edu/abs/2012NatCo...3.1304G}
}

@ARTICLE{Volonteri_formation_2012,
       author = {{Volonteri}, M.},
        title = "{The Formation and Evolution of Massive Black Holes}",
      journal = {Science},
         year = 2012,
        month = aug,
       volume = {337},
       number = {6094},
        pages = {544},
          doi = {10.1126/science.1220843},
archivePrefix = {arXiv},
       eprint = {1208.1106},
 primaryClass = {astro-ph.CO},
       adsurl = {https://ui.adsabs.harvard.edu/abs/2012Sci...337..544V}
}

@ARTICLE{Inayoshi_assembly_2020,
       author = {{Inayoshi}, Kohei and {Visbal}, Eli and {Haiman}, Zolt{\'a}n},
        title = "{The Assembly of the First Massive Black Holes}",
      journal = {\araa},
         year = 2020,
        month = aug,
       volume = {58},
        pages = {27-97},
          doi = {10.1146/annurev-astro-120419-014455},
archivePrefix = {arXiv},
       eprint = {1911.05791},
 primaryClass = {astro-ph.GA},
       adsurl = {https://ui.adsabs.harvard.edu/abs/2020ARA&A..58...27I}
}

@ARTICLE{Volonteri_rapid_2005,
       author = {{Volonteri}, Marta and {Rees}, Martin J.},
        title = "{Rapid Growth of High-Redshift Black Holes}",
      journal = {\apj},
         year = 2005,
        month = nov,
       volume = {633},
       number = {2},
        pages = {624-629},
          doi = {10.1086/466521},
archivePrefix = {arXiv},
       eprint = {astro-ph/0506040},
 primaryClass = {astro-ph},
       adsurl = {https://ui.adsabs.harvard.edu/abs/2005ApJ...633..624V}
}

@ARTICLE{Alexander_what_2025,
       author = {{Alexander}, D.~M. and {Hickox}, R.~C. and {Aird}, J. and {Combes}, F. and {Costa}, T. and {Habouzit}, M. and {Harrison}, C.~M. and {Leng}, R.~I. and {Morabito}, L.~K. and {Uckelman}, S.~L. and {Vickers}, P.},
        title = "{What drives the growth of black holes: A decade of progress}",
      journal = {\nar},
         year = 2025,
        month = dec,
       volume = {101},
          eid = {101733},
        pages = {101733},
          doi = {10.1016/j.newar.2025.101733},
archivePrefix = {arXiv},
       eprint = {2506.19166},
 primaryClass = {astro-ph.GA},
       adsurl = {https://ui.adsabs.harvard.edu/abs/2025NewAR.10101733A}
}

@ARTICLE{Reines_hunting_2022,
       author = {{Reines}, Amy E.},
        title = "{Hunting for massive black holes in dwarf galaxies}",
      journal = {Nature Astronomy},
         year = 2022,
        month = jan,
       volume = {6},
        pages = {26-34},
          doi = {10.1038/s41550-021-01556-0},
archivePrefix = {arXiv},
       eprint = {2201.10569},
 primaryClass = {astro-ph.GA},
       adsurl = {https://ui.adsabs.harvard.edu/abs/2022NatAs...6...26R}
}

@ARTICLE{Reines_observational_2016,
       author = {{Reines}, Amy E. and {Comastri}, Andrea},
        title = "{Observational Signatures of High-Redshift Quasars and Local Relics of Black Hole Seeds}",
      journal = {\pasa},
         year = 2016,
        month = oct,
       volume = {33},
          eid = {e054},
        pages = {e054},
          doi = {10.1017/pasa.2016.46},
archivePrefix = {arXiv},
       eprint = {1609.03562},
 primaryClass = {astro-ph.GA},
       adsurl = {https://ui.adsabs.harvard.edu/abs/2016PASA...33...54R}
}

@ARTICLE{Habouzit_blossoms_2017,
       author = {{Habouzit}, M{\'e}lanie and {Volonteri}, Marta and {Dubois}, Yohan},
        title = "{Blossoms from black hole seeds: properties and early growth regulated by supernova feedback}",
      journal = {\mnras},
         year = 2017,
        month = jul,
       volume = {468},
       number = {4},
        pages = {3935-3948},
          doi = {10.1093/mnras/stx666},
archivePrefix = {arXiv},
       eprint = {1605.09394},
 primaryClass = {astro-ph.GA},
       adsurl = {https://ui.adsabs.harvard.edu/abs/2017MNRAS.468.3935H}
}

@ARTICLE{Evans+24,
       author = {{Evans}, Ian N. and {Evans}, Janet D. and {Mart{\'\i}nez-Galarza}, J. Rafael and {Miller}, Joseph B. and {Primini}, Francis A. and {Azadi}, Mojegan and {Burke}, Douglas J. and {Civano}, Francesca M. and {D'Abrusco}, Raffaele and {Fabbiano}, Giuseppina and {Graessle}, Dale E. and {Grier}, John D. and {Houck}, John C. and {Lauer}, Jennifer and {McCollough}, Michael L. and {Nowak}, Michael A. and {Plummer}, David A. and {Rots}, Arnold H. and {Siemiginowska}, Aneta and {Tibbetts}, Michael S.},
        title = "{The Chandra Source Catalog Release 2 Series}",
      journal = {\apjs},
         year = 2024,
        month = oct,
       volume = {274},
       number = {2},
          eid = {22},
        pages = {22},
          doi = {10.3847/1538-4365/ad6319},
archivePrefix = {arXiv},
       eprint = {2407.10799},
 primaryClass = {astro-ph.HE},
       adsurl = {https://ui.adsabs.harvard.edu/abs/2024ApJS..274...22E}
}

@ARTICLE{Luo+14,
       author = {{Luo}, B. and {Brandt}, W.~N. and {Alexander}, D.~M. and {Stern}, D. and {Teng}, S.~H. and {Ar{\'e}valo}, P. and {Bauer}, F.~E. and {Boggs}, S.~E. and {Christensen}, F.~E. and {Comastri}, A. and {Craig}, W.~W. and {Farrah}, D. and {Gandhi}, P. and {Hailey}, C.~J. and {Harrison}, F.~A. and {Koss}, M. and {Ogle}, P. and {Puccetti}, S. and {Saez}, C. and {Scott}, A.~E. and {Walton}, D.~J. and {Zhang}, W.~W.},
        title = "{Weak Hard X-Ray Emission from Broad Absorption Line Quasars: Evidence for Intrinsic X-Ray Weakness}",
      journal = {\apj},
         year = 2014,
        month = oct,
       volume = {794},
       number = {1},
          eid = {70},
        pages = {70},
          doi = {10.1088/0004-637X/794/1/70},
archivePrefix = {arXiv},
       eprint = {1408.3633},
 primaryClass = {astro-ph.HE},
       adsurl = {https://ui.adsabs.harvard.edu/abs/2014ApJ...794...70L}
}

@ARTICLE{Lyu+17,
       author = {{Lyu}, Jianwei and {Rieke}, G.~H. and {Shi}, Yong},
        title = "{Dust-deficient Palomar-Green Quasars and the Diversity of AGN Intrinsic IR Emission}",
      journal = {\apj},
         year = 2017,
        month = feb,
       volume = {835},
       number = {2},
          eid = {257},
        pages = {257},
          doi = {10.3847/1538-4357/835/2/257},
archivePrefix = {arXiv},
       eprint = {1612.06857},
 primaryClass = {astro-ph.GA},
       adsurl = {https://ui.adsabs.harvard.edu/abs/2017ApJ...835..257L}
}

@ARTICLE{Lyu+18,
       author = {{Lyu}, Jianwei and {Rieke}, George H.},
        title = "{Polar Dust, Nuclear Obscuration, and IR SED Diversity in Type-1 AGNs}",
      journal = {\apj},
         year = 2018,
        month = oct,
       volume = {866},
       number = {2},
          eid = {92},
        pages = {92},
          doi = {10.3847/1538-4357/aae075},
archivePrefix = {arXiv},
       eprint = {1809.03080},
 primaryClass = {astro-ph.GA},
       adsurl = {https://ui.adsabs.harvard.edu/abs/2018ApJ...866...92L}
}

@ARTICLE{Brandt+15,
       author = {{Brandt}, W.~N. and {Alexander}, D.~M.},
        title = "{Cosmic X-ray surveys of distant active galaxies. The demographics, physics, and ecology of growing supermassive black holes}",
      journal = {\aapr},
         year = 2015,
        month = jan,
       volume = {23},
          eid = {1},
        pages = {1},
          doi = {10.1007/s00159-014-0081-z},
archivePrefix = {arXiv},
       eprint = {1501.01982},
 primaryClass = {astro-ph.HE},
       adsurl = {https://ui.adsabs.harvard.edu/abs/2015A&ARv..23....1B}
}

@ARTICLE{Laurenti+22,
       author = {{Laurenti}, M. and {Piconcelli}, E. and {Zappacosta}, L. and {Tombesi}, F. and {Vignali}, C. and {Bianchi}, S. and {Marziani}, P. and {Vagnetti}, F. and {Bongiorno}, A. and {Bischetti}, M. and {del Olmo}, A. and {Lanzuisi}, G. and {Luminari}, A. and {Middei}, R. and {Perri}, M. and {Ricci}, C. and {Vietri}, G.},
        title = "{X-ray spectroscopic survey of highly accreting AGN}",
      journal = {\aap},
         year = 2022,
        month = jan,
       volume = {657},
          eid = {A57},
        pages = {A57},
          doi = {10.1051/0004-6361/202141829},
archivePrefix = {arXiv},
       eprint = {2110.06939},
 primaryClass = {astro-ph.GA},
       adsurl = {https://ui.adsabs.harvard.edu/abs/2022A&A...657A..57L}
}

@ARTICLE{Pacucci+24,
       author = {{Pacucci}, Fabio and {Narayan}, Ramesh},
        title = "{Mildly Super-Eddington Accretion onto Slowly Spinning Black Holes Explains the X-Ray Weakness of the Little Red Dots}",
      journal = {\apj},
         year = 2024,
        month = nov,
       volume = {976},
       number = {1},
          eid = {96},
        pages = {96},
          doi = {10.3847/1538-4357/ad84f7},
archivePrefix = {arXiv},
       eprint = {2407.15915},
 primaryClass = {astro-ph.HE},
       adsurl = {https://ui.adsabs.harvard.edu/abs/2024ApJ...976...96P}
}

@ARTICLE{Ishibashi+25,
       author = {{Ishibashi}, W. and {Fabian}, A.~C. and {Maiolino}, R. and {Gursahani}, Y. and {Reynolds}, C.~S.},
        title = "{Another view into JWST-discovered X-ray weak AGNs via radiative dusty feedback}",
      journal = {\mnras},
         year = 2025,
        month = nov,
       volume = {544},
       number = {1},
        pages = {726-734},
          doi = {10.1093/mnras/staf1439},
archivePrefix = {arXiv},
       eprint = {2509.05423},
 primaryClass = {astro-ph.GA},
       adsurl = {https://ui.adsabs.harvard.edu/abs/2025MNRAS.544..726I}
}

@ARTICLE{Teng+14,
       author = {{Teng}, Stacy H. and {Brandt}, W.~N. and {Harrison}, F.~A. and {Luo}, B. and {Alexander}, D.~M. and {Bauer}, F.~E. and {Boggs}, S.~E. and {Christensen}, F.~E. and {Comastri}, A. and {Craig}, W.~W. and {Fabian}, A.~C. and {Farrah}, D. and {Fiore}, F. and {Gandhi}, P. and {Grefenstette}, B.~W. and {Hailey}, C.~J. and {Hickox}, R.~C. and {Madsen}, K.~K. and {Ptak}, A.~F. and {Rigby}, J.~R. and {Risaliti}, G. and {Saez}, C. and {Stern}, D. and {Veilleux}, S. and {Walton}, D.~J. and {Wik}, D.~R. and {Zhang}, W.~W.},
        title = "{NuSTAR Reveals an Intrinsically X-Ray Weak Broad Absorption Line Quasar in the Ultraluminous Infrared Galaxy Markarian 231}",
      journal = {\apj},
         year = 2014,
        month = apr,
       volume = {785},
       number = {1},
          eid = {19},
        pages = {19},
          doi = {10.1088/0004-637X/785/1/19},
archivePrefix = {arXiv},
       eprint = {1402.4811},
 primaryClass = {astro-ph.HE},
       adsurl = {https://ui.adsabs.harvard.edu/abs/2014ApJ...785...19T}
}

@ARTICLE{Guainazzi+09,
       author = {{Guainazzi}, M. and {Risaliti}, G. and {Nucita}, A. and {Wang}, J. and {Bianchi}, S. and {Soria}, R. and {Zezas}, A.},
        title = "{AGN/starburst connection in action: the half million second RGS spectrum of NGC 1365}",
      journal = {\aap},
         year = 2009,
        month = oct,
       volume = {505},
       number = {2},
        pages = {589-600},
          doi = {10.1051/0004-6361/200912758},
archivePrefix = {arXiv},
       eprint = {0908.0268},
 primaryClass = {astro-ph.CO},
       adsurl = {https://ui.adsabs.harvard.edu/abs/2009A&A...505..589G}
}

@INPROCEEDINGS{Comastri04,
       author = {{Comastri}, A.},
        title = "{Compton-Thick AGN: The Dark Side of the X-Ray Background}",
    booktitle = {Supermassive Black Holes in the Distant Universe},
         year = 2004,
       editor = {{Barger}, A.~J.},
       series = {Astrophysics and Space Science Library},
       volume = {308},
        month = aug,
        pages = {245},
          doi = {10.1007/978-1-4020-2471-9_8},
archivePrefix = {arXiv},
       eprint = {astro-ph/0403693},
 primaryClass = {astro-ph},
       adsurl = {https://ui.adsabs.harvard.edu/abs/2004ASSL..308..245C}
}

@ARTICLE{Maiolino+98,
       author = {{Maiolino}, R. and {Salvati}, M. and {Bassani}, L. and {Dadina}, M. and {della Ceca}, R. and {Matt}, G. and {Risaliti}, G. and {Zamorani}, G.},
        title = "{Heavy obscuration in X-ray weak AGNs}",
      journal = {\aap},
         year = 1998,
        month = oct,
       volume = {338},
        pages = {781-794},
          doi = {10.48550/arXiv.astro-ph/9806055},
archivePrefix = {arXiv},
       eprint = {astro-ph/9806055},
 primaryClass = {astro-ph},
       adsurl = {https://ui.adsabs.harvard.edu/abs/1998A&A...338..781M}
}

@ARTICLE{Noboriguchi+19,
       author = {{Noboriguchi}, Akatoki and {Nagao}, Tohru and {Toba}, Yoshiki and {Niida}, Mana and {Kajisawa}, Masaru and {Onoue}, Masafusa and {Matsuoka}, Yoshiki and {Yamashita}, Takuji and {Chang}, Yu-Yen and {Kawaguchi}, Toshihiro and {Komiyama}, Yutaka and {Nobuhara}, Kodai and {Terashima}, Yuichi and {Ueda}, Yoshihiro},
        title = "{Optical Properties of Infrared-bright Dust-obscured Galaxies Viewed with Subaru Hyper Suprime-Cam}",
      journal = {\apj},
         year = 2019,
        month = may,
       volume = {876},
       number = {2},
          eid = {132},
        pages = {132},
          doi = {10.3847/1538-4357/ab1754},
archivePrefix = {arXiv},
       eprint = {1803.09951},
 primaryClass = {astro-ph.GA},
       adsurl = {https://ui.adsabs.harvard.edu/abs/2019ApJ...876..132N}
}

@ARTICLE{Assef+16,
       author = {{Assef}, R.~J. and {Walton}, D.~J. and {Brightman}, M. and {Stern}, D. and {Alexander}, D. and {Bauer}, F. and {Blain}, A.~W. and {Diaz-Santos}, T. and {Eisenhardt}, P.~R.~M. and {Finkelstein}, S.~L. and {Hickox}, R.~C. and {Tsai}, C.-W. and {Wu}, J.~W.},
        title = "{Hot Dust Obscured Galaxies with Excess Blue Light: Dual AGN or Single AGN Under Extreme Conditions?}",
      journal = {\apj},
         year = 2016,
        month = mar,
       volume = {819},
       number = {2},
          eid = {111},
        pages = {111},
          doi = {10.3847/0004-637X/819/2/111},
archivePrefix = {arXiv},
       eprint = {1511.05155},
 primaryClass = {astro-ph.GA},
       adsurl = {https://ui.adsabs.harvard.edu/abs/2016ApJ...819..111A}
}

@ARTICLE{Riguccini+15,
       author = {{Riguccini}, L. and {Le Floc'h}, E. and {Mullaney}, J.~R. and {Men{\'e}ndez-Delmestre}, K. and {Aussel}, H. and {Berta}, S. and {Calanog}, J. and {Capak}, P. and {Cooray}, A. and {Ilbert}, O. and {Kartaltepe}, J. and {Koekemoer}, A. and {Lutz}, D. and {Magnelli}, B. and {McCracken}, H. and {Oliver}, S. and {Roseboom}, I. and {Salvato}, M. and {Sanders}, D. and {Scoville}, N. and {Taniguchi}, Y. and {Treister}, E.},
        title = "{The composite nature of Dust-Obscured Galaxies (DOGs) at z {\ensuremath{\sim}} 2-3 in the COSMOS field - I. A far-infrared view}",
      journal = {\mnras},
         year = 2015,
        month = sep,
       volume = {452},
       number = {1},
        pages = {470-485},
          doi = {10.1093/mnras/stv1297},
archivePrefix = {arXiv},
       eprint = {1506.05475},
 primaryClass = {astro-ph.GA},
       adsurl = {https://ui.adsabs.harvard.edu/abs/2015MNRAS.452..470R}
}

@ARTICLE{Melbourne+12,
       author = {{Melbourne}, J. and {Soifer}, B.~T. and {Desai}, Vandana and {Pope}, Alexandra and {Armus}, Lee and {Dey}, Arjun and {Bussmann}, R.~S. and {Jannuzi}, B.~T. and {Alberts}, Stacey},
        title = "{The Spectral Energy Distributions and Infrared Luminosities of z {\ensuremath{\approx}} 2 Dust-obscured Galaxies from Herschel and Spitzer}",
      journal = {\aj},
         year = 2012,
        month = may,
       volume = {143},
       number = {5},
          eid = {125},
        pages = {125},
          doi = {10.1088/0004-6256/143/5/125},
archivePrefix = {arXiv},
       eprint = {1203.3199},
 primaryClass = {astro-ph.CO},
       adsurl = {https://ui.adsabs.harvard.edu/abs/2012AJ....143..125M}
}

@ARTICLE{Assef+20,
       author = {{Assef}, R.~J. and {Brightman}, M. and {Walton}, D.~J. and {Stern}, D. and {Bauer}, F.~E. and {Blain}, A.~W. and {D{\'\i}az-Santos}, T. and {Eisenhardt}, P.~R.~M. and {Hickox}, R.~C. and {Jun}, H.~D. and {Psychogyios}, A. and {Tsai}, C.-W. and {Wu}, J.~W.},
        title = "{Hot Dust-obscured Galaxies with Excess Blue Light}",
      journal = {\apj},
         year = 2020,
        month = jul,
       volume = {897},
       number = {2},
          eid = {112},
        pages = {112},
          doi = {10.3847/1538-4357/ab9814},
archivePrefix = {arXiv},
       eprint = {1905.04320},
 primaryClass = {astro-ph.GA},
       adsurl = {https://ui.adsabs.harvard.edu/abs/2020ApJ...897..112A}
}

@ARTICLE{Reines+15,
       author = {{Reines}, Amy E. and {Volonteri}, Marta},
        title = "{Relations between Central Black Hole Mass and Total Galaxy Stellar Mass in the Local Universe}",
      journal = {\apj},
         year = 2015,
        month = nov,
       volume = {813},
       number = {2},
          eid = {82},
        pages = {82},
          doi = {10.1088/0004-637X/813/2/82},
archivePrefix = {arXiv},
       eprint = {1508.06274},
 primaryClass = {astro-ph.GA},
       adsurl = {https://ui.adsabs.harvard.edu/abs/2015ApJ...813...82R}
}

@ARTICLE{Kennicutt+98,
       author = {{Kennicutt}, Jr., Robert C.},
        title = "{The Global Schmidt Law in Star-forming Galaxies}",
      journal = {\apj},
         year = 1998,
        month = may,
       volume = {498},
       number = {2},
        pages = {541-552},
          doi = {10.1086/305588},
archivePrefix = {arXiv},
       eprint = {astro-ph/9712213},
 primaryClass = {astro-ph},
       adsurl = {https://ui.adsabs.harvard.edu/abs/1998ApJ...498..541K}
}

@ARTICLE{Gillman+20,
       author = {{Gillman}, S. and {Tiley}, A.~L. and {Swinbank}, A.~M. and {Harrison}, C.~M. and {Smail}, Ian and {Dudzevi{\v{c}}i{\={u}}t{\.{e}}}, U. and {Sharples}, R.~M. and {Cortese}, L. and {Obreschkow}, D. and {Bower}, R.~G. and {Theuns}, T. and {Cirasuolo}, M. and {Fisher}, D.~B. and {Glazebrook}, K. and {Ibar}, Edo and {Mendel}, J. Trevor and {Sweet}, Sarah M.},
        title = "{From peculiar morphologies to Hubble-type spirals: the relation between galaxy dynamics and morphology in star-forming galaxies at z {\ensuremath{\sim}} 1.5}",
      journal = {\mnras},
         year = 2020,
        month = feb,
       volume = {492},
       number = {1},
        pages = {1492-1512},
          doi = {10.1093/mnras/stz3576},
archivePrefix = {arXiv},
       eprint = {1911.12375},
 primaryClass = {astro-ph.GA},
       adsurl = {https://ui.adsabs.harvard.edu/abs/2020MNRAS.492.1492G}
}

@ARTICLE{Genzel+11,
       author = {{Genzel}, R. and {Newman}, S. and {Jones}, T. and {F{\"o}rster Schreiber}, N.~M. and {Shapiro}, K. and {Genel}, S. and {Lilly}, S.~J. and {Renzini}, A. and {Tacconi}, L.~J. and {Bouch{\'e}}, N. and {Burkert}, A. and {Cresci}, G. and {Buschkamp}, P. and {Carollo}, C.~M. and {Ceverino}, D. and {Davies}, R. and {Dekel}, A. and {Eisenhauer}, F. and {Hicks}, E. and {Kurk}, J. and {Lutz}, D. and {Mancini}, C. and {Naab}, T. and {Peng}, Y. and {Sternberg}, A. and {Vergani}, D. and {Zamorani}, G.},
        title = "{The Sins Survey of z \raisebox{-0.5ex}\textasciitilde 2 Galaxy Kinematics: Properties of the Giant Star-forming Clumps}",
      journal = {\apj},
         year = 2011,
        month = jun,
       volume = {733},
       number = {2},
          eid = {101},
        pages = {101},
          doi = {10.1088/0004-637X/733/2/101},
archivePrefix = {arXiv},
       eprint = {1011.5360},
 primaryClass = {astro-ph.CO},
       adsurl = {https://ui.adsabs.harvard.edu/abs/2011ApJ...733..101G}
}

@ARTICLE{Swinbank+12,
       author = {{Swinbank}, A.~M. and {Smail}, Ian and {Sobral}, D. and {Theuns}, T. and {Best}, P.~N. and {Geach}, J.~E.},
        title = "{The Properties of the Star-forming Interstellar Medium at z = 0.8-2.2 from HiZELS: Star Formation and Clump Scaling Laws in Gas-rich, Turbulent Disks}",
      journal = {\apj},
         year = 2012,
        month = dec,
       volume = {760},
       number = {2},
          eid = {130},
        pages = {130},
          doi = {10.1088/0004-637X/760/2/130},
archivePrefix = {arXiv},
       eprint = {1209.1396},
 primaryClass = {astro-ph.CO},
       adsurl = {https://ui.adsabs.harvard.edu/abs/2012ApJ...760..130S}
}

@ARTICLE{Fisher+17,
       author = {{Fisher}, David B. and {Glazebrook}, Karl and {Damjanov}, Ivana and {Abraham}, Roberto G. and {Obreschkow}, Danail and {Wisnioski}, Emily and {Bassett}, Robert and {Green}, Andy and {McGregor}, Peter},
        title = "{DYNAMO-HST survey: clumps in nearby massive turbulent discs and the effects of clump clustering on kiloparsec scale measurements of clumps}",
      journal = {\mnras},
         year = 2017,
        month = jan,
       volume = {464},
       number = {1},
        pages = {491-507},
          doi = {10.1093/mnras/stw2281},
archivePrefix = {arXiv},
       eprint = {1608.08241},
 primaryClass = {astro-ph.GA},
       adsurl = {https://ui.adsabs.harvard.edu/abs/2017MNRAS.464..491F}
}

@ARTICLE{Calabro+17,
       author = {{Calabr{\`o}}, A. and {Amor{\'\i}n}, R. and {Fontana}, A. and {P{\'e}rez-Montero}, E. and {Lemaux}, B.~C. and {Ribeiro}, B. and {Bardelli}, S. and {Castellano}, M. and {Contini}, T. and {De Barros}, S. and {Garilli}, B. and {Grazian}, A. and {Guaita}, L. and {Hathi}, N.~P. and {Koekemoer}, A.~M. and {Le F{\`e}vre}, O. and {Maccagni}, D. and {Pentericci}, L. and {Schaerer}, D. and {Talia}, M. and {Tasca}, L.~A.~M. and {Zucca}, E.},
        title = "{Characterization of star-forming dwarf galaxies at 0.1 {\ensuremath{\lesssim}}z {\ensuremath{\lesssim}} 0.9 in VUDS: probing the low-mass end of the mass-metallicity relation}",
      journal = {\aap},
         year = 2017,
        month = may,
       volume = {601},
          eid = {A95},
        pages = {A95},
          doi = {10.1051/0004-6361/201629762},
archivePrefix = {arXiv},
       eprint = {1701.04418},
 primaryClass = {astro-ph.GA},
       adsurl = {https://ui.adsabs.harvard.edu/abs/2017A&A...601A..95C}
}

@ARTICLE{Amorin+14,
       author = {{Amor{\'\i}n}, R. and {Sommariva}, V. and {Castellano}, M. and {Grazian}, A. and {Tasca}, L.~A.~M. and {Fontana}, A. and {Pentericci}, L. and {Cassata}, P. and {Garilli}, B. and {Le Brun}, V. and {Le F{\`e}vre}, O. and {Maccagni}, D. and {Thomas}, R. and {Vanzella}, E. and {Zamorani}, G. and {Zucca}, E. and {Bardelli}, S. and {Capak}, P. and {Cassar{\'a}}, L.~P. and {Cimatti}, A. and {Cuby}, J.~G. and {Cucciati}, O. and {de la Torre}, S. and {Durkalec}, A. and {Giavalisco}, M. and {Hathi}, N.~P. and {Ilbert}, O. and {Lemaux}, B.~C. and {Moreau}, C. and {Paltani}, S. and {Ribeiro}, B. and {Salvato}, M. and {Schaerer}, D. and {Scodeggio}, M. and {Talia}, M. and {Taniguchi}, Y. and {Tresse}, L. and {Vergani}, D. and {Wang}, P.~W. and {Charlot}, S. and {Contini}, T. and {Fotopoulou}, S. and {L{\'o}pez-Sanjuan}, C. and {Mellier}, Y. and {Scoville}, N.},
        title = "{Discovering extremely compact and metal-poor, star-forming dwarf galaxies out to z \raisebox{-0.5ex}\textasciitilde 0.9 in the VIMOS Ultra-Deep Survey}",
      journal = {\aap},
         year = 2014,
        month = aug,
       volume = {568},
          eid = {L8},
        pages = {L8},
          doi = {10.1051/0004-6361/201423816},
archivePrefix = {arXiv},
       eprint = {1403.3692},
 primaryClass = {astro-ph.GA},
       adsurl = {https://ui.adsabs.harvard.edu/abs/2014A&A...568L...8A}
}

@ARTICLE{Efstathiou+14,
       author = {{Efstathiou}, A. and {Pearson}, C. and {Farrah}, D. and {Rigopoulou}, D. and {Graci{\'a}-Carpio}, J. and {Verma}, A. and {Spoon}, H.~W.~W. and {Afonso}, J. and {Bernard-Salas}, J. and {Clements}, D.~L. and {Cooray}, A. and {Cormier}, D. and {Etxaluze}, M. and {Fischer}, J. and {Gonz{\'a}lez-Alfonso}, E. and {Hurley}, P. and {Lebouteiller}, V. and {Oliver}, S.~J. and {Rowan-Robinson}, M. and {Sturm}, E.},
        title = "{Herschel observations and a model for IRAS 08572+3915: a candidate for the most luminous infrared galaxy in the local (z < 0.2) Universe}",
      journal = {\mnras},
         year = 2014,
        month = jan,
       volume = {437},
       number = {1},
        pages = {L16-L20},
          doi = {10.1093/mnrasl/slt131},
archivePrefix = {arXiv},
       eprint = {1309.7005},
 primaryClass = {astro-ph.CO},
       adsurl = {https://ui.adsabs.harvard.edu/abs/2014MNRAS.437L..16E}
}

@ARTICLE{Marleau+17,
       author = {{Marleau}, Francine R. and {Clancy}, Dominic and {Habas}, Rebecca and {Bianconi}, Matteo},
        title = "{Infrared signature of active massive black holes in nearby dwarf galaxies}",
      journal = {\aap},
         year = 2017,
        month = jun,
       volume = {602},
          eid = {A28},
        pages = {A28},
          doi = {10.1051/0004-6361/201629832},
archivePrefix = {arXiv},
       eprint = {1411.3844},
 primaryClass = {astro-ph.GA},
       adsurl = {https://ui.adsabs.harvard.edu/abs/2017A&A...602A..28M}
}

@ARTICLE{Messick+25,
       author = {{Messick}, Alexander and {Baldassare}, Vivienne and {Jones}, David O. and {French}, K. Decker and {Raimundo}, Sandra I. and {Earl}, Nicholas and {Auchettl}, Katie and {Coulter}, David A. and {Huber}, Mark E. and {Verrico}, Margaret E. and {de Boer}, Thomas and {Chambers}, Kenneth C. and {Gao}, Hua and {Lin}, Chien-Cheng and {Wainscoat}, Richard J.},
        title = "{A Large-scale Search for Photometrically Variable Active Galactic Nuclei in Dwarf Galaxies Using the Young Supernova Experiment}",
      journal = {\apj},
         year = 2025,
        month = jun,
       volume = {985},
       number = {2},
          eid = {223},
        pages = {223},
          doi = {10.3847/1538-4357/adcdff},
archivePrefix = {arXiv},
       eprint = {2504.00971},
 primaryClass = {astro-ph.GA},
       adsurl = {https://ui.adsabs.harvard.edu/abs/2025ApJ...985..223M}
}

@ARTICLE{Wasleske+24,
       author = {{Wasleske}, Erik J. and {Baldassare}, Vivienne F.},
        title = "{Active Dwarf Galaxy Database. I. Overlap between Active Galactic Nuclei Selected by Different Techniques}",
      journal = {\apj},
         year = 2024,
        month = aug,
       volume = {971},
       number = {1},
          eid = {68},
        pages = {68},
          doi = {10.3847/1538-4357/ad5442},
archivePrefix = {arXiv},
       eprint = {2405.20312},
 primaryClass = {astro-ph.GA},
       adsurl = {https://ui.adsabs.harvard.edu/abs/2024ApJ...971...68W}
}

@ARTICLE{Mezcua+18,
       author = {{Mezcua}, M. and {Civano}, F. and {Marchesi}, S. and {Suh}, H. and {Fabbiano}, G. and {Volonteri}, M.},
        title = "{Intermediate-mass black holes in dwarf galaxies out to redshift {\ensuremath{\sim}}2.4 in the Chandra COSMOS-Legacy Survey}",
      journal = {\mnras},
         year = 2018,
        month = aug,
       volume = {478},
       number = {2},
        pages = {2576-2591},
          doi = {10.1093/mnras/sty1163},
archivePrefix = {arXiv},
       eprint = {1802.01567},
 primaryClass = {astro-ph.GA},
       adsurl = {https://ui.adsabs.harvard.edu/abs/2018MNRAS.478.2576M}
}

@ARTICLE{Bertin+96,
       author = {{Bertin}, E. and {Arnouts}, S.},
        title = "{SExtractor: Software for source extraction.}",
      journal = {\aaps},
         year = 1996,
        month = jun,
       volume = {117},
        pages = {393-404},
          doi = {10.1051/aas:1996164},
       adsurl = {https://ui.adsabs.harvard.edu/abs/1996A&AS..117..393B}
}

@ARTICLE{Barbary+16, doi = {10.21105/joss.00058}, url = {https://doi.org/10.21105/joss.00058}, year = {2016}, publisher = {The Open Journal}, volume = {1}, number = {6}, pages = {58}, author = {Kyle Barbary}, title = {SEP: Source Extractor as a library}, journal = {Journal of Open Source Software} }

@ARTICLE{Kron+80,
       author = {{Kron}, R.~G.},
        title = "{Photometry of a complete sample of faint galaxies.}",
      journal = {\apjs},
         year = 1980,
        month = jun,
       volume = {43},
        pages = {305-325},
          doi = {10.1086/190669},
       adsurl = {https://ui.adsabs.harvard.edu/abs/1980ApJS...43..305K}
}

@INPROCEEDINGS{Perrin+14,
       author = {{Perrin}, Marshall D. and {Sivaramakrishnan}, Anand and {Lajoie}, Charles-Philippe and {Elliott}, Erin and {Pueyo}, Laurent and {Ravindranath}, Swara and {Albert}, Lo{\"\i}c.},
        title = "{Updated point spread function simulations for JWST with WebbPSF}",
    booktitle = {Space Telescopes and Instrumentation 2014: Optical, Infrared, and Millimeter Wave},
         year = 2014,
       editor = {{Oschmann}, Jr., Jacobus M. and {Clampin}, Mark and {Fazio}, Giovanni G. and {MacEwen}, Howard A.},
       series = {Society of Photo-Optical Instrumentation Engineers (SPIE) Conference Series},
       volume = {9143},
        month = aug,
          eid = {91433X},
        pages = {91433X},
          doi = {10.1117/12.2056689},
       adsurl = {https://ui.adsabs.harvard.edu/abs/2014SPIE.9143E..3XP}
}

@ARTICLE{Gordon+23,
       author = {{Gordon}, Karl D. and {Clayton}, Geoffrey C. and {Decleir}, Marjorie and {Fitzpatrick}, E.~L. and {Massa}, Derck and {Misselt}, Karl A. and {Tollerud}, Erik J.},
        title = "{One Relation for All Wavelengths: The Far-ultraviolet to Mid-infrared Milky Way Spectroscopic R(V)-dependent Dust Extinction Relationship}",
      journal = {\apj},
         year = 2023,
        month = jun,
       volume = {950},
       number = {2},
          eid = {86},
        pages = {86},
          doi = {10.3847/1538-4357/accb59},
archivePrefix = {arXiv},
       eprint = {2304.01991},
 primaryClass = {astro-ph.GA},
       adsurl = {https://ui.adsabs.harvard.edu/abs/2023ApJ...950...86G}
}

@ARTICLE{Calzetti+94,
       author = {{Calzetti}, Daniela and {Kinney}, Anne L. and {Storchi-Bergmann}, Thaisa},
        title = "{Dust Extinction of the Stellar Continua in Starburst Galaxies: The Ultraviolet and Optical Extinction Law}",
      journal = {\apj},
         year = 1994,
        month = jul,
       volume = {429},
        pages = {582},
          doi = {10.1086/174346},
       adsurl = {https://ui.adsabs.harvard.edu/abs/1994ApJ...429..582C}
}

\begin{appendix}

\section{Dataset}
\label{app:dataset}
In the following section, we describe in more detail the dataset collected for both targets and adopted in this work.
We also report the estimated \hst\ and \jwst\ photometry of the targets presented in this work in Table \ref{app-tab:photometry_combined}, while constraints from the multi-wavelength dataset are presented in Table \ref{app-tab:ancillary_photometry}.
We finally describe the spectroscopic redshift estimate of both targets.  

\subsection{\namemiric\ multi-wavelength data}
In addition to the deep \jwst/MIRI imaging at 7.7 and 10~\um\ from {\it MIRIC} (Iani et al., {\it in prep.}, reduced with the same pipeline adopted in \citealt{Rinaldi+25}), we collect all the available ancillary imaging from the \hst~ Advanced Camera for Surveys (ACS) and Wide Field Camera 3 (WFC3) instruments as well as the \jwst\ NIRCam.
In particular, we resort to fully reduced and publicly available \hst~ imaging from the programs {\it the Cluster Lensing And Supernova survey with Hubble} \cite[CLASH,][]{Postman+12}, {\it the  Hubble Frontier Fields} \cite[HFF,][]{Lotz+17} and {\it Beyond Ultra-deep Frontier Fields And Legacy Observations} \cite[BUFFALO, ][]{Steinhardt+20}.
Despite the large photometric dataset available for this field, \namemiric'coverage is limited to the 7 ACS Wide Field Camera (WFC) filters F435W, F475W, F606W, F625W, F775W, F814W and F850LP. 

For \jwst/NIRCam imaging, we downloaded the observations from PID 1176 \cite[][]{Windhorst+17jwst.prop.1176}, 1208 \cite[CANUCS;][]{Willott+2017jwst.prop.1208, Sarrouh+25}, 2883 \cite[MAGNIF;][]{Sun+23jwst.prop.2883, Fu+25} and 3538 \cite[][]{Iani+23jwst.prop.3538} from the \textsc{Mikulski Archive for Space Telescopes} (\textsc{MAST})\footnote{\url{https://mast.stsci.edu/portal/Mashup/Clients/Mast/Portal.html}}.
These programs provide coverage in 15 filters out of which 7 wide bands (F090W, F115W, F150W, F200W, F277W, F356W, F444W) and 8 medium bands (F182M, F210M, F300M, F335M, F360M, F410M, F460M, F480M).
The reduction of this dataset was produced using v. 1.14.0 of the \textsc{JWST pipeline}\footnote{\url{https://jwst-docs.stsci.edu/jwst-science-calibration-pipeline\#gsc.tab=0}} and \textsc{Calibration References Data System} (\textsc{CRDS}) context \texttt{jwst\_1210.pmap}. 
We adopt the same data-reduction methodology outlined in recent works \cite[e.g.,][]{Rinaldi+23, Iani+24, Oestlin+25}.

In addition to photometric data, \jwst\ has provided spectroscopic coverage for our target in Wide Field Slitless Spectroscopy (WFSS) mode with the NIRCam and the Near Infrared Imager and Slitless Spectrograph \cite[NIRISS,][]{Willott+22, Doyon+23}.
We reduce the NIRCam/WFSS observations in F300M, F335M, F360M, F410M, F460M, F480M from PID 3538 and MAGNIF by means of the reduction pipeline {\it Allegro} (Kramarenko et al., {\it in prep.}) initially developed for the EIGER program \citep{Kashino+23, Matthee+23}. 
The full-reduced NIRISS dataset has been kindly provided by the CANUCS collaboration (\textit{private comm.}) with the data reduction performed as detailed in \cite{Sarrouh+25} (see also Noirot et al., \textit{in prep.}). 

Since the dataset presented here has been collected over different years, we test if possible effects of variability due to transient phenomena affecting our targets (e.g. AGN, supernovae) could bias our analysis and the interpretation of our targets' SED, see \S~\ref{app:variability}.
Our test, however, excludes effects of variability to the level of our photometry precision.

To further constrain the properties of \namemiric, we look for additional data coming from observations at both shorter and longer wavelengths.
In particular, we look for possible constraints from the X-rays to the radio domain. 

We check whether X-ray observations in the M0416 field report our target's detection \cite[e.g.,][]{Ogrean+15}.
For the flux at rest-frame 2~keV $f({\rm 2keV})$ (corresponding to observed 1.2~keV at \zspecmiric), we check the {\it Chandra} Source Catalog\footnote{\url{https://cxc.cfa.harvard.edu/csc/}} \cite[v. 2.1,][]{Evans+24}.
The catalog does not report a detection at the target coordinates nor in its immediate vicinity ($\delta = 0.''5$) but tabulates a sensitivity limit ($3\sigma$) at the source coordinates for a point-like source in the {\it Chandra}/ACIS broad energy band 0.5 - 7.0 keV of $1.22 \times 10^{-15}\rm ~ erg~s^{-1}~cm^{-2}$. 
Assuming a photon index for the power law of the X-ray spectrum of $\Gamma = 1.7$, we obtain a $3\sigma$ upper limit on the observed X-ray flux $f({\rm 2keV}) \leq 1.4 \times 10^{-4}~\mu$Jy. 

The M0416 galaxy cluster was also observed by the Photodetector Array Camera and Spectrometer \cite[PACS;][]{Poglitsch+10} on board of the \textit{Herschel} observatory at 100~$\mu$m (green channel) and 160~$\mu$m (red channel) as well as from the Spectral and Photometric Imaging Receiver \cite[SPIRE; ][]{Griffin+10} at 250, 350 and 500~$\mu$m as part of the Herschel Lensing Survey \cite[HLS; PIDs: 1342250291–2, 1342241122, PI: E. Egami;][]{Egami+10}.
Due to confusion noise and large PSF, we do not take into account the SPIRE observations.
We download the fully-reduced PACS images from the European Space Agency (ESA) website\footnote{\url{https://archives.esac.esa.int/hsa/whsa/?ACTION=PUBLICATION&ID=2016MNRAS.459.1626R}}.
At the coordinates of our target, we do not detect any \textit{Herschel} counterpart. 
This is also confirmed by the fact that no detection has been reported by \cite{Rawle+16, Sun+22, Kokorev+22}.
We therefore estimate $3\sigma$ local upper-limits of $2.6\times 10^3$~$\mu$Jy at 100~$\mu$m and $4.8\times 10^3$~$\mu$Jy at 160~$\mu$m.

The Atacama Large Sub-Millimeter Array (ALMA) mosaicked the M0416 area at $\approx$~1 mm (band 6; PID 2013.1.00999.S, PI: F. Bauer; see e.g. \citealt{Gonzalez-Lopez+17a, Gonzalez-Lopez+17b}) and 2~mm (band 4; PID 2022.1.01356.S, PI: E. Egami).
In both maps, our target is undetected. 
The local $3\sigma$ sensitivity level is 557~$\mu$Jy at 1.1~mm and 343~$\mu$Jy at 2.1~mm.

We finally look for possible detections from the  Karl G. Jansky Very Large Array (JVLA) in the 6 cm (5 - 7 GHz, C-band) and 13 cm (2 - 4 GHz, S-band) maps \citep{Heywood+21}.
We download the JVLA fully-reduced maps from the National Radio Astronomy Observatory (NRAO) archive\footnote{\url{https://science.nrao.edu/science/surveys/vla-ff/data}.}.
In both maps our target is undetected; thus we set local $3\sigma$ sensitivity levels of 3.1~$\mu$Jy at 6~cm and ~9.1~$\mu$Jy at 13~cm.

We report the constraints coming from these panchromatic ancillary datasets in Table~\ref{app-tab:ancillary_photometry}.

\subsubsection{NIRCam variability test in \namemiric}
\label{app:variability}
To investigate the possible presence of variability in the rest-frame optical-to-NIR photometry of \namemiric, we collect from the MAST all the fully-reduced single-epoch images available in all the NIRCam filters that cover our target and that were taken by PIDs 1176, 1208, 2883 and 3538. 
Out of 15 NIRCam filters, nine have multi-epochs observations: F090W, F115W, F150W, F182M, F200W, F210M, F277W, F410M and F444W.
Depending on the filter, the time separation between observations spans from a few weeks to several months.
For each filter with multi-epoch coverage, we extract our target's photometry from every single-epoch mosaic available following the methodology presented in \S~\ref{sec:photometry}. 

To quantify the level of variability of our target, we adopt the methodology introduced by \cite{Vaughan+03}: the so-called fractional variability $F_{\rm var}$.
To estimate this quantity, \cite{Vaughan+03} first introduces the parameters of sample variance of the light curve $S^2$ and its mean square error $<\sigma^2>$.
These parameters are defined as:
\begin{equation}
S^2 = \frac{1}{N-1}\cdot \sum_i(f_i - \bar{f})^2
\end{equation}
\begin{equation}
<\sigma^2> = \frac{1}{N}\cdot \sum_i \sigma_i^2
\end{equation}
where $f_i$ and $\sigma_i$ are the flux and error (in a given filter) at a given epoch, $N$ is the total number of epochs available and $\bar{f}$ is the mean value of $f_i$ over the different epochs. 
Knowing the sample variance and mean square error it is possible to derive the excess (or intrinsic) variance $\sigma_{\rm int}^2$ as:
\begin{equation}
\sigma_{\rm int}^2 = S^2 - <\sigma^2> 
\end{equation}
After having estimated the excess variance, our target fractional variability in a given filter is given by:
\begin{equation}
F_{\rm var} = \frac{\sqrt{\sigma_{\rm int}^2}}{\bar{f}}
\end{equation}
If this quantity is found to be negative, it means that there is no evidence for intrinsic variability.
In this case, we estimate a conservative $3\sigma$ upper limit on the fractional variability assuming that the observed scatter is entirely driven by measurement uncertainties as:
\begin{equation}
F_{\rm var, 3\sigma} = \frac{3\sqrt{<\sigma^2>}}{\bar{f}}
\end{equation}
We report the values of sample variance, mean square error, excess variance, fractional variability (and $3\sigma$ upperlimit) of our target \namemiric\ in Table~\ref{app-tab:fractional_var}.
The resulting $3\sigma$ upper limits on the intrinsic fractional variability range from 
$F_{\rm var, 3\sigma}<0.18$ (F115W) to $F_{\rm var, 3\sigma} < 0.47$ (F277W), implying no evidence for significant variability at the $\leq 20 - 50$\% level, depending on filter, thus ruling out any possible impact of transient phenomena at the origin of our target SED shape.

\subsubsection{\namemiric\ redshift estimate}
\label{app-sec:miric_redshift}
The NIRISS/WFSS observations of \namemiric\ in the F090W, F115W and F200W filters reveal the presence of the \hb, \oiii$\lambda\lambda4959,5007$, H$\alpha$ and the \hei$\lambda10830$.
We fit these lines with Gaussians by means of the \textsc{Python} library \texttt{scipy.optimize}. 
For each line, we estimate its flux and corresponding error (see Table~\ref{tab:lines_combined}) by drawing 1000 Monte Carlo realizations of the NIRISS spectrum, perturbing the observed monochromatic flux densities according to their errors following a Gaussian distribution.
We adopt as final flux of each line the 50th percentile of its total flux distribution while as error the half-distance between the 16th and 84th percentiles.

From a weighted average of the observed central wavelength of these lines with respect to their rest-frame emission and after including an absolute wavelength calibration floor of about 20~\AA\ for NIRISS/WFSS \citep{Noirot+23}, we estimate a spectroscopic (systemic) redshift $z_{\rm spec} = $~\zspecmiric$\pm 0.0018$.
In addition to NIRISS slitless spectroscopy, we also detected the \paa\ and \brg\ emission lines from the NIRCam F335M (PID 3538) and F360M (MAGNIF) grism spectra, respectively.
For the \paa\ we find $z = 0.7098 \pm 0.0005$, while for the \brg\ transition $z = 0.7093 \pm 0.0001$, consistent with the NIRISS redshift estimate within $\approx 2\sigma$ confidence interval. 
In the following, we therefore assume $z = $~\zspecmiric\ as the reference value for our analysis of \namemiric.

According to the adopted cosmology, \namemiric\ ($z = 0.7050$) is at a luminosity distance of $d_L \approx 4.30\rm~ Gpc$ with the age of the universe at this redshift being about 7.1 Gyr.

\begin{table*}
\centering
\footnotesize
\setlength{\tabcolsep}{3.5pt}
\caption{Photometry of \namemiric\ (M0416) and \namejades\ (GOODS-N)}
\label{app-tab:photometry_combined}
\begin{tabular}{ll | cccc | cccc}
\hline
\hline
 &  & \multicolumn{4}{c}{\namemiric} & \multicolumn{4}{c}{\namejades} \\
Instrument & Filter
& $\mu \cdot f_\nu$ & $\mu \cdot$ err$_{f_{\nu}}$ & $f_{\rm ext}$ & $\mu \cdot f_{\nu,\rm eml}$ 
& $f_\nu$ & err$_{f_{\nu}}$ & $f_{\rm ext}$ & $f_{\nu,\rm eml}$ \\
 &  & [$\mu$Jy] & [$\mu$Jy] &  & [$\mu$Jy] & [$\mu$Jy] & [$\mu$Jy] &  & [$\mu$Jy] \\
\hline
HST/WFC3\_UVIS & F275W  & --    & --    & --    & --    
                          & 0.199 & 0.043 & 1.060 & --    \\

HST/ACS\_WFC & F435W  & 0.174 & 0.032 & 1.144 & --    
                        & 0.194 & 0.010 & 1.037 & --    \\

HST/ACS\_WFC & F475W  & 0.224 & 0.040 & 1.131 & --    
                        & --    & --    & --    & --    \\

HST/ACS\_WFC & F606W  & 0.209 & 0.038 & 1.100 & --    
                        & 0.218 & 0.037 & 1.026 & 0.041 \\

HST/ACS\_WFC & F625W  & 0.239 & 0.044 & 1.089 & --    
                        & --    & --    & --    & --    \\

HST/ACS\_WFC & F775W  & 0.214 & 0.037 & 1.066 & 0.073  
                        & 0.141 & 0.008 & 1.017 & 0.058 \\

HST/ACS\_WFC & F814W  & 0.304 & 0.046 & 1.062 & 0.113 
                        & 0.261 & 0.052 & 1.017 & 0.172 \\

HST/ACS\_WFC & F850LP & 0.527 & 0.059 & 1.050 & 0.268 
                        & 0.501 & 0.099 & 1.013 & 0.416 \\

HST/WFC3\_IR & F105W  & --    & --    & --    & --   
                        & 0.265 & 0.086 & 1.010 & 0.099 \\

JWST/NIRCam  & F090W  & 0.417 & 0.073 & 1.051 & 0.170
                        & 0.395 & 0.021 & 1.014 & 0.254 \\

JWST/NIRCam  & F115W  & 0.241 & 0.012 & 1.032 & 0.092
                        & 0.267 & 0.032 & 1.008 & 0.129 \\

HST/WFC3\_IR & F125W  & --    & --    & --    & --   
                        & 0.274 & 0.029 & 1.008 & 0.132 \\

JWST/NIRCam  & F150W  & 0.137 & 0.007 & 1.020 & 0.004
                        & 0.135 & 0.012 & 1.006 & 0.009 \\

HST/WFC3\_IR & F160W  & --    & --    & --    & --   
                        & 0.069 & 0.032 & 1.005 & 0.011 \\

JWST/NIRCam  & F182M  & 0.166 & 0.046 & 1.014 & 0.017
                        & --    & --    & --    & --    \\

JWST/NIRCam  & F200W  & 0.179 & 0.039 & 1.013 & --   
                        & 0.174 & 0.009 & 1.003 & 0.055 \\

JWST/NIRCam  & F210M  & 0.117 & 0.046 & 1.011 & --   
                        & --    & --    & --    & --    \\

JWST/NIRCam  & F277W  & 0.153 & 0.046 & 1.007 & --   
                        & 0.136 & 0.013 & 1.002 & 0.009 \\

JWST/NIRCam  & F300M  & 0.157 & 0.028 & 1.006 & --   
                        & --    & --    & --    & --    \\

JWST/NIRCam  & F335M  & 0.374 & 0.038 & 1.005 & --   
                        & 0.340 & 0.074 & 1.001 & 0.095 \\

JWST/NIRCam  & F356W  & 0.381 & 0.032 & 1.005 & --   
                        & 0.361 & 0.041 & 1.001 & 0.034 \\

JWST/NIRCam  & F360M  & 0.337 & 0.063 & 1.004 & --   
                        & --    & --    & --    & --    \\

JWST/NIRCam  & F410M  & 0.607 & 0.030 & 1.004 & --   
                        & 0.578 & 0.029 & 1.001 & --    \\

JWST/NIRCam  & F444W  & 0.877 & 0.044 & 1.003 & --   
                        & 0.928 & 0.046 & 1.001 & --    \\

JWST/NIRCam  & F460M  & 0.938 & 0.169 & 1.003 & --   
                        & --    & --    & --    & --    \\

JWST/NIRCam  & F480M  & 1.145 & 0.179 & 1.003 & --   
                        & --    & --    & --    & --    \\

JWST/MIRI    & F770W  & 5.528 & 1.039 & 1.003 & --   
                        & 5.937 & 1.415 & 1.001 & --    \\

JWST/MIRI    & F1000W & 12.240& 2.909 & 1.007 & --   
                        & --    & --    & --    & --    \\

JWST/MIRI    & F1280W & --    & --    & --    & --   
                        & 14.293& 1.367 & 1.001 & --    \\
\hline
\hline
\end{tabular}
\tablefoot{
Flux densities are corrected for Galactic extinction using $f_{\rm ext}$.
For \namemiric, fluxes are not corrected for lensing magnification.
$f_{\nu,\rm eml}$ is the estimated contamination from strong emission lines.
Compactness is $c \equiv f(0.''2)/f(0.''1)$.
}
\end{table*}

\begin{table}
\centering
\footnotesize
\setlength{\tabcolsep}{3.5pt}
\caption{Ancillary multi-wavelength data for \namemiric\ (M0416) and \namejades\ (GOODS-N)}
\label{app-tab:ancillary_photometry}
\begin{tabular}{ll | cc | cc}
\hline
\hline
 &  & \multicolumn{2}{c}{\namemiric} & \multicolumn{2}{c}{\namejades} \\
Instrument & Band
& $\mu\cdot f_\nu$ & $\mu \cdot~$err$_{f_{\nu}}$ 
& $f_\nu$ & err$_{f_{\nu}}$  \\
 &  & [$\mu$Jy] & [$\mu$Jy] 
 & [$\mu$Jy] & [$\mu$Jy] \\
\hline
Chandra & 0.5 - 7 keV & <$14\times 10^{-5}$ & - & <$8\times 10^{-5}$ & - \\
Spitzer/IRS PUI & 16~$\mu$m & - & - & <22.5 & - \\
Spitzer/MIPS & 24~$\mu$m & - & - & 25.6 & 8 \\
Herschel/PACS & 100~$\mu$m & <2600 & - & 424.6 & 367.4 \\
Herschel/PACS & 160~$\mu$m & <4800 & - & <2755.2 & - \\
ALMA & band 6 & <557 & - & - & - \\
ALMA & band 4 & <343 & - & - & - \\
JVLA & C & <3.1 & - & - & - \\
JVLA & S & <9.1 & - & - & - \\
JVLA & L & - & - & <7.5 & - \\
\hline
\hline
\end{tabular}
\tablefoot{
For \namemiric, fluxes are not corrected for lensing magnification.
}
\end{table}

\begin{table}[]
    \centering
    \caption{Fractional Variability Parameters for \namemiric\ NIRCam photometry}
    \footnotesize
    \begin{tabular}{lcccccc}
    \hline
    \hline
    Filter & \# Obs. & PIDs & $S^2$ & $<\sigma^2>$ & $\sigma^2_{\rm int}$ & $F_{\rm var, 3\sigma}$\\
    \hline
        F090W & 4 & 1176, 1208       & 4.2E-4 & 5.7E-4 & -1.4E-4 & 0.22 \\    
        F115W & 4 & 1176, 1208       & 2.0E-6 & 1.3E-4 & -1.3E-4 & 0.18 \\    
        F150W & 4 & 1176, 1208       & 2.5E-5 & 3.8E-5 & -1.3E-5 & 0.24 \\    
        F182M & 3 & 2883, 3538       & 1.5E-5 & 7.8E-5 & -6.3E-5 & 0.22 \\    
        F200W & 4 & 1176, 1208       & 4.6E-5 & 5.8E-5 & -1.2E-5 & 0.21 \\    
        F210M & 3 & 2883, 3538       & 6.0E-6 & 8.7E-5 & -8.1E-5 & 0.36 \\    
        F277W & 2 & 1176, 1208       & 8.2E-5 & 3.3E-4 & -2.5E-4 & 0.47 \\    
        F410M & 3 & 1176, 1208, 3538 & 6.5E-4 & 1.6E-3 & -9.0E-4 & 0.22 \\   
        F444W & 2 & 1176, 1208       & <1E-6 & 5.9E-3 & -5.9E-3 & 0.27 \\ 
    \hline
    \hline
    \end{tabular}
    \label{app-tab:fractional_var}
\end{table}

\subsection{\namejades\ multi-wavelength data}
The area of GOODS-N has been extensively observed by \hst\ through the years and recently all the publicly available datasets have been reduced and made accessible as part of the {\it Hubble Legacy Fields} program \cite[HLF, see ][]{Illingworth+16, Whitaker+19}.
In particular, with the HLF Data Release v. 2.5\footnote{\url{https://archive.stsci.edu/prepds/hlf/}} \hst\ imaging taken over nearly 18 years and from 29 approved programs have been incorporated in the released dataset of 11 filters consisting of two WFC3/UVIS, five ACS/WFC and four WFC3/IR bands, out of which nine filters cover the area of \namejades: F275W, F435W, F606W, F775W, F814W, F850LP, F105W, F125W and F160W.
We download the fully reduced \hst\ dataset from the HLF website.

The GOODS-N area has also been targeted by the \jwst\ program PID 1181 \citep{Eisenstein+17jwst.prop.1181} in 9 NIRCam filters ( F090W, F115W, F150W, F200W, F277W, F335M, F356W, F410M, F444W) and with MIRI observations at 7.7 (F770W) and 12.8~\um\ (F1280W).
We download the dataset from MAST and reduce it following the same methodology applied in the reduction performed for \namemiric. 

Besides photometric imaging, \namejades\ was observed with \jwst\ NIRSpec PRISM and grisms G235H+F170LP and G395H+F290LP as part of the NIRSpec Wide GTO Survey program PID 1211 \citep{2017jwst.prop.1211F, Maseda+24}.
We download the fully reduced dataset (v. 4) directly from the DJA.

Also in the case of \namejades, our target is not detected in X-ray {\it Chandra} observations with the {\it Chandra} Source Catalog v. 2.1 reporting a $3\sigma$ sensitivity limit in the ACIS broad energy band 0.5 - 7.0~keV of $6.96\times 10^{-16}\rm ~erg~s^{-1}~cm^{-2}$. 
Following the same reasoning described above, the observed X-ray flux $f(2\rm keV) \leq 8\times 10^{-5}\rm~\mu Jy$. 
We also check if any X-ray counterpart was reported by the dedicated analysis of \textit{Chandra} data by \cite{Xue+16} and found no X-ray source at our target coordinates. 

GOODS-N was also observed by the \textit{Spitzer}, \textit{Herschel} and JVLA observatories.
In particular, \cite{Liu+18} presented a `super-deblended' catalog collecting data from all these observatories.
For \textit{Spitzer} observations, the dataset comprises observations that were carried out with the Infrared Spectrograph Peak-Up imaging \cite[IRS/PUI; ][]{Teplitz+04} at 16~$\mu$m and the Multiband Imaging Photometer \cite[MIPS; ][]{Rieke+04} at 24~$\mu$m \cite[GOODS-\textit{Spitzer} program, PI: M. Dickinson; ][]{Dickinson+03}.
For \textit{Herschel}, the observations are based on the co-added PACS images from \cite{Magnelli+13}.
Also in this case, while SPIRE observations are available, we do not use them following the same motivations adopted for the SPIRE dataset in M0416.
Finally, for JVLA, a map at $\approx 20$~cm (L-band, 1-2 GHz; program ID: VLA/13B-038, PI: M. Aravena) was adopted \cite[e.g.,][]{Owen+18}.
By matching the catalog by \cite{Liu+18} to our target spatial coordinates, we find a counterpart (ID17474) with reported $z = 0.748$, in perfect agreement with our target's spectroscopic redshift.
Our object results having detection at 16~$\mu$m ($25.6 \pm 8~\mu$Jy) and 100~$\mu$m ($424.6 \pm 367.4~\mu$Jy).
For the other bands we have the following $3\sigma$ upper-limits: $22.5~\mu$Jy at 16~$\mu$m, $2755.2~\mu$Jy at 160~$\mu$m and $7.5~\mu$Jy at 20~cm.

Also in this case, we report the constraints coming from these panchromatic ancillary datasets in Table~\ref{app-tab:ancillary_photometry}.

\subsubsection{\namejades\ redshift estimate} 
\label{sec:z_mu}
In the case of \namejades, the analysis of the available NIRSpec/MSA observations, both PRISM and in the high-resolution grism G235H, 
reveal the detection of multiple emission lines.
We model the emission lines of \namejades{} in the NIRSpec/PRISM spectrum with Gaussian profiles and the continuum with a power law ($f(\lambda) \propto \lambda^{\beta}$) over the relevant wavelength intervals. 
The line flux measurements are listed in Table~\ref{tab:lines_combined}. 
Because the PRISM resolution in the rest-optical regime of \namejades{} is low ($R \sim 30$--100), some strong lines are heavily blended. The fitting windows are chosen case by case, depending on the lines and blends.  
For blended features, we impose a common line width and include at least one isolated strong line in the fit to anchor the width. We fix the [\ion{O}{iii}] $\lambda\lambda4960,5008$ amplitude ratio to 2.98, doublet that is also blended with \hb. 
The fluxes of \ion{He}{i} $\lambda10830$ and Pa$\gamma$ are measured from the high-resolution G235H spectrum, where they can be measured separately.
In particular the analysis of the G235H data, allows us to derive a spectroscopic redshift of $z_{\rm spec} = $~\zspecjades~$\pm 0.0022$.

We measure the emission line profiles of \pab, \pag\ and \ion{He}{i} $\lambda 10830$, the strongest in the G235H data. 
All three lines have widths in velocity space smaller than the nominal resolution from JDox\footnote{\url{https://jwst-docs.stsci.edu}} (125~km~s$^{-1}$ and 140~km~s$^{-1}$ at the observed wavelengths of \pab\ and \ion{He}{i} $\lambda10830$, respectively). 
\citet{shajib2025} showed that the effective in-flight resolution of NIRSpec MOS\footnote{\citet{shajib2025} provide measurements for NIRSpec in fixed slit mode, but they argue that the results are directly applicable to multi-object spectroscopy mode due to the similar slit aperture.} is up to 53\% greater than the nominal value, which assumes a fully illuminated shutter \citep[see also][]{degraaff2024}. 
Modelling the NIRSpec LSF as a Gaussian kernel with a width corresponding to, conservatively, 1.5$\times$ the nominal value, we find that the lines are marginally resolved, with FWHM(\pab)~$= 98 \pm 12$~km\,s$^{-1}$, FWHM(\pag)~$= 136 \pm 26$~km~s$^{-1}$ and FWHM(\ion{He}{i} $\lambda 18033$)$~= 72 \pm 17$~km~s$^{-1}$.

According to the adopted cosmology, \namejades\ ($z = 0.7474$) has $d_L \approx 4.62\rm~ Gpc$ and the age of the universe at its redshift was about 6.9 Gyr.

\begin{table}
\centering
\small
\caption{Observed Emission Line Fluxes of \namemiric\ and \namejades}
\begin{tabular}{l c | cc | cc}
\hline
\hline
 &  & \multicolumn{2}{c}{\namemiric} & \multicolumn{2}{c}{\namejades} \\
Line & $\lambda_0$ & $\mu\cdot f$ & $\mu\cdot$err$_f$ & $f$ & err$_f$ \\
 & [\um] & [cgs]$^a$ & [cgs]$^a$ & [cgs]$^a$ & [cgs]$^a$ \\
\hline
{[\ion{O}{ii}]}$^b$      & 0.3727 & --    & --    & 6.0   & 2.0 \\
{[\ion{Ne}{iii}]}        & 0.3870 & --    & --    & 10.0  & 3.0 \\
{H$\delta$}              & 0.4103 & --    & --    & 2.5   & 0.8 \\
{H$\gamma$}              & 0.4342 & --    & --    & 5.1   & 0.9 \\
{H$\beta$}               & 0.4863 & 11.9  & 4.9   & 25.0  & 5.0 \\
{[\ion{O}{iii}]}         & 0.4960 & 22.7  & 5.5   & 43.0  & 2.0 \\
{[\ion{O}{iii}]}         & 0.5008 & 102.2 & 7.8   & 128.0 & 6.0 \\
{H$\alpha$}              & 0.6565 & 58.1  & 6.3   & 72.0  & 3.0 \\
{\ion{He}{i}}            & 1.0833 & 9.1   & 0.6   & 10.8  & 0.4 \\
{Pa$\gamma$}             & 1.0941 & --    & --    & 3.5   & 0.5 \\
{Pa$\beta$}              & 1.2822 & --    & --    & 4.6   & 0.3 \\
{Pa$\alpha$}             & 1.8756 & 7.9    & 0.4    & 9.6   & 0.4 \\
{Br$\delta$}             & 1.9451 & --    & --    & 0.4   & 0.2 \\
{Br$\gamma$}             & 2.1661 & 0.8    & 0.2    & 0.6   & 0.2 \\
{Br$\beta$}              & 2.6259 & --    & --    & 1.7   & 0.2 \\
\hline
\hline
\end{tabular}
\tablefoot{
$^a$ Fluxes and errors are in units of $\rm 10^{-18}\,erg\,s^{-1}\,cm^{-2}$. In the case of \namemiric, the reported values are not corrected for lensing magnification $\mu$.\\
$^b$ [O\,\textsc{ii}] refers to the blended doublet.
}
\label{tab:lines_combined}
\end{table}

\section{parameters from 2D morphological fit}
\begin{figure*}
\centering
    \includegraphics[width = \textwidth]{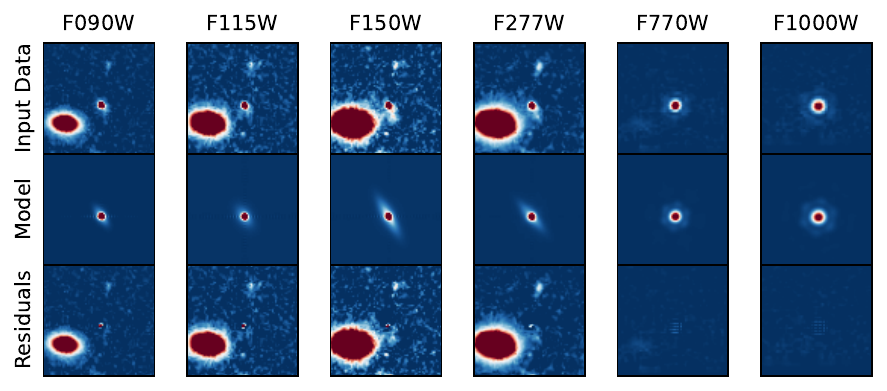}
    \includegraphics[width = \textwidth]{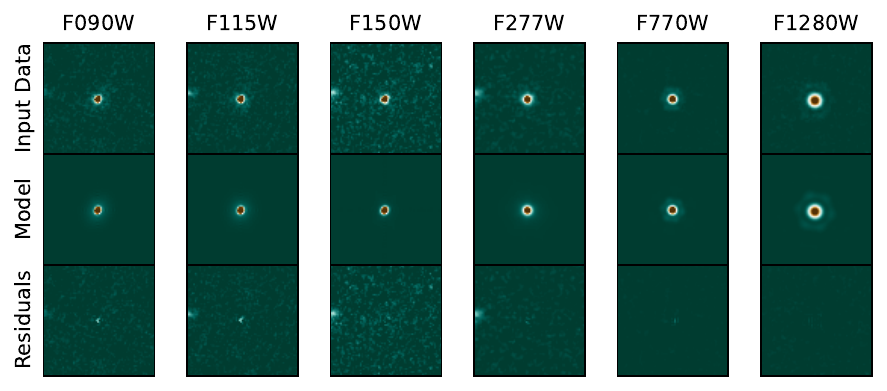}
    \caption{Examples of the 2D morphological fitting of our targets \namemiric\ (top panels) and \namejades\ (bottom panels) performed with \textsc{pysersic}. For each filter we show the original image (top), in the best-fit model (center) and residuals (bottom). The colorbar is fixed for a given filter in all the presented panels.}
    \label{app-fig:2d_morph_fit}
\end{figure*}

In Figure \ref{app-fig:2d_morph_fit} we present the 2D morphological decomposition obtained with \textsc{pysersic} for both \namemiric\ and \namejades\ for a selection of \jwst\ NIRCam and MIRI filters. 
We report the morphological parameters derived from {\sc pysersic} in Table~\ref{app-tab:2d_morph}.

\begin{table*}
\caption{Morphological parameters for \namemiric\ (M0416) and \namejades\ (GOODS-N).}
\label{app-tab:2d_morph}
\centering
\begin{tabular}{lccccccccc}
\hline\hline
& Filter & $n_1$ & $r_{\rm eff, circ ~1}$ & $q_1$ & $n_2$ & $r_{\rm eff, circ ~2}$ & $q_2$ & $f_{1/2}$ & $\chi^2_{\rm red}$\\
& & & [arcsec] & & & [arcsec] & & &\\
\hline
\multicolumn{10}{l}{\it \namemiric} \\
\hline
& F090W  & $0.66^{+0.02}_{-0.01}$ & $0.06 \pm 0.01$ & $0.98^{+0.01}_{-0.03}$ & $0.84^{+0.18}_{-0.09}$ & $0.17^{+0.04}_{-0.03}$ & $0.43^{+0.07}_{-0.07}$ & $0.73^{+0.03}_{-0.03}$ & 1.08\\
& F115W  & $0.71^{+0.09}_{-0.04}$ & $0.06 \pm 0.01$ & $0.92^{+0.04}_{-0.06}$ & $1.13^{+0.97}_{-0.34}$ & $0.26^{+0.11}_{-0.09}$ & $0.67^{+0.17}_{-0.23}$ & $0.67^{+0.06}_{-0.07}$ & 1.06\\
& F150W  & $1.09^{+0.33}_{-0.19}$ & $0.05 \pm 0.01$ & $0.76^{+0.06}_{-0.07}$ & $0.94^{+0.39}_{-0.18}$ & $0.27^{+0.07}_{-0.06}$ & $0.35^{+0.07}_{-0.06}$ & $0.45^{+0.07}_{-0.07}$ & 1.20\\
& F182M  & $0.69^{+0.06}_{-0.02}$ & $0.06 \pm 0.01$ & $0.85^{+0.03}_{-0.05}$ & $1.52^{+0.86}_{-0.47}$ & $0.26^{+0.19}_{-0.12}$ & $0.53^{+0.23}_{-0.25}$ & $0.88^{+0.06}_{-0.09}$ & 1.07\\
& F200W  & $0.91^{+0.22}_{-0.11}$ & $0.05 \pm 0.01$ & $0.87^{+0.06}_{-0.08}$ & - & - & - & - & 1.00\\
& F210M  & $0.87^{+0.19}_{-0.10}$ & $0.06 \pm 0.01$ & $0.79^{+0.08}_{-0.09}$ & - & - & - & - & 1.10 \\
& F277W  & $0.97^{+0.32}_{-0.19}$ & $0.06 \pm 0.01$ & $0.79^{+0.06}_{-0.08}$ & $0.96^{+0.52}_{-0.23}$ & $0.28^{+0.11}_{-0.09}$ & $0.33^{+0.09}_{-0.08}$ & $0.49^{+0.10}_{-0.10}$ & 0.95\\
& F300M  & $0.94^{+0.28}_{-0.15}$ & $0.05 \pm 0.01$ & $0.86^{+0.06}_{-0.08}$ & - & - & - & - & 1.01\\
& F335M  & $0.88^{+0.22}_{-0.12}$ & $0.05 \pm 0.01$ & $0.83^{+0.07}_{-0.08}$ & - & - & - & - & 1.00 \\
& F356W  & $0.71^{+0.09}_{-0.06}$ & $0.05 \pm 0.01$ & $0.86^{+0.06}_{-0.08}$ & $1.11^{+0.56}_{-0.30}$ & $0.29^{+0.11}_{-0.09}$ & $0.33^{+0.09}_{-0.08}$ & $0.61^{+0.08}_{-0.08}$ & 1.03 \\
& F360M  & $0.85^{+0.25}_{-0.13}$ & $0.05 \pm 0.01$ & $0.82^{+0.07}_{-0.09}$ & - & - & - & - & 1.00\\
& F410M  & $0.70^{+0.10}_{-0.06}$ & $0.05 \pm 0.01$ & $0.86^{+0.06}_{-0.08}$ & $1.17^{+0.63}_{-0.33}$ & $0.30^{+0.11}_{-0.09}$ & $0.32^{+0.09}_{-0.08}$ & $0.59^{+0.08}_{-0.08}$ & 1.03\\
& F444W  & $0.90^{+0.28}_{-0.14}$ & $0.05 \pm 0.01$ & $0.81^{+0.07}_{-0.08}$ & - & - & - & - & 1.07 \\
& F460M  & $0.91^{+0.27}_{-0.14}$ & $0.05 \pm 0.01$ & $0.79^{+0.07}_{-0.08}$ & - & - & - & - & 0.99\\
& F480M  & $0.95^{+0.53}_{-0.20}$ & $0.04 \pm 0.01$ & $0.86^{+0.07}_{-0.09}$ & - & - & - & - & 1.01\\
& F770W  & $1.31^{+0.78}_{-0.37}$ & $0.06 \pm 0.01$ & $0.85^{+0.07}_{-0.10}$ & - & - & - & - & 1.02\\
& F1000W & $1.47^{+1.31}_{-0.54}$ & $0.07^{+0.02}_{-0.01}$ & $0.74^{+0.09}_{-0.11}$ & - & - & - & - & 0.83\\
\hline
\multicolumn{10}{l}{\it \namejades} \\
\hline
& F090W   & $0.66 \pm 0.01$ & $0.07 \pm 0.01$ & $0.84^{+0.02}_{-0.02}$ & $1.32^{+0.38}_{-0.27}$ & $0.34\pm 0.05$ & $0.89^{+0.06}_{-0.10}$ & $0.67\pm 0.03$ & 0.98\\
& F115W  & $0.66^{+0.02}_{-0.01}$ & $0.07 \pm 0.01$ & $0.87 \pm 0.02$ & $1.52^{+0.39}_{-0.28}$ & $0.32 \pm 0.05$ & $0.93^{+0.04}_{-0.08}$  & $0.68 \pm 0.03$ & 0.97\\
& F150W  & $0.66 \pm 0.02$ & $0.06\pm 0.01$ & $0.87\pm 0.03$ & $1.52^{+0.39}_{-0.27}$ & $0.31^{+0.04}_{-0.03}$ & $0.93^{+0.04}_{-0.07}$ & $0.68\pm 0.03$ & 0.98\\
& F200W  & $0.67^{+0.03}_{-0.01}$ & $0.07 \pm 0.01$ & $0.91^{+0.04}_{-0.06}$ & $1.54^{+0.64}_{-0.37}$ & $0.22 \pm 0.07$ & $0.78^{+0.14}_{-0.23}$ & $0.72^{+0.06}_{-0.04}$ & 0.98\\
& F277W  & $0.69^{+0.05}_{-0.02}$ & $0.11 \pm 0.01$ & $0.98^{+0.01}_{-0.02}$ & $1.26^{+0.37}_{-0.26}$ & $0.45^{+0.12}_{-0.10}$ & $0.74^{+0.09}_{-0.11}$ & $0.70^{+0.03}_{-0.04}$ & 0.98\\
& F335M  & $0.66^{+0.02}_{-0.01}$ & $0.09 \pm 0.01$ & $0.96^{+0.02}_{-0.03}$ & $0.83^{+0.16}_{-0.08}$ & $0.26\pm 0.03$ & $0.91^{+0.04}_{-0.06}$ & $0.70^{+0.04}_{-0.03}$ & 1.02\\
& F356W  & $0.66 \pm 0.01$ & $0.09 \pm 0.01$ & $0.95^{+0.01}_{-0.02}$ & $0.74^{+0.10}_{-0.05}$ & $0.32\pm 0.03$ & $0.90^{+0.04}_{-0.05}$ & $0.76\pm 0.02$ & 1.00\\
& F410M  & $0.66 \pm 0.02$ & $0.11 \pm 0.01$ & $0.98^{+0.01}_{-0.03}$ & - & - & - & - & 0.96 \\
& F444W  & $0.66 \pm 0.01$ & $0.10 \pm 0.01$ & $0.94\pm 0.02$ & $0.75^{+0.10}_{-0.05}$ & $0.39\pm 0.03$ & $0.83^{+0.04}_{-0.05}$ & $0.81 \pm 0.01$ & 2.55\\
& F770W & $0.79^{+0.11}_{-0.06}$ & $0.07 \pm 0.01$ & $0.89 \pm 0.02$ & - & - & - & - & 1.42\\
&F1280W & $0.78^{+0.11}_{-0.06}$ & $0.07 \pm 0.01$ & $0.93 \pm 0.04$ & - & - & - & - & 0.87\\
\hline
\hline
\end{tabular}
\tablefoot{Effective radii are circularised and converted to arcseconds assuming $1\,{\rm pix}=60\,{\rm mas}$; $q = 1-\mathrm{ellip}$. For single-S\'ersic fits, component~2 is not applicable. For \namemiric , the reported values have been corrected for lensing magnification $\mu$.}
\end{table*}

\section{Corrected optical and NIR slopes}
\label{app:slopes}
For the optical slope of \namemiric, the F814W, F850LP, F090W and F115W filters are contaminated by the presence of strong emission lines (e.g., \hb+\oiii, \ha). 
Figure~\ref{fig:sed_cutouts_miric} shows a photometric bump at the observed wavelengths of the complex \hb+\oiii~($\approx 0.855$~\um) and the \ha~transition ($\approx 1.122$~\um). 
If not properly taken into account, the clear boost in the observed photometry due to these strong optical emission lines would bias our optical continuum slope towards bluer colors.
However, instead of discarding them, we account for the emission lines contamination by convolving the NIRISS spectrum (only emission lines) for the transmission curve of the different filters and remove the inferred fluxes from our total photometry.
Despite the fact that the emission lines observed with NIRISS are only part of the optical emission lines that can boost the observed flux of our galaxy, they are the main and most prominent optical transitions.
In Table~\ref{app-tab:photometry_combined}, we report the flux corrections we obtain by convolving the principal \hst\ and \jwst\ filters covering the rest-frame optical emission of our target.
From the fitting of the corrected photometry, we derive a $\beta_{\rm opt} = -2.18\pm 0.14$.
We note that without corrections the beta slope measured would have been $\beta_{\rm opt} = -3.07\pm 0.11$.

In the case of the NIR slope, the \paa~($\approx 3.208$~\um) would contaminate the F335M and F356W.
While we drop the F335M filter since medium bands are more prone to be affected by emission lines \cite[e.g.,][]{Papovich+23}, we do not discard the wide band filters since the \paa\ transition is typically one order of magnitude fainter than the \ha\ \cite[\paa/\ha~$=0.118$, assuming Case B recombination,][]{Osterbrock+06}. 
From the fitting, we derive $\beta_{\rm NIR} = 2.05\pm 0.29$.

In the case of \namejades\, the NIRSpec MSA/PRISM spectrum detects the target's continuum emission.
Therefore, we perform the measurements directly on the spectrum after masking the emission lines.
By doing so, we find $\beta_{\rm opt} = -2.00\pm 0.15$ and $\beta_{\rm NIR} = 3.12\pm 0.04$.
Nonetheless, as for \namemiric, we also estimate the contamination to the observed photometry due to the emission lines. 
To do so, we first model and remove the continuum emission from the spectrum, and then we convolve it with the different filter passbands.
As sanity check, we also decide to derive the $\beta$-slopes from the corrected photometry following the same steps as for \namemiric.
From the fitting, we derive $\beta_{\rm opt} = -1.80\pm 0.10$ and $\beta_{\rm NIR} = 2.31\pm 0.25$.
Without correcting the photometry, the derived rest-frame optical slope would have been $\beta_{\rm opt} = -2.40\pm 0.07$.

While the $\beta_{\rm opt}$ derived from the spectrum and from the corrected photometry are quite in good agreement with each other (within 2$\sigma$), the difference between the rest-frame NIR slopes is significant. 
As shown by Figure~\ref{fig:sed_cutouts_jades}, this flattening of the photometric slope is mainly driven by the F277W filter that, at the target's redshift, covers the V-shaped turnover.
For this reason the F277W magnitude is boosted towards higher values due to the partial coverage of the rising part of the target' SED.
In fact, if we remove F277W from the fitting we obtain $\beta_{\rm NIR} = 3.54\pm 0.60$, in agreement (within 1$\sigma$) with the spectral estimate.

\section{Comparison to empirical templates}
\label{app:empirical_templates}
We compare the extracted photometry of our targets with 166 empirical galaxy templates of nearby/low-z galaxies reported by \cite{Polletta+07, Brown+14, Brown+19,  Akins+25, Trefoloni+25, Lyu+17, Lyu+18}.
We also add the comparison with the single stellar population models from {\sc bpass} \cite[v. 2.2;][]{Stanway+18}.
The {\sc bpass} models were selected having a Chabrier IMF with cut-off mass at 300 $M_\odot$, ages of 1, 5, 10, 100, 1000 Myr, solar and sub-solar metallicities (0.2, 1 $Z_\odot$) and allowing for binary models. 
Before applying redshift and normalising each model to the photometry of \namemiric\ in the F277W filter, we also generate models applying dust extinction according the attenuation law by \cite{Calzetti+00} and assuming $E(B-V) = 0.1, 0.5, 1, 1.5$.
In total we generate 50 {\sc bpass} single stellar population models.

\begin{figure*}
    \centering
    \includegraphics[width=\textwidth]{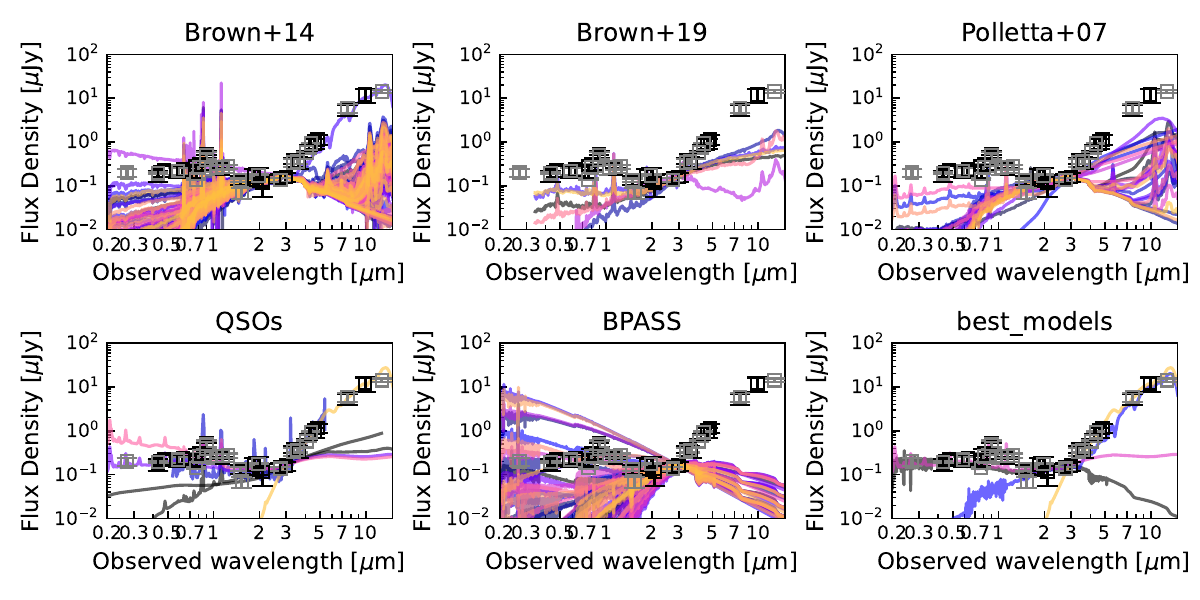}
    \caption{Panels showing the comparison between the photometry of our targets \namemiric\ (in black) and \namejades\ (in grey) and a set of observed empirical galaxy templates among which: the nearby/low-$z$ quiescent, star-forming, starburst and AGN-hosting galaxies by \cite{Brown+14, Brown+19, Polletta+07}, a selection of QSO templates from \cite{Lyu+17, Lyu+18, Trefoloni+25}, the maximal LRD SED model by \cite{Akins+25} and a set of 50 single stellar population models from \textsc{bpass} \citep{Stanway+18} with different ages, metallicities and dust obscuration. The bottom right panel shows a selection of the best fitting templates to our targets photometry.}
    \label{app-fig:empirical_templates}
\end{figure*}

A visual comparison (see Figure~\ref{app-fig:empirical_templates}) between the photometry of our two targets (black and grey squared markers) and the empirical models show that the only models that can reproduce the steep red slope of our objects are the ULIRG IRAS 08572+3915 from \cite{Brown+14} and the strongly obscured QSO model by \cite{Lyu+18} with an optical depth $\tau = 20$ implying an heavily buried type II AGN. 
Interestingly, among the models that better reproduce the shortest wavelengths ($< 3~\mu$m observed) one of the best templates is a QSO model with an optical depth $\tau = 0.5$ representative of an optically thin and marginally obscured type I AGN. 
Also the \textsc{bpass} model of a 5~Myr old single stellar population with solar metallicity and $E(B-V) = 0.1$ can well reproduce the observed short-wavelength \hst\ + \jwst/NIRCam photometry of \namemiric\ and \namejades.

\section{Dust extinction estimates}
In Table \ref{tab-app:Av_estimates} we present the estimates of the nebular reddening due to dust derived for our targets assuming both the \cite{Calzetti+00} attenuation law and the SMC Bar law by \cite{Gordon+03}.

\begin{table*}
\caption{Nebular reddening estimates for \namemiric\ (M0416) and \namejades\ (GOODS-N).}
\label{tab-app:Av_estimates}
\centering
\small
\begin{tabular}{llcccc}
\hline\hline
 & & \multicolumn{2}{c}{Calzetti+00} & \multicolumn{2}{c}{SMC Bar (Gordon+03)} \\
  & & $E(B - V)_{\rm neb}$ & $A_V$ [mag]
       & $E(B - V)_{\rm neb}$ & $A_V$ [mag] \\
\hline
\multicolumn{6}{l}{\it \namemiric} \\
\hline
& H$\alpha$/H$\beta$   & $0.46 \pm 0.36$ & $1.85 \pm 1.47$
                     & $0.53 \pm 0.42$ & $1.44 \pm 1.14$ \\
& Pa$\alpha$/H$\alpha$ & $0.06 \pm 0.05$ & $0.23 \pm 0.19$
                     & $0.10 \pm 0.09$ & $0.28 \pm 0.23$ \\
& Br$\gamma$/H$\alpha$ & $0.12 \pm 0.11$ & $0.50 \pm 0.43$
                     & $0.21 \pm 0.17$ & $0.56 \pm 0.48$ \\
\hline
& {\bf Combined (NIR)} & $\mathbf{0.07 \pm 0.04}$ & $\mathbf{0.27 \pm 0.17}$
                     & $\mathbf{0.12 \pm 0.08}$ & $\mathbf{0.33 \pm 0.21}$ \\
\hline
\multicolumn{6}{l}{\it \namejades} \\
\hline
& H$\alpha$/H$\beta$   & $0.01 ^{+0.02}_{-0.01}$ & $0.02^{+0.71}_{-0.02}$
                     & $0.01^{+0.20}_{-0.01}$ & $0.02^{+0.55}_{-0.02}$ \\
& Pa$\alpha$/H$\alpha$ & $0.05 \pm 0.02$ & $0.19 \pm 0.10$
                     & $0.09 \pm 0.04$ & $0.24 \pm 0.12$ \\
& Br$\gamma$/H$\alpha$ & $ \leq 0.06$ & $\leq 0.23$
                     & $\leq 0.09$ & $\leq 0.25$ \\
\hline
& {\bf Adopted (NIR)} & $\mathbf{0.05 \pm 0.02}$ & $\mathbf{0.19 \pm 0.10}$
                     & $\mathbf{0.09 \pm 0.04}$ & $\mathbf{0.24 \pm 0.12}$ \\
\hline
\end{tabular}
\tablefoot{Nebular reddening estimates derived from different hydrogen recombination-line decrements assuming Case~B intrinsic ratios.
For each decrement we report the inferred $E(B-V)_{\rm neb}$ and the corresponding visual attenuation $A_V$ obtained using either the \citet{Calzetti+00} attenuation law ($R_V=4.05$) or the SMC Bar extinction curve of \citet{Gordon+03} ($R_V=2.74$).
The last rows list the inverse-variance weighted combination of the NIR decrements (Pa$\alpha$/H$\alpha$ and Br$\gamma$/H$\alpha$), adopted as the best estimate for correcting nebular line fluxes.}
\end{table*}

\end{appendix}

\end{document}